# Characterisation of the ALice Tpc ReadOut chip

Dissertation
zur Erlangung des Doktorgrades
der Naturwissenschaften

vorgelegt beim Fachbereich 13 (Physik)
der Johann Wolfgang von Goethe - Universität
in Frankfurt am Main

von
Roland Bramm

Frankfurt 2005
(D F 1)

# Content

# Introduction

# Introduction

At CERN, a new accelerator is built which opens up the access to higher energies, higher luminosities and more manpower in consequence of increased size, increased technical requirements, increased synergy of projects and increased use of human resources. This machine is compressed in the short name »LHC« (**L**arge **H**adron **C**ollider) and is the successor over the LEP (**L**arge **E**lectron **P**ositron Collider), which was removed for building the LHC.
The genealogy of heavy ion accelerators starts with the completed Bevatron/Bevalac at LBL (**L**awrence **B**erkeley **L**abs), over the AGS (**A**lternating **G**radient **S**ynchrotron) at BNL (**B**rookhaven **N**ational **L**abs), followed by the still running SPS (**S**uper **P**roton **S**ynchrotron) at CERN and RHIC (**R**elativistic **H**eavy **I**on **C**ollider) at BNL and the currently built LHC as the newest and most powerful generation. Updates on the AGS and SPS added the capability to accelerate heavy ions, which the RHIC and the LHC can do on purpose. The interest in heavy ion collisions is driven by their capability to produce compressed baryonic matter at densities several times higher then the ground state density. Additionally, the search for the QGP (**Q**uark **G**luon **P**lasma) as the only phase transition predicted by the standard model which is reachable by laboratory experiments is of large interest not only by heavy ion physicist but also for the cosmology. The existence of the QGP was observed by several experiments at the SPS [1] and later confirmed by RHIC [2] experiments.
As the energy is increasing over the stated accelerators also one of the most fundamental observables, the particle multiplicity, is increasing. At LHC, a multiplicity of 1500 to 6000 charged particles per unit of rapidity is expected. As this quantity will only definitely be known with the inspection of the first events, the detectors have to be built according to the highest expected multiplicity. The large uncertainties in the multiplicity are due to the fact that many models with different theoretical bases are existing and their predicted multiplicity deviates. To measure this huge amount of tracks a TPC (**T**ime **P**rojection **C**hamber) is a nicely fitting detector as it was already used in heavy ion experiments before with NA49 at the SPS and STAR at RHIC, as examples. It provides tracking capability of charged particles and particle identification over a large volume without a big amount of material.
At the LHC, ALICE is the experiment dedicated to study heavy ion collisions with a TPC as main tracking detector embedded into other detectors like a silicon based vertex tracker or a transition radiation detector. The two endcaps of the TPC are realised as multiwire proportional chambers in a segmentation of over 560000 pads producing up to 100 GByte of raw data per second. This huge data amount has to be reduced directly on the detector without loss of information. Therefore each pad is connected to a chain of signal processing tasks as the signal has to be integrated, amplified, shaped, digitised, processed, compressed and transferred. This huge amount of tasks is all done by the on detector electronics, mainly by the PASA (**Pre**amplifier/**S**h**a**per) and ALTRO (**AL**ICE **T**PC **R**ead**o**ut) chips. The task of the ALTRO is the digitalisation of the signal, several processing steps like baseline correction, ion tail cancellation, zero suppression, data formatting and temporary storage, altogether combined in one chip consisting of 16 channels. The transfer and first data stream merging is done with the RCU (Readout Controller Unit) and sent via an optical fibre to the data acquisition.
The central point of this thesis is the ALTRO in conjunction with the TPC. Due to the fact that the final detector is not yet finished, a prototype of the TPC with the final electronics was the central data source as well as the test object to implement tasks like online monitoring, configuration and data acquisition. The focus is set on one hand to develop and test procedures to extract the configuration data for the ALTRO and check their performance and on the other hand to understand detector effects.
This thesis starts in the chapter »The Experiment« with a short description on the ALICE detector, followed by a more detailed characterisation of the working principle of a TPC in correlation with the present implementation of the ALICE TPC, but with particular interest in supporting the gas choice.
The next topic »Front End Electronics« is the delineation of the complete on-detector electronics subdivided logically and physically. Physically, starting with the front end card, the subsequent backplane and the RCU with its daughter boards. Logically, starting with a small sketch of the PASA and an extensive definition of the ALTRO and its internal processing units.
In »Jitter«, a simulation on the necessary clock accuracy of the front end electronics is described including the expected error in the signal measurement.
The chapter »Prototype Environment« outlines the different setups of the prototype TPC in cosmic ray and beam running, the progression in the data format and storage, the development of an online TPC pad monitor for all setups and the configuration procedures.
The extraction of the configuration parameters of the ALTRO are stated in »ALTRO Parameter Optimisation«. Schemes for the extraction of the needed parameters of the ALTRO are developed. For the parameter calculation of the pedestal memory and the tail cancellation these schemes were implemented and tested. This includes a calculation of the requirements in CPU time to extract the parameters and bandwidth to configure the ALTRO. A bit-exact software emulation of the ALTRO digital chain for the testing of the parameters is implemented and described in »ALTRO++«.
Finally, an analysis of the complete signal including the ion tail, which is induced by ions drifting to different targets in the readout chamber, is shown. This is the first analysis which could quantify the spread in the variations of the avalanche to avalanche creation, and educed that the angle of incidence of individual primary electrons are not playing a significant role in the signal shape.
In the end, the »Résumé« presents a summary on the results in parameter extraction and signal creation as well as a perspective for the future plans.

# The Experiment

## The Experiment

ALICE (**A** **L**arge **I**on **C**ollider **E**xperiment) [1-5] is an experiment at the LHC (**L**arge **H**adron **C**ollider) with the goal to study heavy ion collisions up to the top energy available. It is designed to study the physics of strongly interacting matter and the QGP (**Q**uark **G**luon **P**lasma). The experimental setup with its various subdetectors is shown in the picture below.
In general, the detectors are highly capable to measure and identify hadrons, leptons and photons around mid rapidity over a broad range from very low (100 MeV) up to fairly high (100 GeV) momenta. In addition, there is one myon arm ⑩ [6], which covers the detection of myons at large rapidities ($-4 < \eta < -2.4$). In a moderate magnetic field of up to 0.5 T provided by the reused and modernised solenoidal L3 Magnet ① [7] are the central detectors positioned which are covering the mid rapidity region ($-0.9 < \eta < 0.9$). A big part of ALICE also covers 360° in $\phi$ in this region. For the tracking the main detectors the ITS (**I**nner **T**racking **S**ystem, ②) [8] as a silicon based detector, the TPC (**T**ime **P**rojection **C**hamber, ③) [9] and a highly granular TRD (**T**ransition **R**adiation **D**etector, ④) [10] are used. This set of detectors is called »central barrel«. For particle identification the TPC measurement of the energy loss (dE/dx), the transition radiation of the TRD and the time of flight of the TOF (**T**ime **O**f **F**light, ⑤) [11] is used. In addition, there is the HMPID (**H**igh **M**omentum **P**article **Id**entification **D**etector, ⑥) [12] for high momentum particles, and a photon spectrometer PHOS (**Pho**ton **S**pectrometer, ⑦) [13] for photon measurements. These two detectors only cover a small fraction in $\phi$ of 60° and 100° respectively. There are fast detectors for the trigger at large rapidities like the FMD (**F**orward **M**ultiplicity **D**etector, ⑧) [14,15], V0 and T0 at ($-3.4 < \eta < -5.1$) [15-17] for measuring charged particles and a narrower band for photons ($2.3 < \eta < 3.5$) with the PMD (**P**hoton **M**ultiplicity **D**etector, ⑨) [18]. At last, there are two ZDC (**Z**ero **D**egree **C**alorimeters) [19], located 120 m away from the interaction point, to measure the spectator nucleons at beam rapidity.

## TPC

A time projection chamber (TPC) provides a complete, 3D picture of the ionisation deposited in a gas volume. It acts similar like a bubble chamber, however with a fast and purely electronic readout. This 3D »imaging« capability defines the usefulness as a tracking device in a high track density environment and for the identification of particles through their ionisation energy loss (dE/dx). Therefore it is the main tracking detector in the central barrel of the ALICE experiment. The usage as a large acceptance tracking and particle identification detector in heavy ion experiments starts with NA49 [20] and STAR (**S**olenoidal **T**racker **A**t **R**HIC) [21] at the SPS (**S**uper **P**roton **S**ynchrotron) and RHIC (**R**elativistic **H**eavy **I**on **C**ollider), respectively.
A TPC consists of mainly three parts, the drift chamber volume, the readout chambers and the front end electronics. The field cage surrounds the detector gas and provides a homogeneous electrical field to transport the electrons

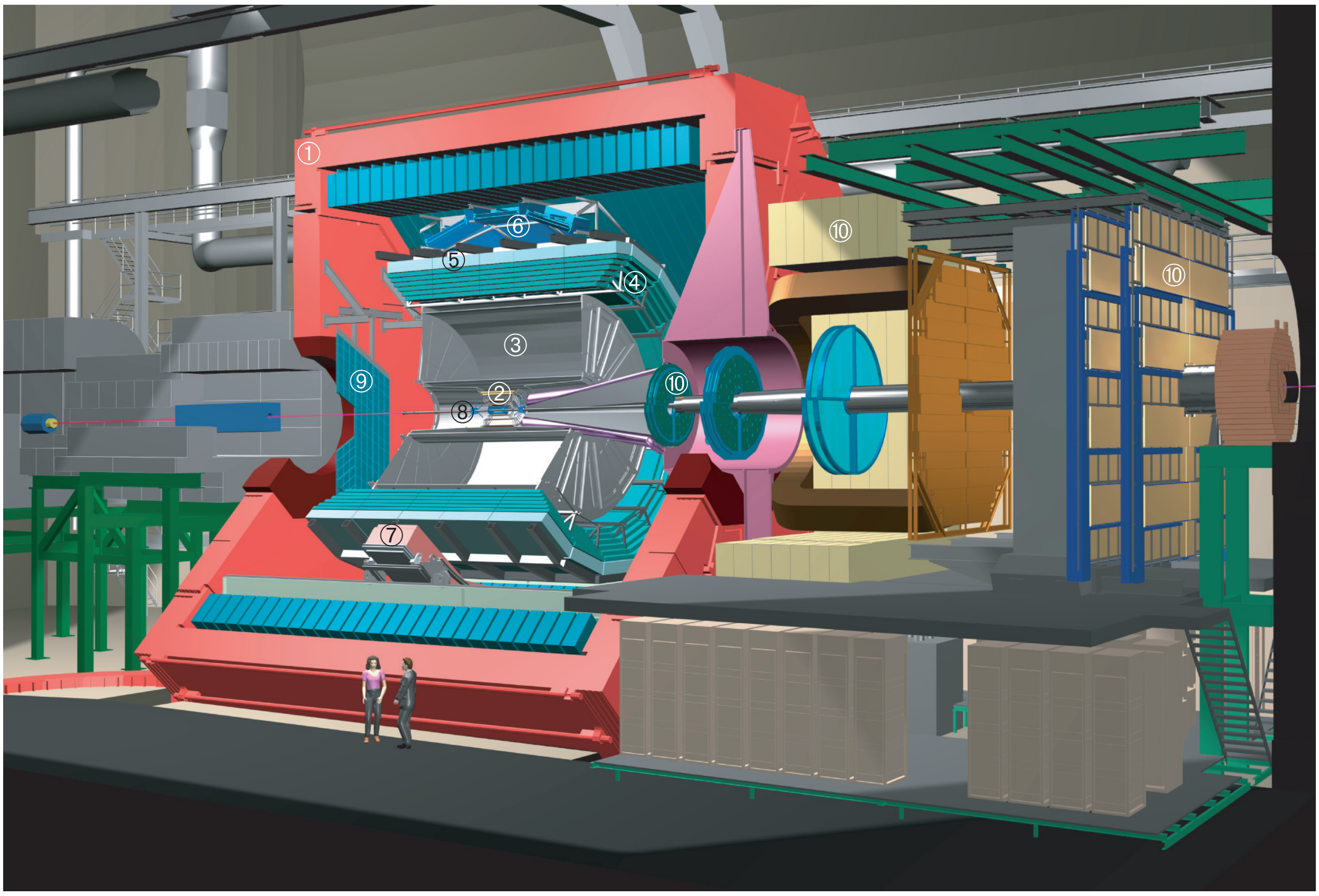

The ALICE Experiment: ① L3 Magnet, ② ITS, ③ TPC, ④ TRD, ⑤ TOF, ⑥ HMPID, ⑦ PHOS, ⑧ FMD, ⑨ PMD, ⑩ Myon Arm

of the ionisation to the readout chambers. This is also the sensitive volume of a TPC. The readout chamber strengthens the signal and provides the coupling in between the gas and the front end electronics. Here the signal is again amplified, shaped, digitised, processed, stored and then transferred to the data acquisition.

### Requirements

The physics program foreseen [5,9] determines the exigencies at the ALICE TPC alone and in conjunction with other detectors. The hadron physics demands:

- Two Track Resolution: The two track resolution has to be sufficient to allow a HBT [22] measurement with a resolution in relative momentum of a few (< 5) MeV.
- dE/dx Resolution: The dE/dx resolution should be at least 8% or better to properly identify hadrons.
- Track Matching: For fast decaying particles a proper (85% - 95%) matching capability of the TPC to ITS or TOF or both is needed.

For leptonic observables the demands are partially differing:

- Tracking Efficiency: Since electron pairs are most interesting, a tracking efficiency of at least 90% for tracks at $p_t > 1$ GeV should be achieved.
- Momentum Resolution: To get a good mass resolution (< 100 MeV) for heavy mesons like the Y, the momentum resolution for electrons of about 4 GeV should be at least 2.5%.
- dE/dx Resolution: For electrons the dE/dx resolution should be better than 10%. In cooperation with the TRD, this leads to a electron-to-pion separation of more than a factor of 1000.
- Rate Capability: For the inspection of electrons the TPC should work at 200 Hz when taking heavy ion collisions.

For the proton running of ALICE the demands are partially lower because of the low multiplicity but on the other hand higher since the TPC has to run at a higher rate due to the high luminosity and the need of high statistics for rare signals.

- Rate Capability: Due to the high luminosity in the proton running the TPC has to operate at 1 kHz or more.

These demands lead to a design of a quite conventional TPC, but with many new solutions in detail. The major facets are:

- Material Budget: To minimise the effect of multiple scattering and secondary particle production the material amount should be minimised. This determines the light field cage material as well as the gas choice.
- Field Cage: To match the rate exigencies the field cage has to provide a high field of 400 V/cm, which implies a voltage greater than 100 kV at the central membrane.
- Acceptance: The acceptance of the TPC matches the one of the ITS, TRD and TOF. For event-by-event studies as well as for all rare observables a reasonably big acceptance is necessary to collect enough statistics. This leads to a size as shown in the table at the end of this section.
- Readout Chambers: The readout chambers cover an area of 33 $m^2$ at the two endcaps of the field cage and are built as conventional multiwire proportional chamber. To fulfill the necessary accuracy in dE/dx and position resolution, as well as double track resolution, there will be about 560000 readout pads.
- Electronics: The electronics for these 560000 pads has to reside as close as possible to the readout chambers to avoid transporting the analog signals over big distances, this demands a highly integrated system.
- Intelligent Readout: Even after the zero suppression directly in the detector electronics an event is still 60 MByte in size. The data throughput, when reading out at the highest detector readout rate, exceeds the allowed throughput to a permanent storage by roughly a factor of 10. To get the highest acquisition rate for special events (e.g. high momentum jets [23], Y particle [24], away side correlations [25-27]) a HLT (**H**igh **L**evel **T**rigger) [28-33] is foreseen to find candidates for these events online.

| | | |
|---|---|---|
| Size | Length | 5 m |
| | Inner Radius | 80 cm |
| | Outer Radius | 250 cm |
| Gas | Composition | Ne/$CO_2$ 90/10 |
| | Volume | 88 $m^3$ |
| | Drift Field | 400 V/cm |
| | Drift Velocity | 2.85 cm/$\mu$s |
| | Drift Time | 88 $\mu$s |
| FEE | #channels | 557568 |
| | Signal/Nosie | 30:1 |
| | Dynamic Range | 900:1 |
| | Noise (ENC) | 1000 $e^-$ |
| | Crosstalk | < 0,3% / -60 db |
| | Power Consumption | < 100 mW |
| | Max. Dead Time | 10% |
| Event Size | Pb-Pb central | 85 MByte |
| | p-p | 1-2 MByte |
| Trigger Rate | Pb-Pb central | 200 Hz |
| | p-p | 1000 Hz |

Technical data of the ALICE TPC [5,9]

## Working principle

Starting from a particle traversing the gas of the drift volume it ionises the gas molecules, so that a track of ions remains along the particle trajectory. The electrical field applied by the fieldcage now lets the electron cloud drift with a constant velocity in field direction, away from the central membrane towards the two readout planes. There, the signal will be amplified by avalanche creation and read out at the pad plane. The front end electronics then electrically amplifies, shapes and digitises the signal. The x and y coordinate are defined by the pad and the row coordinate in the readout chamber. The z coordinate is defined as the drift time of the electron cloud.

### Gas Ionisation

A charged particle travelling a gas can ionise gas atoms and thereby produce primary electrons. The statistics of the primary interactions implies a Poisson distribution of a number of primary electrons as shown below.

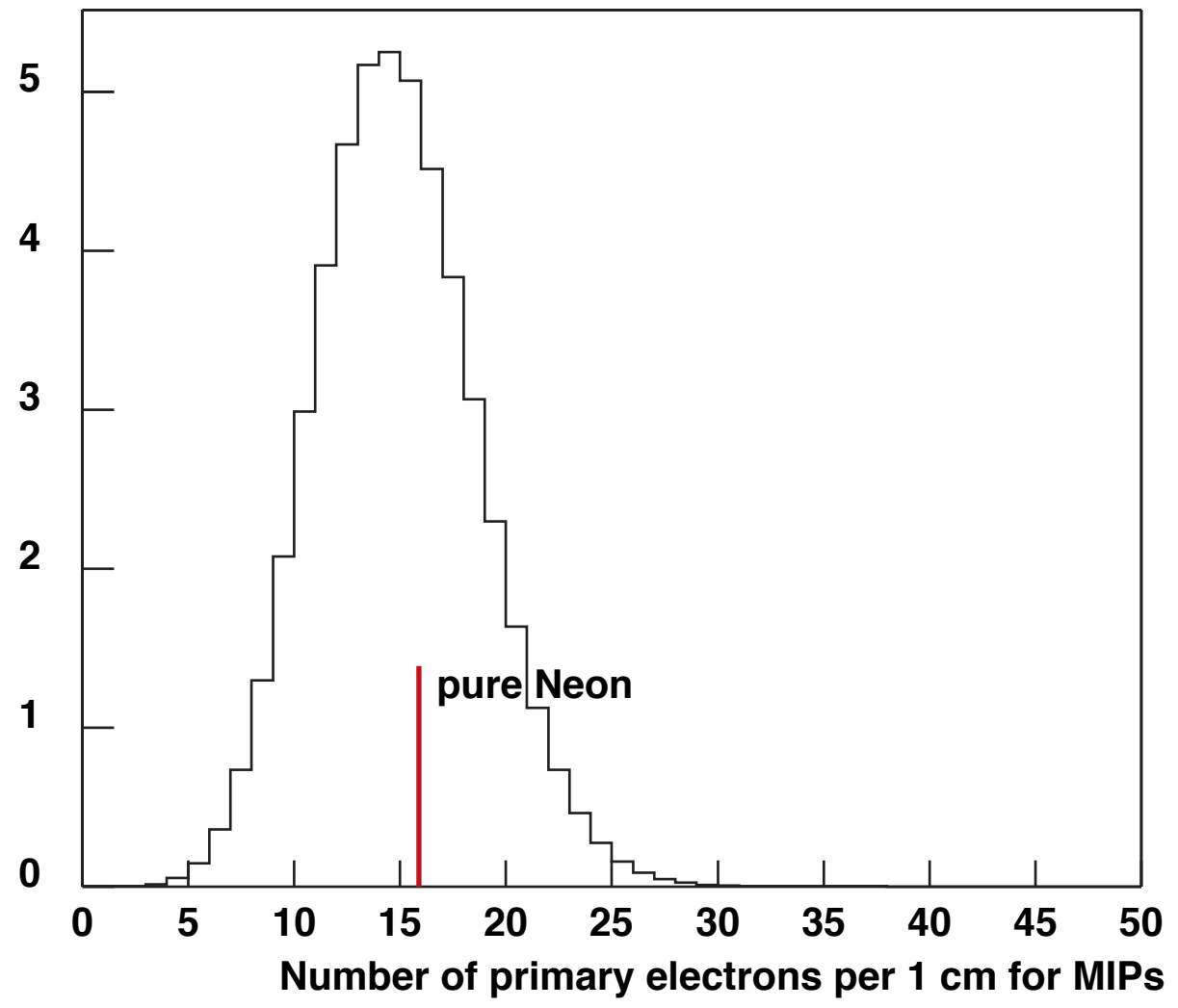

Distribution of the number of primary electrons per 1 cm for a Minimum Ionising Particle in $Ne/CO_2$ 90/10 [9]. The red line indicates the most probable number of primary electrons of pure neon as shown in the table to the right.

The distance between collisions is described by an exponential [34]

$$P(l) = \frac{1}{\lambda} e^{\frac{-l}{\lambda}} \tag{1}$$

with l as the distance between two successive collisions and λ as the mean distance between primary ionisations

$$\lambda = \frac{1}{N_{prim} \cdot f(\beta\gamma)} \tag{2}$$

where $N_{prim}$ is the number of primary electrons per centimetre produced by a MIP (**M**inimum **I**onising **P**article) and f(βγ) the Bethe Bloch curve [35,36]:

$$\frac{dE}{dx} = \frac{4\pi N e^4}{mc^2\beta^2} z^2 \left( ln \frac{2mc^2\beta^2\gamma^2}{I} - \beta^2 \right) \tag{3}$$

Based on the parametrisation proposed by the ALEPH (**A**pparatus for **LE**P **Ph**ysics) [34] collaboration

$$f(\beta\gamma) = \frac{P_1}{\beta^{P_4}} \cdot \left\{ P_2 - \beta^{P_4} - ln\left[ P_3 + \frac{1}{(\beta\gamma)^{P_5}} \right] \right\} \tag{4}$$

with the parameters $P_1 = 0.762 \bullet 10^{-1}$, $P_2 = 10.632$, $P_3 = 0.134 \bullet 10^{-4}$, $P_4 = 1.863$ and $P_5 = 1.948$, the energy loss data for the gas mixture of $Ar/C_{H4}$ 90/10 (90% Argon and 10% Methane ($CH_4$)) [34,37] is shown in the figure below.

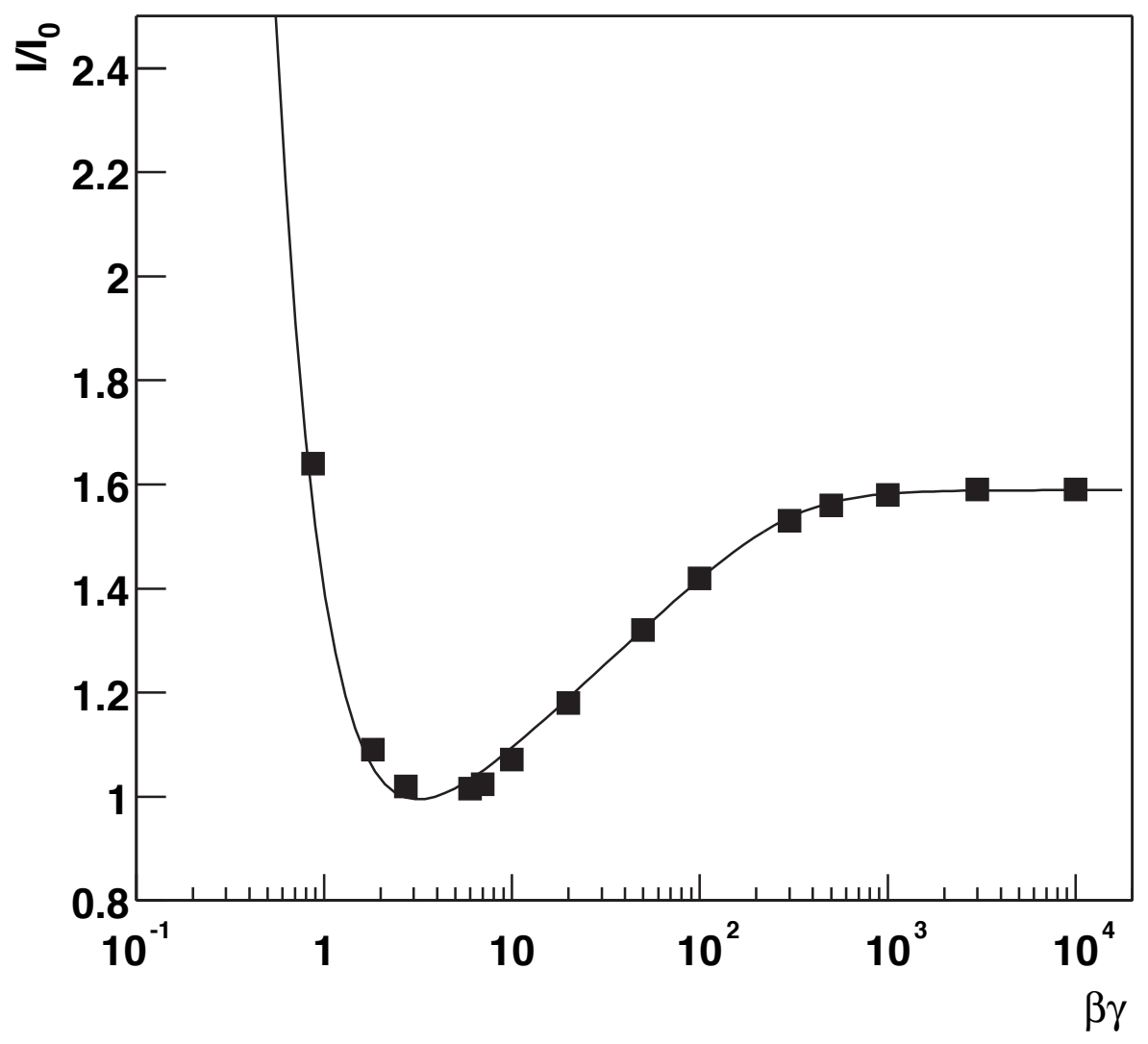

Bethe Bloch curve for $Ar/CH_4$ 90/10 [9], data from [34]

This is also used in the simulation due to the lack of energy loss data in the $1/\beta^2$ region and of the behaviour of Neon being quite similar to the Argon based mixtures [37].
With sufficient energy the primary electron can ionise atoms and therefore produce additional secondary electrons. The total number of electrons in an electron cluster is described by:

$$N_{tot} = \frac{E_{tot} - I_{pot}}{W_i} + 1 \tag{5}$$

with $E_{tot}$ as the energy loss in a given collision, $W_i$ the effective energy required to produce an electron-ion pair and $I_{pot}$ is the first ionisation potential. These clusters are treated pointlike, so that primary and secondary electrons are treated indifferently. This is justified because the effective range of low energy electrons is small [9].

| Gas | ρ[g/l] | $X_0$[m] | $n_{mp}$[1/cm] |
|---|---|---|---|
| Helium | 0.1785 | 5280 | 2.7 |
| Neon | 0.89990 | 322 | 16 |
| Argon | 1.784 | 110 | 38 |
| Krypton | 3.733 | 30.4 | 63 |
| Xenon | 5.887 | 14.4 | 115 |

Parameters of the noble gases used in TPCs [38], ρ is the density, $X_0$ is the radiation length and $n_{mp}$ as the most probable number of primary electrons per cm in the gas.

To optimise the signal-to-noise ratio the number of produced electrons should be as high as possible, which would lead to the usage of a heavier gas, but also an increasing space charge due to the larger number of electron/ion pairs produced and a lower ion mobility which is leading to a higher space charge.

Taking the maximum multiplicity expected at LHC of dn/dy ~ 6000 [23] into account the heavy gases are ruled out. Finally, since the TPC is the second innermost detector, the material budget should be minimised, so this speaks in favour for a light gas. Additionally, a light gas shows a lower multiple scattering.

### Electron/Ion Drift

Due to the influence of the homogeneous electric field provided by the fieldcage, the electron cloud moves with a constant speed towards the readout chambers. The drift speed $v_D$ is a dynamical equilibrium of the acceleration due to the drift field and the deceleration due to the collisions with the gas atoms. The drift speed $v_D$ is:

$$v_D = \frac{e}{\sqrt{2m_e}} \cdot \frac{1}{\sigma(\epsilon)\sqrt{\epsilon}} \cdot \frac{E}{N} \tag{6}$$

with E the electric field and N the density of the gas. The drift speed changes with the effective cross section $\sigma(\epsilon)$ depending of the kinetic energy of the electrons.

The drift speed as a function of the field is shown in the figure below for different noble gases. The drift speed has to be quite high to allow the needed readout rates of the TPC. For a drift time of 88 µs the drift field would be far beyond 1 kV/cm when using a noble gas alone.

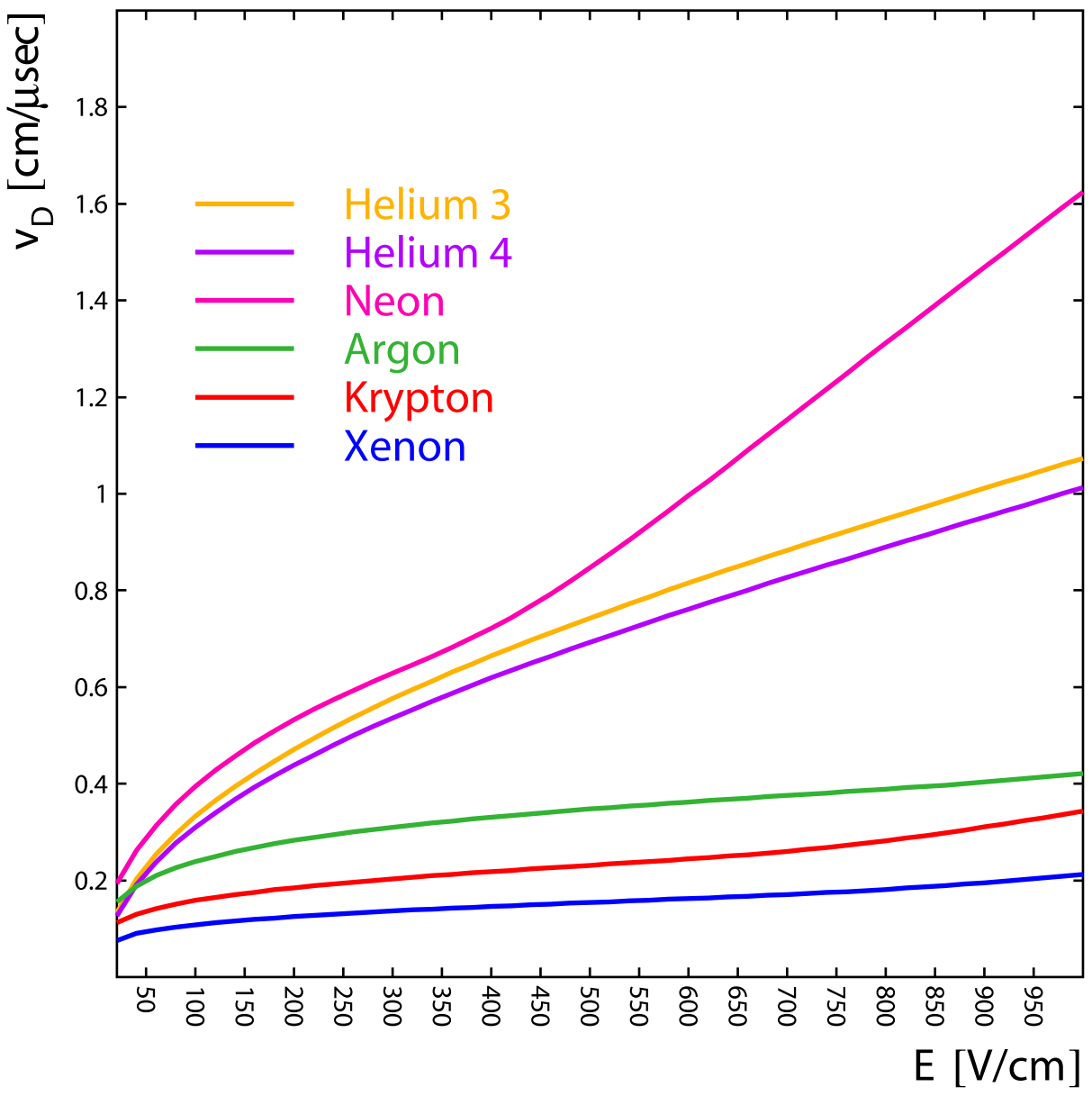

Pure noble gases have a low drift speed [38]

When adding minute amounts of $CO_2$ the drift speed increases, since the cross section increases. This is described by the peculiarities of the cross sections of the components as described in [38]. The plot in the next column is showing the velocity increase when adding $CO_2$ to neon. This leads to a field of 400 V/cm to reach the drift time of 88 µs as indicated for Ne/$CO_2$ 90/10.

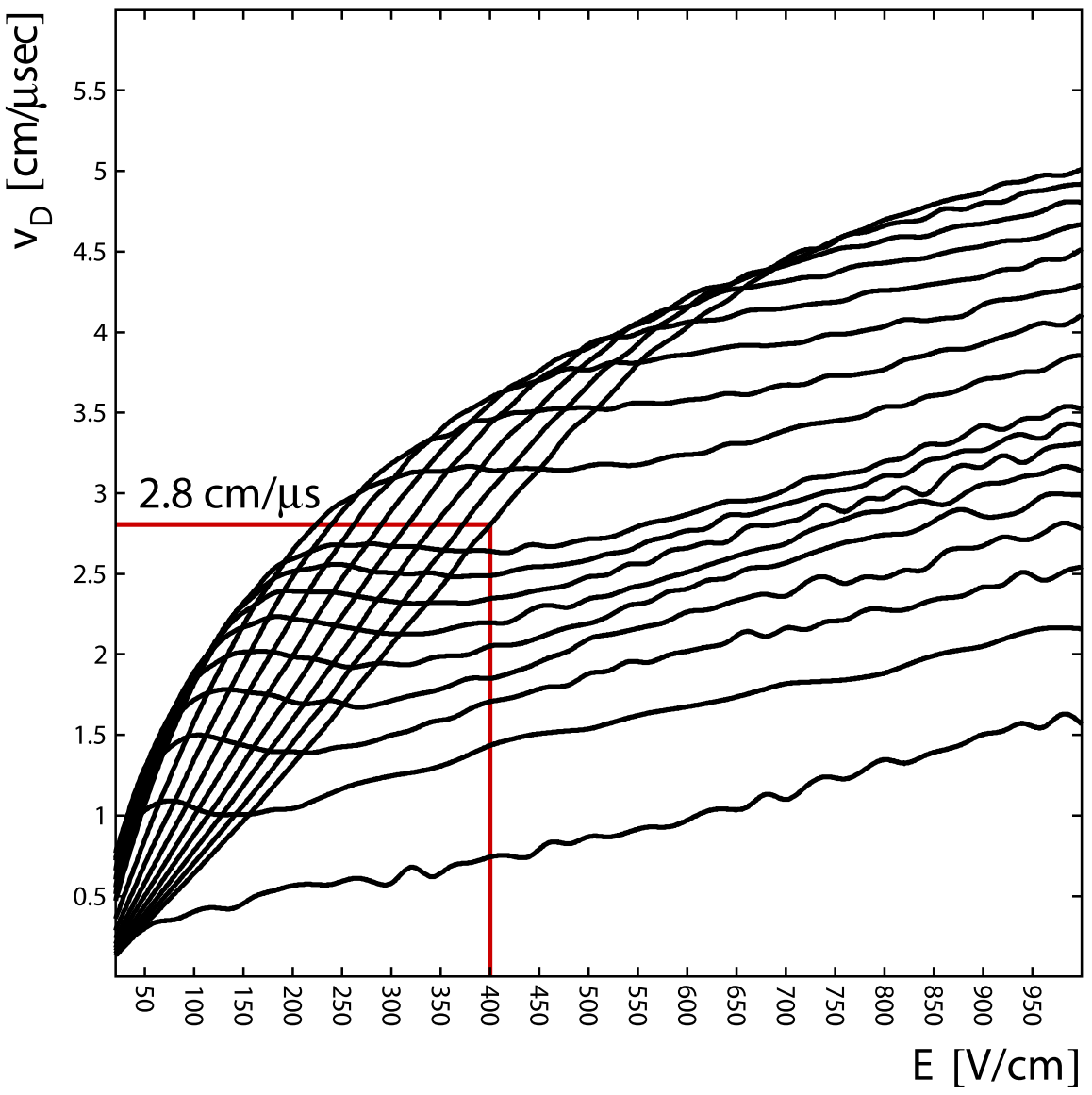

When adding minimal amounts of $CO_2$ to neon, the drift velocity at low fields of this gas increases rapidly. The lowest curve is for the pure gas, the following for 0.25%, 0.5%, ..., 1.75%, 2%, 3%, ..., 9%, 10% [38], respectively.

Ions drift at a much lower speed (several orders of magnitude slower than the electrons) in the opposite direction towards the central membrane. The mobility of Neon is 2.5 times larger than of Argon. Helium has an extremely high drift speed due to its light mass, but it is difficult to be contained in a detector due to its high leak rate.

### Electron Diffusion

The drift speed of one single electron differs from the mean motion of the electron cloud due to the statistical process of the scattering. These electrons follow a thermal energy distribution (Maxwell distribution) [39]:

$$F(e) = \sqrt{\frac{4\epsilon}{\pi k^3 T^3}} \cdot e^{\frac{-\epsilon}{kT}} \tag{7}$$

The mean thermal energy is defined by the integral:

$$\langle e \rangle = \int_0^\infty \epsilon F(\epsilon)\, d\epsilon = \frac{3}{2} kT = \bar{\epsilon} \tag{8}$$

According to this effect, the electron cloud will widen up during the drift time. Starting with a point-like electron cloud at t = 0 and the assumption of constant broadening the cloud will get a Gaussian-shaped density distribution:

$$n = \left( \frac{1}{\sqrt{4\pi Dt}} \right)^3 \cdot e^{\frac{-r^2}{4Dt}} \tag{9}$$

with

$$r^2 = x^2 + y^2 + (z - vt)^2 \tag{10}$$

and D as the diffusion coefficient calculated via the use of the mean free path:

$$D = \frac{1}{3}\,\overline{v}\;\lambda(\epsilon) \tag{11}$$

The width of (9) is

$$\sigma_x = \sqrt{2Dt} = \sqrt{\frac{2DL}{\mu E}} = \sqrt{\frac{4\epsilon L}{3eE}} \tag{12}$$

when using

$$v = \mu E \tag{13}$$

and the Nernst-Townsend formula [40]:

$$\frac{D}{\mu} = \frac{kT}{e} \tag{14}$$

To get a small $\sigma_x$ at high drift fields, small electron energies are required. In Argon or Neon, a field strength of 1 V/cm already produces electron energies larger than the thermal energy, so Argon is called a »Hot Gas«. On the contrary, for $CO_2$ this behaviour occurs at fields of 2 kV/cm, so it is a »Cold Gas«. The reason is a large energy loss due to the internal degrees of freedom which are already accessible at low collision energies. In the ALICE gas mixture Neon is foreseen. To reduce the effect of the diffusion $CO_2$ is added. The longitudinal and transversal diffusion is shown in the following plots.

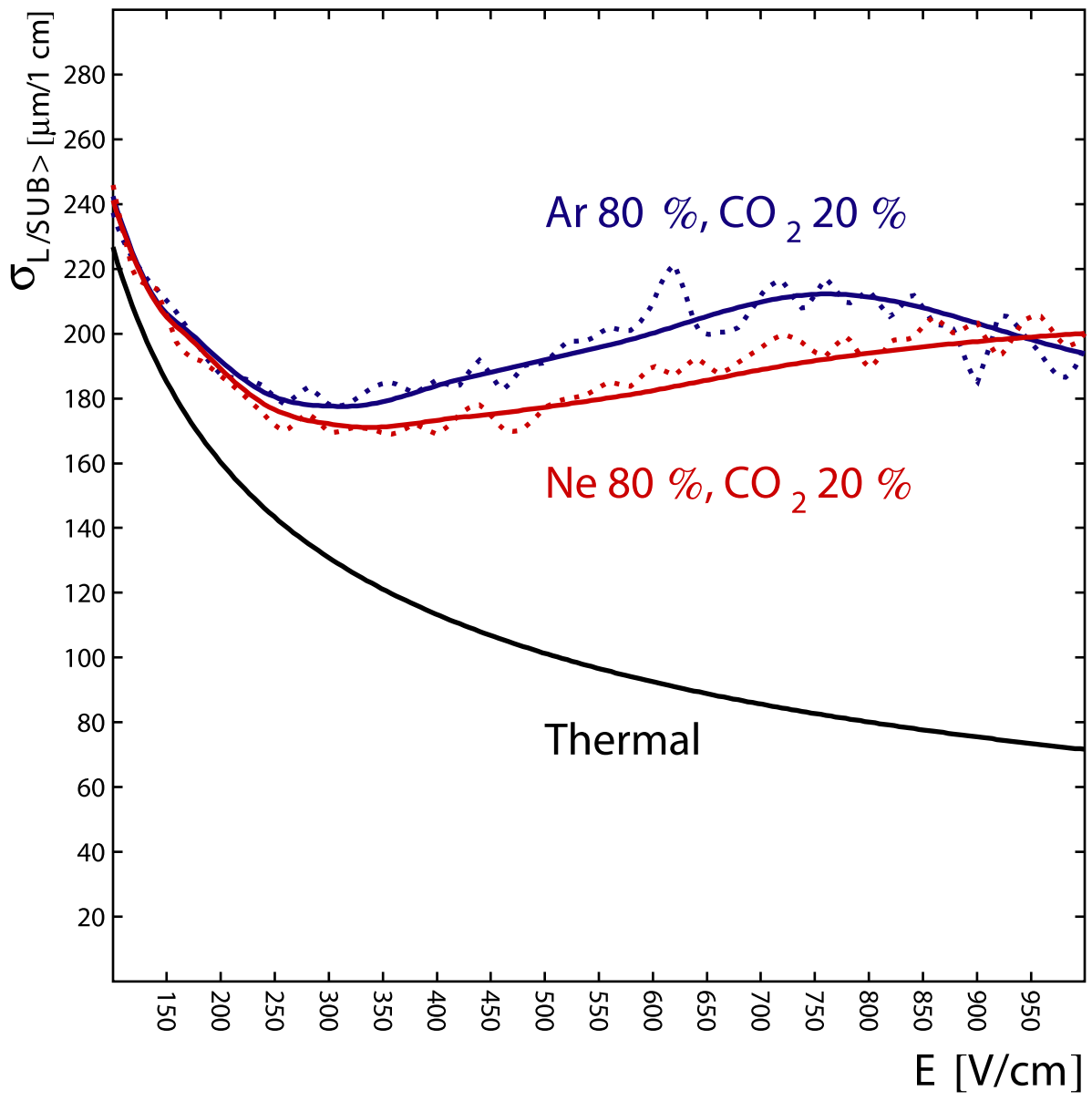

Longitudinal diffusion coefficient in 80% Neon 20% $CO_2$ approaches the thermal limit at low fields. Dashed lines are for B = 0 T and solid lines for B = 0.5 T [38].

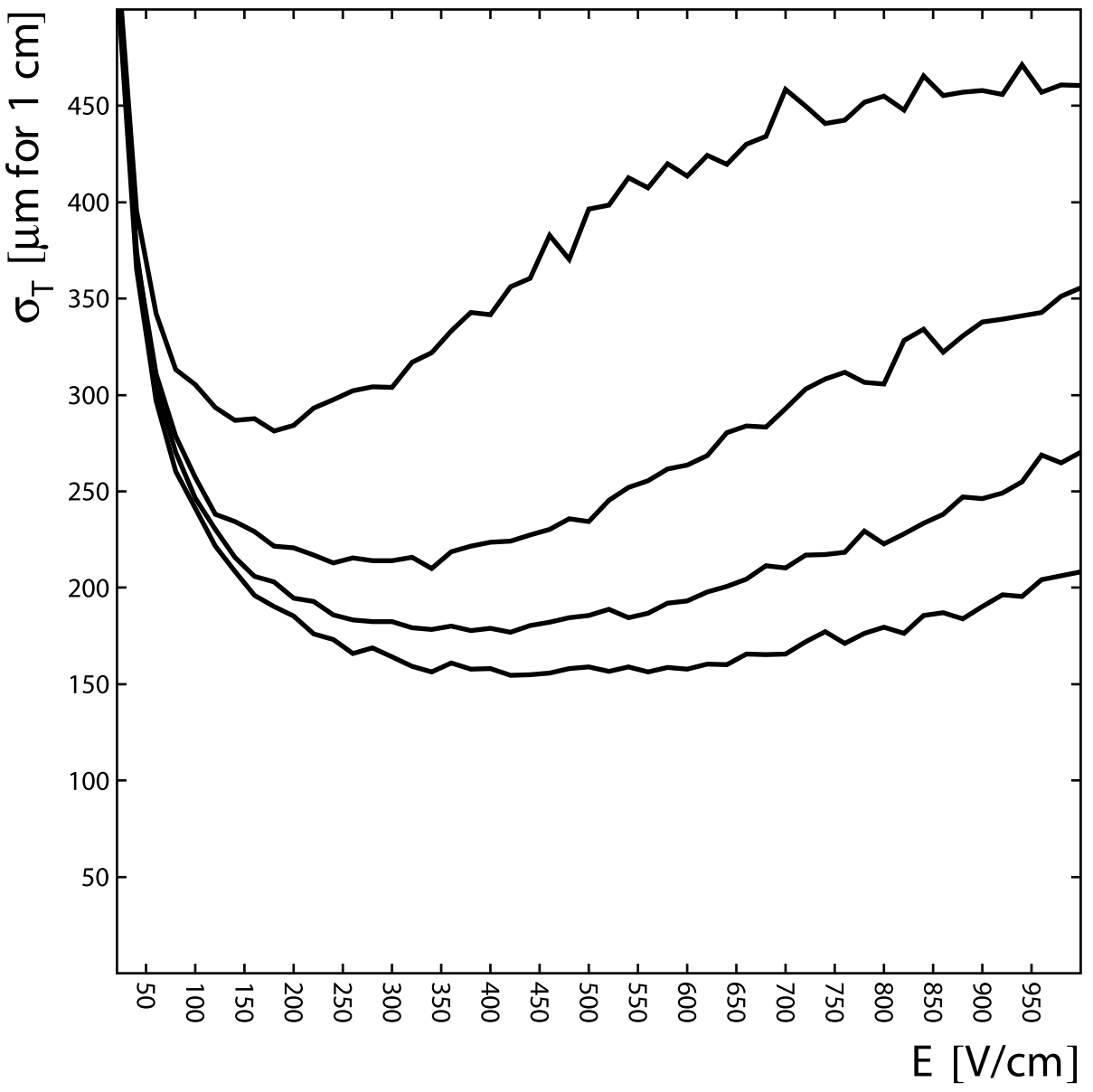

Transverse diffusion for Neon mixed with, from top to bottom, 5%, 10%, 15%, 20% $CO_2$. These curves are calculated without magnetic field [38].

### Readout Chamber

The readout chamber is based on the commonly used scheme of an anode wire grid above the pad plane, a cathode wire grid and a gating wire grid. An electron which approaches the anode wire plane, after passing the cathode plane, will be accelerated by the strong field induced by this plane. The energy transferred to the electron gets high enough to ionise the gas, so that at this point the opposite behaviour as in the drift region is desired. The newly produced electron is also accelerated and ionises another gas atom so that, as the number of electrons multiplies in successive generations, the avalanche continues to grow until all electrons are collected by the anode wire. The remaining ions in between the cathode and anode plane drift towards the cathode wire grid and are mostly collected there. The rest is absorbed by the gating grid. The processes in detail are quite complicated, as there is ionisation, multiple ionisation, optical and metastable excitations and recombinations and energy transfer by collisions between atoms. The signal reaching the pads is proportional to the number of produced electrons. The readout chamber in this type of TPC is also known as MWPC (**M**ulti **W**ire **P**roportional **C**hamber). The multiplication of ionisation is described by the Townsend coefficient $\alpha$. The increase of the number of electrons is given by:

$$dN = N\alpha\;ds \tag{15}$$

Due to the various processes which are included in $\alpha$, no fundamental description exists and it has to be measured or simulated for every gas mixture. For the Ne/$CO_2$ mixture, $\alpha$ is calculated using Magboltz [41] and shown in the following graph on the next page.

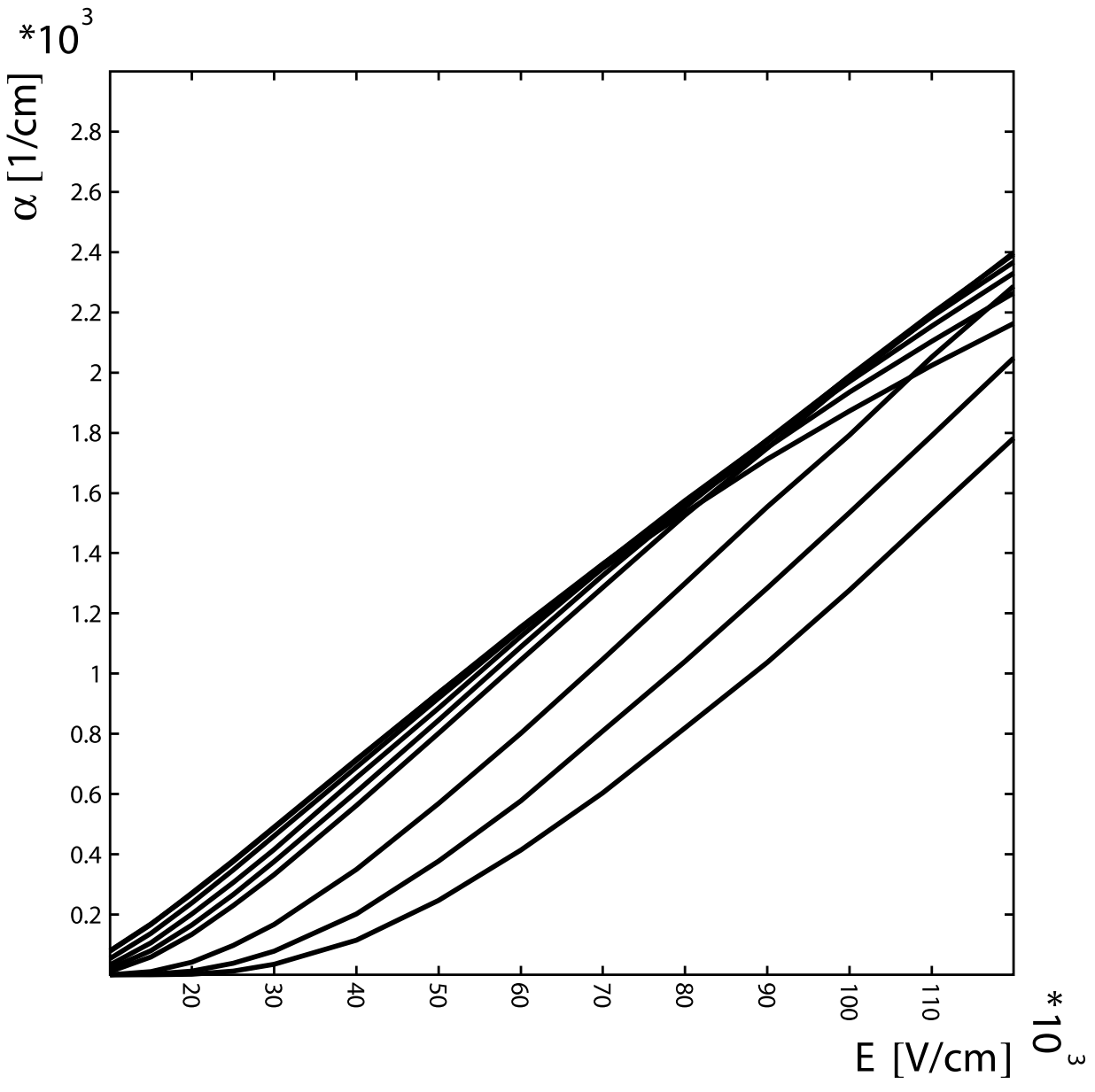

The Townsend coefficient for neon mixed with, from top to bottom, 0%, 5%, 10%, 15%, 20%, 25%, 50%, 75%, 100% $CO_2$ [38].

The gain in connection with the applied potential is a key feature of a proportional chamber. The gain factor M describes the ratio of the produced electrons n to the initial electrons $n_0$. When using the Townsend coefficient, M can be expressed by:

$$M = \frac{n}{n_0} = exp\left(\int_{x_0}^{x_1} \alpha(x)\ dx\right) \tag{16}$$

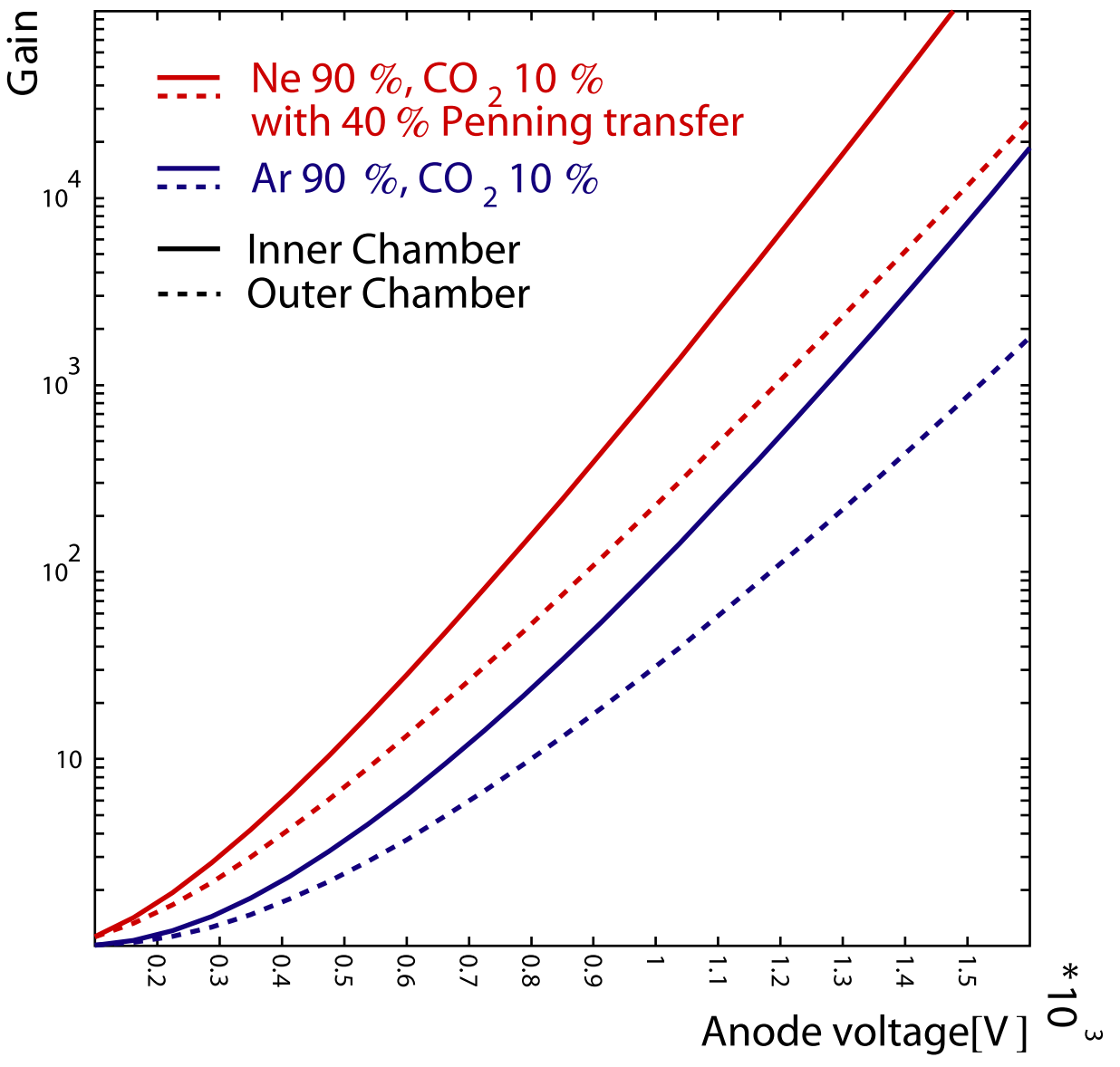

Gas gain of the ALICE TPC of the inner (solid line) and outer chambers (dashed line). Here assumed that 40% of the excited neon atoms produce $CO_2$ ions [38].

There is one effect which would spoil the space resolution when using a noble gas, as during the avalanche creation also photons are produced which have a bigger cruising range and can have energies which are sufficient to ionise atoms. So they could create another avalanche at a different place, which would result in a fake cluster not belonging to a particle track. In addition, this load, when exceeding the Raether limit [42], could generate spark discharges producing aging effects or possibly destroying the readout chamber. When adding a gas with a high photo absorption cross section these photons are captured early and the readout chamber can be driven with a higher field and therefore with an higher amplification factor. A quencher gas is an organic gas due to the high number of degrees of freedom. In ALICE, $CO_2$ is used as a quencher, which avoids the aging effects induced by using a organic gas like $CH_4$.

### Signal creation

The electrons drift towards the anode wire grid and are collected there. This induces a signal with a fast rise of a few picoseconds on the pad plane. The ions are drifting much slower (several orders of magnitude) towards the cathode plane, away from the pad plane. They are inducing a mirror charge on the pad plane which is the measured signal.

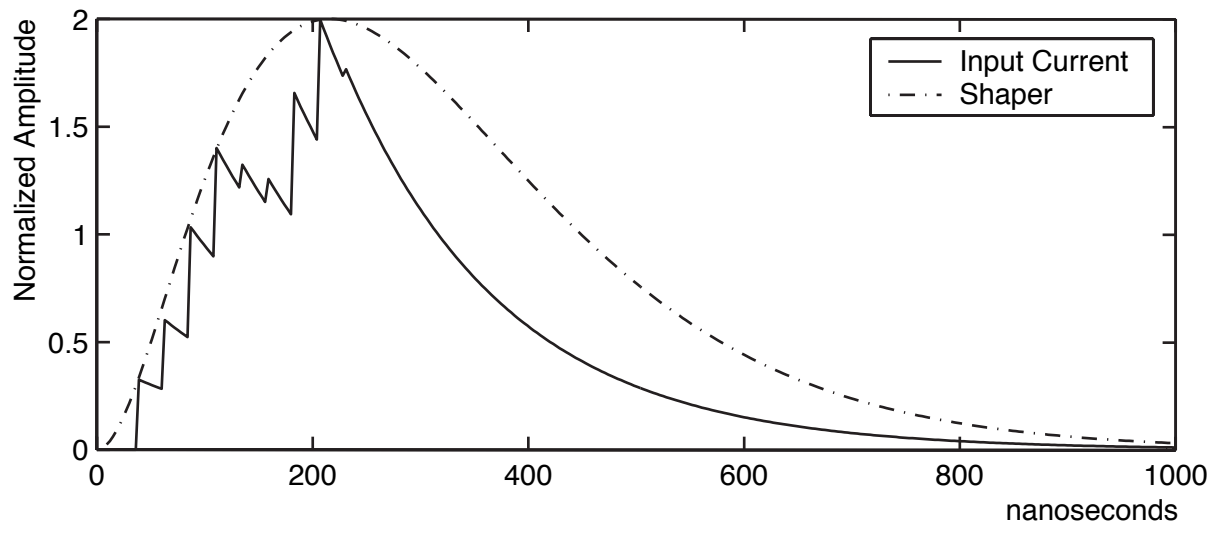

Integrated shape of several avalanche processes including the ion tail [43].

Each avalanche signal is the result of the contribution of many positive ions leaving the anode wires in various angles. An amplifier shaper integrates and differentiates the signal over several avalanche processes producing a pulse with a long falling tail as shown in the plot above. The width of the pulse depends on the track inclination, the drift length and the diffusion [43].

The signal shape for individual ions has been simulated using Garfield [44] as shown in the plot on the following page. When the ions reach either the vicinity of the cathode wires or the gate wires they suffer an acceleration, inducing a secondary spike. They can also reach the pads directly, inducing a slow signal change. The fraction of ions drifting in any of the aforementioned directions depends on the angle of incidence of the primary electron on the anode wires, and how the avalanche spreads around it [45]. The tail shape of the ALICE TPC prototype is discussed in »Ion Tail Analysis« on page 47.

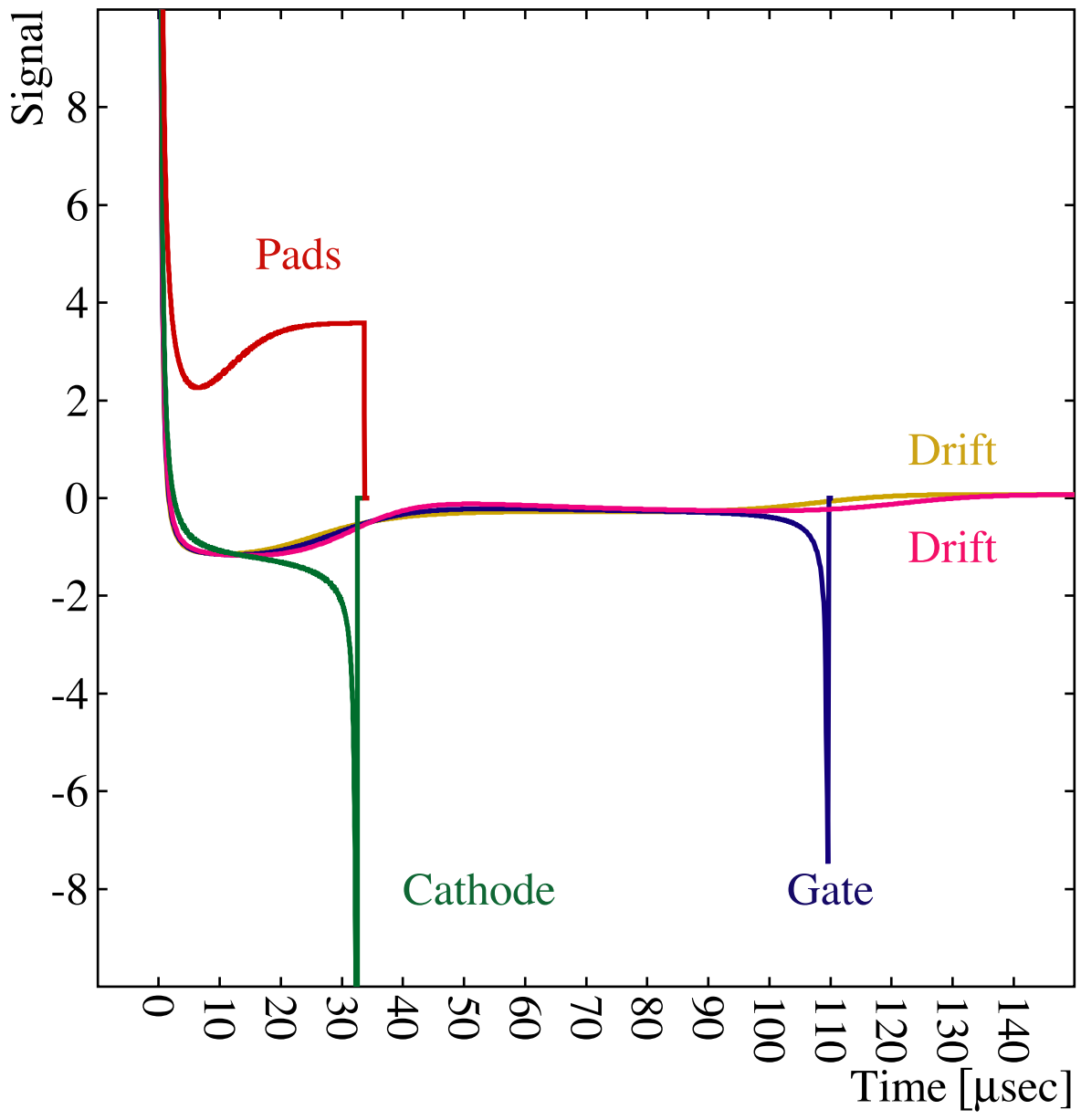

Contribution to the pad signal from different drift paths of the positive ions according to a Garfield simulation [38].

## Prototype

A small prototype was built to do a TPC performance test. It consists of one IROC (**I**nner **R**ead**o**ut **C**hamber) module on one side and a complete fieldcage with a central membrane in a gas tight aluminium box as shown in the photo below.

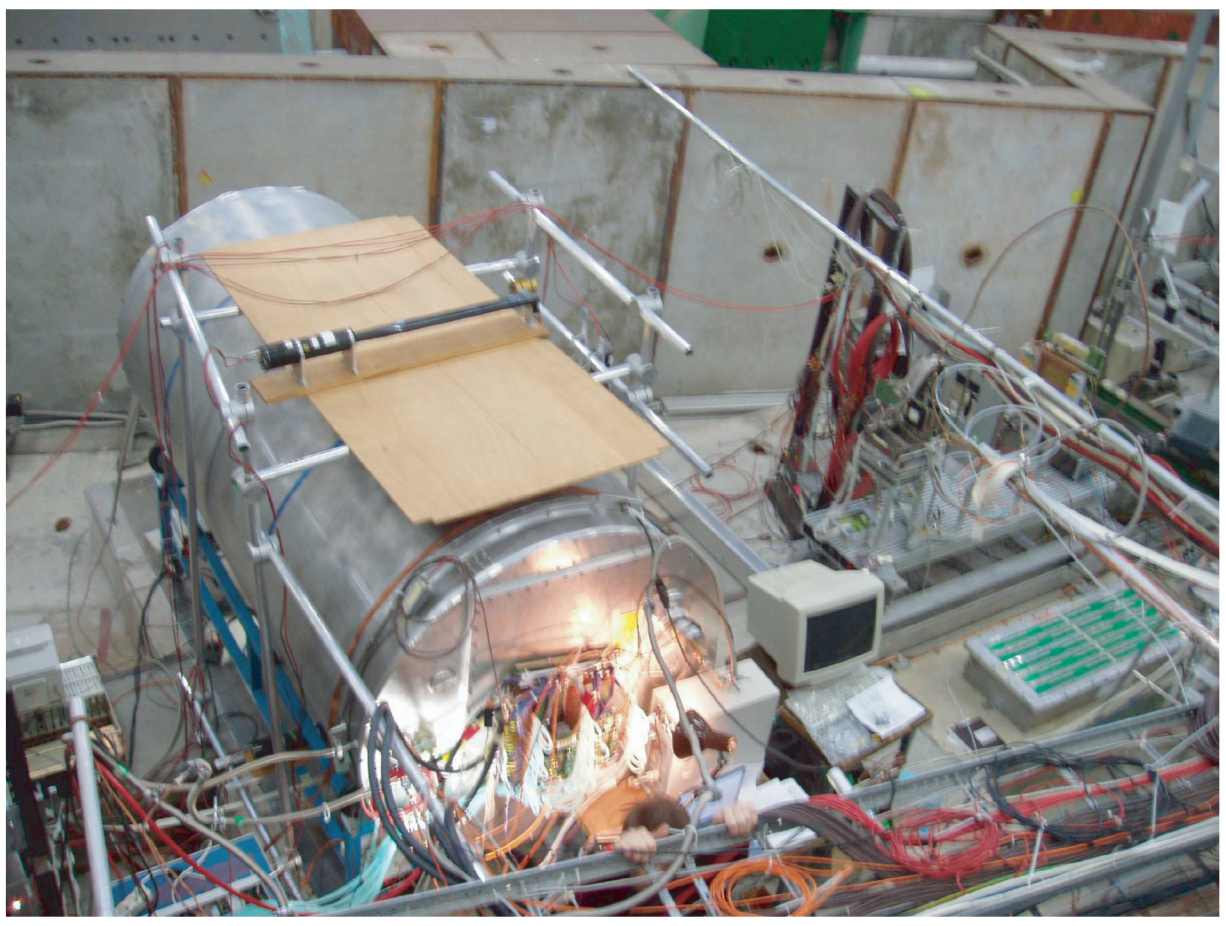

Picture of the prototype setup at the PS testbeam.

First tests were done to verify the electrostatic behaviour without a readout system. Later, for the complete TPC performance test, it was equipped with four FEC boards and a triggering setup using scintillators for cosmic rays. At this moment the front end cards were not cooled and the test TPC was filled with $Ar/CO_2$. Later a simple cooling setup and four additional cards were added and the Gas was changed to $Ne/CO_2$. The cooling system is based on a underpressure liquid cooling system and copper shielding plates around the front end card as shown in the chapter »Front End Electronics on page 15. This setup was used to gather some statistics of cosmic particles. Mostly MIPs are seen in the data but showers were used as an estimate for the high multiplicity environment. By setting a threshold in the data acquisition, data with pulses with a big ionisation were collected to see saturation or crosstalk effects. Throughout the time, two different gas mixtures were used to study differences in the signals. The data is available at [46]. Later this setup was part of a testbeam. The prototype setup is described in »Prototype Environment« on page 29.

# Front End Electronics

# The Front End Electronics

The FEE (**F**ront **E**nd **E**lectronics) [1-3] of the ALICE TPC has to cope with some strong requirements defined by the foreseen physics program. Due to the needed temperature stability in the TPC the heat dissipation has to be minimised. The huge number of pads requires a highly integrated electronics and additionally, the high readout rate makes an intelligent readout mandatory. To minimise the heat dissipation, the electronics is cooled and the power consumption is minimised. The space consumption was minimised by packing sixteen channels into each integrated circuit and also combining analog and digital electronics in one chip as well as packing many chips on one FEC. To achieve a high rate the sampling speed, the processing power and the transfer bandwidth are maximised. The readout chain is defined as: TPC, FEC, Backplane, RCU, DDL, DAQ and HLT. The on-detector electronics consist of the FEC and the RCU with the daughter boards DCS and SIU. The off-detector electronics consists of the DIU (**D**estination **I**nterface **U**nit), RORC (**R**ead**o**ut **R**eceiver **C**ard), DAQ (**D**ata **Acq**uisition) and HLT [4].

## FEC

The FEC (**F**ront **E**nd **C**ard) [1,2,5] as shown below contains 128 complete readout channels. The signal flow starts at the detector end with the analogue signal transported through six flexible Kapton cables and the connectors. The PASA has short connexion links to these connectors, to minimise the crosstalk caused by the fast input signal from the detector. Afterwards, the ALTROs are directly connected to the PASAs using differential signals. With the ALTRO, the analogue part of the FEC ends and at the same time the digital part starts. The digital outputs are multiplexed through a LVCMOS (**L**ow **V**oltage **CMOS**) bus and translated to the GTL (**G**unning **T**ransceiver **L**ogic) level and linked to the connectors of the backplane.

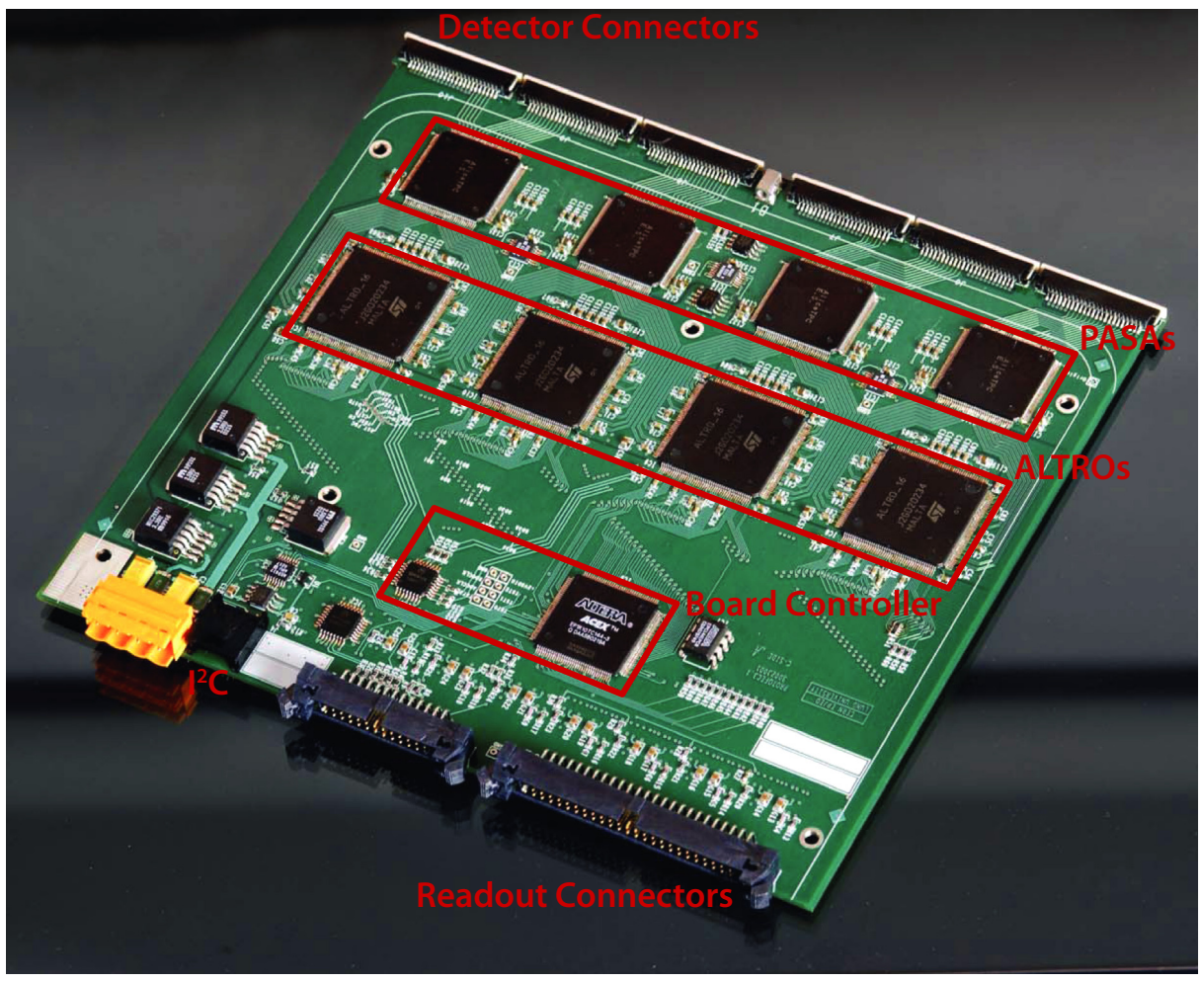

Picture of the FEC PCB with all components. The Signal flows from the top through the connectors, the PASA, the ALTRO and the readout connectors.

In addition, there is a BC (**B**oard **C**ontroller) realised as a FPGA (**F**ield **P**rogrammable **G**ate **A**rray) which provides an independent access to the FEC via the FCB (**F**ront end **C**ontrol **B**us). This is used to control the power state, voltages, currents and temperature of the FEC. Every FEC contains a 10 bit, 5 channel ADC with an on chip temperature sensor which is connected to the BC via a $I^2C$ (**I**nter-**IC**) link. This represents the slow control. The FEC PCB (**P**rinted **C**ircuit **B**oard) contains four signal layers and four power layers divided into two supply layers and two ground layers. The FEC has a width of 19 cm and a length of 17 cm. In total, the FEC has a maximum power consumption of 6 W. The FECs are located directly on the end caps of the TPC and with the strict temperature requirements the heat dissipation of the FEE has to be minimised. For this reason the FEC is embedded in a water cooled enclosure made from copper plates as shown in the picture below.

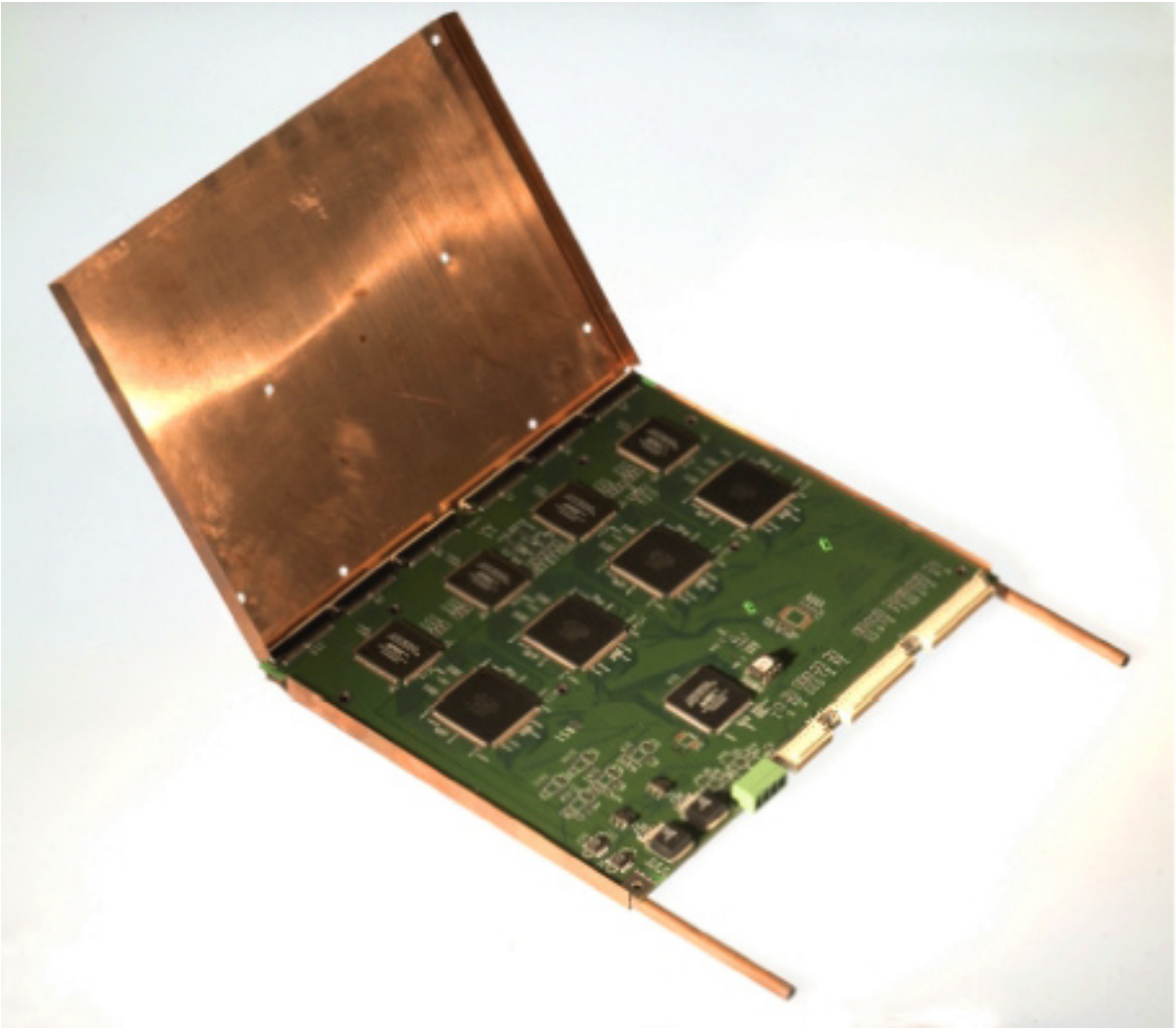

Picture of the FEC in the water cooled copper plates. The FEC shown is a old version, but the dimensions are the same.

## PASA

The charge collected by a TPC pad is integrated, amplified and shaped using the PASA (**P**re**a**mplifier/**Sha**per) [1,6]. The output is connected to the ADC of the ALTRO. The PASA has a low input impedance amplifier which is based on a CSA (**C**harge **S**ensitive **A**mplifier) followed by a semi-Gaussian pulse shaper of the fourth order. The PASA is implemented in the AMS CMOS (**C**omplementary **M**etal **O**xide **S**emiconductor) 0.35 μm technology, and consists like the ALTRO of 16 channels with a power consumption of 11 mW/channel. The conversion gain is 12 mV/fC and the output has a dynamic range of 2 V with a differential non-linearity of 0.2 %. The output is a pulse with a shaping time (FWHM: **F**ull **W**idth **H**alf **M**aximum) of 190 ns. The noise of one single channel is below 570 electrons (RMS: **R**oot **M**ean **S**quare) and a channel to channel crosstalk below -60 db.

## ALTRO

The ALTRO (**AL**ICE **T**PC **r**ea**d**o**u**t) [1-3,7-9] is a chip specially designed for the needs of the ALICE TPC consisting of an analog part in addition to a digital part. A block scheme is shown below. There are 16 channels integrated in one IC (**I**ntegrated **C**ircuit), realised as 0.25 µm CMOS process operating concurrently on the analog signals coming from 16 independent inputs. Each of these channels is composed of an ADC (**A**nalog **D**igital **C**onverter) as the analog part, a BCS1 (**B**aseline **C**orrection and **S**ubtraction **1**), a TCF (**T**ail **C**ancellation **F**ilter), again a BCS2 (**B**aseline **C**orrection and **S**ubtraction **2**), a ZSU (**Z**ero **S**uppression **U**nit), a DFU (**D**ata **F**ormatting **U**nit) and a MEB (**M**ulti **E**vent **B**uffer), as the digital part. In addition, there is a central CCL (**C**ommon **C**ontrol **L**ogic) for the configuration and control for the trigger and bus. There are two frequency domains, one is driven by the bus clock and consists of the Bus Interface in the CCL and the memory in the MEB, and the other is driven by the readout clock and consists of the rest. Since 95% of the ALTRO runs with the sampling clock, the influence of the readout clock on the signal is minimised. The ALTRO is continuously sampling the input, on arrival of a first level trigger (L1) an event is temporarily stored in the memory. The maximum length of an event is 1008 samples. Upon arrival of a second level trigger (L2) the latest acquisition is frozen and kept until readout from the memory by the RCU via the ALTRO bus. The MEB has a capacity of up to eight events. If another level one trigger signal occurs prior to a second level trigger, the first acquisition is discarded and overwritten by the next event.

### ADC

The ADC of the ALTRO is based on a commercial design, the Microelectronics TSA1001 [10], and was slightly modified for the needs defined by ALICE. The TSA1001 was chosen because of the low power consumption, which is quite an important prerequisite of the TPC, since there are extremely tight temperature constraints [2]. The dynamic range is 10 bits and the sampling frequency is up to 25 MSPS (**M**illion **S**amples **P**er **S**econd). Due to the fact that there is an analogue part and a digital part on the ALTRO the electrical coupling has to be minimised to not decrease the quality of the sampling, compared to a design which separates ADC and digital chain in two chips.

### BCS1

In the digital block the first unit is the baseline correction. Its main purpose is the preparation of the signals for the adjacent tail cancellation unit. The TCF demands the removal of the DC level and a relatively stable baseline during the data acquisition and in between. There are several sources of perturbations which can have an impact on the signal. A source of perturbations are low frequency (<1 kHz) variations, which are nearly constant during an acquisition window. The origin of these interferences are temperature variations in the electronics, coupling of AC or DC and the finite detector load [11,12]. The self calibration circuit (AUTOCAL) of the ALTRO removes these disturbances. Since the ALTRO is continuously sampling and processing the input signal, it can detect these slow variations outside of the data acquisition window. This self calibration is stopped on arrival of a level one trigger and the last value is taken as DC level of the baseline which is then removed from the signal. Past a level two trigger the calibration is re-enabled. This is shown in the plot on the next page.
Another source of perturbations are systematic signals like the switching of the gating grid as shown in »BCS1 Parameter extraction« on page 38. The removal of these interferences is based on a LuT (**L**ook **up** **T**able) which is realised as a memory in the ALTRO. This table is extracted

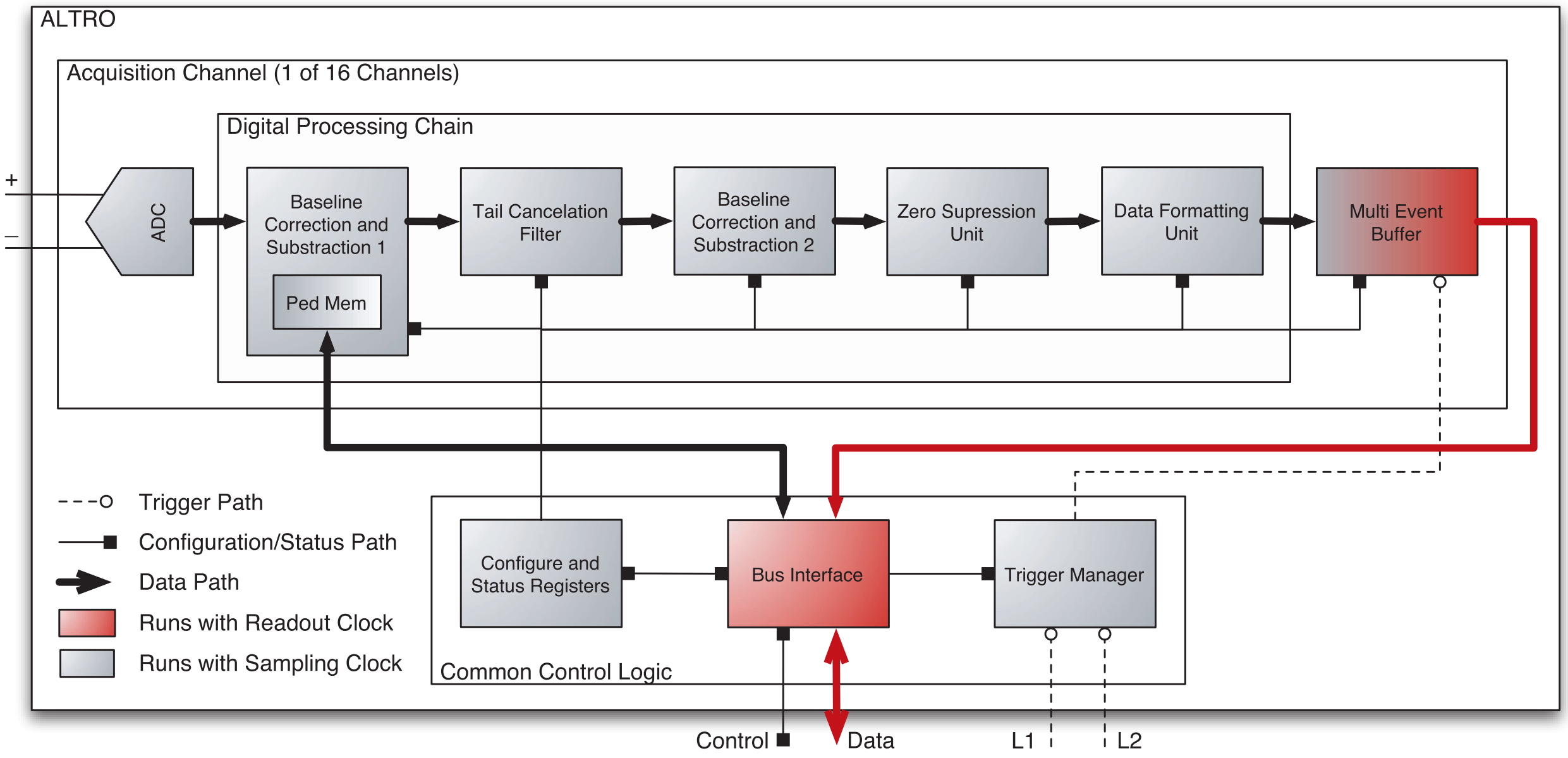

Block scheme of the ALTRO. Only the Bus Interface and partially the Multi Event Buffer are running at the speed of the readout clock.

from acquired empty events which means a normal data acquisition of the TPC just without tracks from a collision. In the second plot below the effect of the LuT correction is shown.

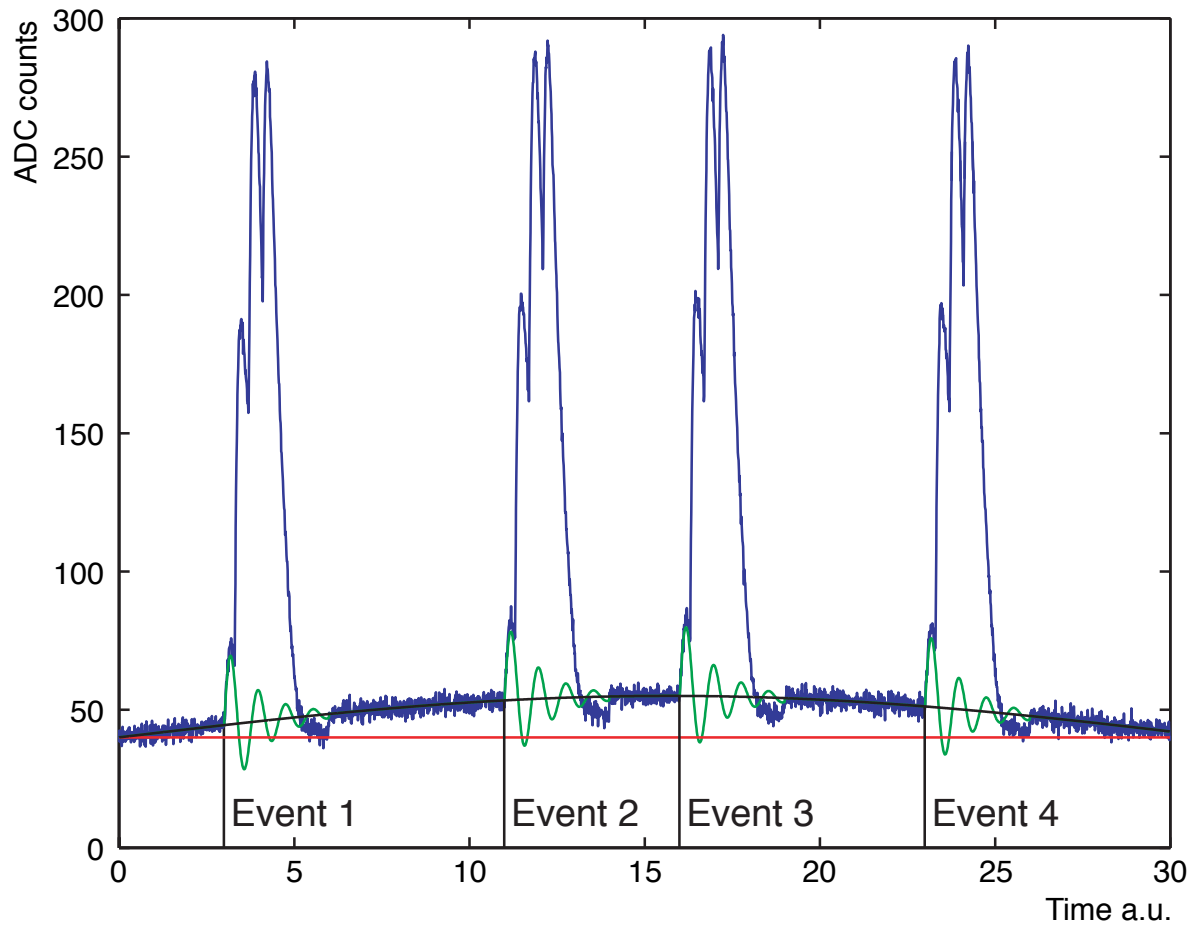

In this picture a low frequency perturbation as the black line is shown, the ALTRO AUTOCAL circuit detects and the BCS1 removes this perturbation from the signal (dark blue line). The red line is the configured fixed pedestal and the green line shows a systematic perturbation.

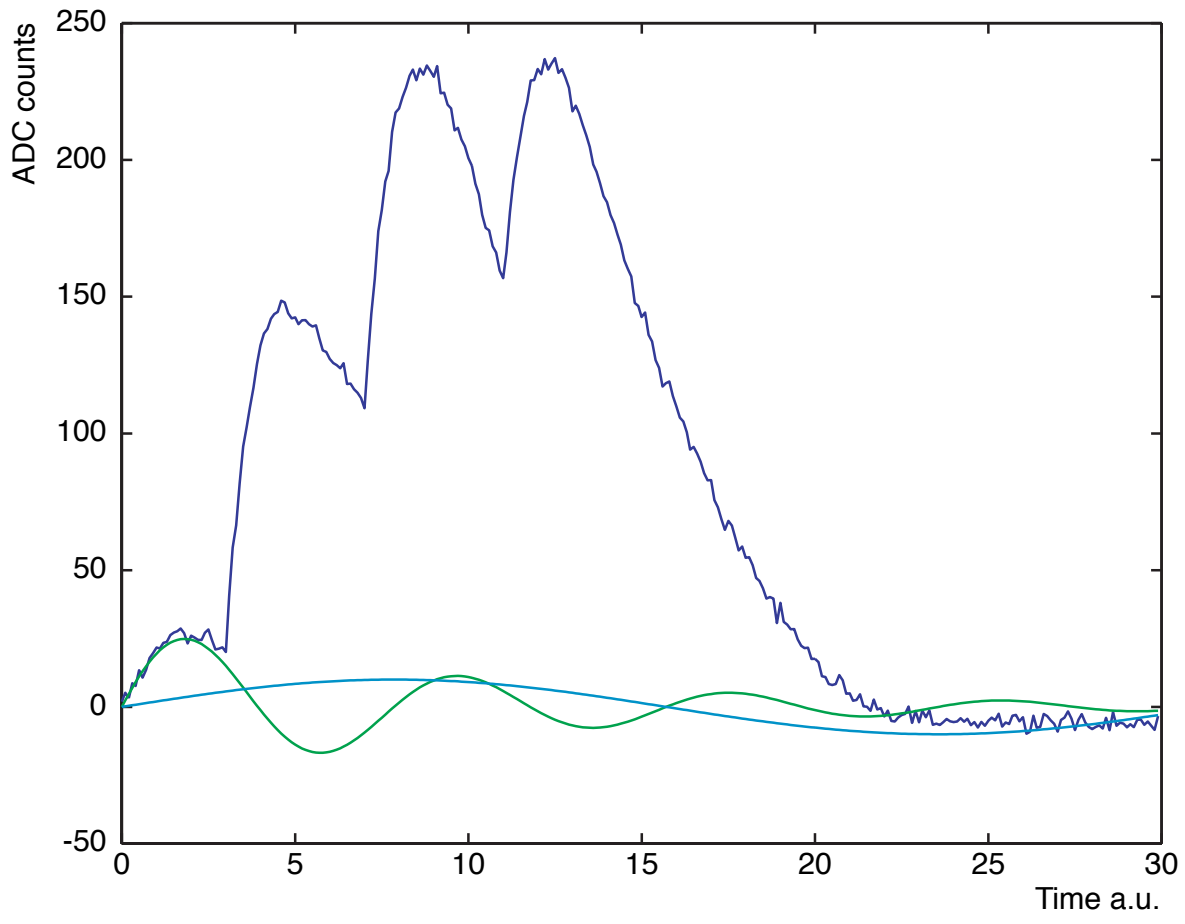

In this picture the zoom on one event with a systematic perturbation (green line) is shown, this is removed by the use of the LuT. The unsystematic perturbation (cyan line) is kept.

In addition, the gain calibration can also be performed by this unit. The activation and the combinations of these different sub entities is configurable by several predefined setups as described in the ALTRO manual [9].

### TCF

The ALTRO was optimised for the TPC type used in ALICE (see chapter »The experiment« on page 5) and the presequent PASA with the semi-Gaussian shaping signal. This combination creates signals with a fast rise time (>1 ns) followed by a long tail. The effectiveness of the later following zero suppression is not efficient, when the expected signal density is taken into account (in the inner rows the occupancy is expected to reach 40%). The problems are the long signal tail by itself, in addition to the pile-up effect when several signals are occurring in a short time. To improve this situation, the signal tail is removed by this entity. It is implemented by a cascade of three first order IIR (**I**nfinite **I**mpulse **R**esponse) filter circuits and described in [13,14]. Each of these circuits has a set of two parameters. So in total there are six parameters to accommodate the TCF to the real signal shape which is described in »TCF Parameters« on page 39.

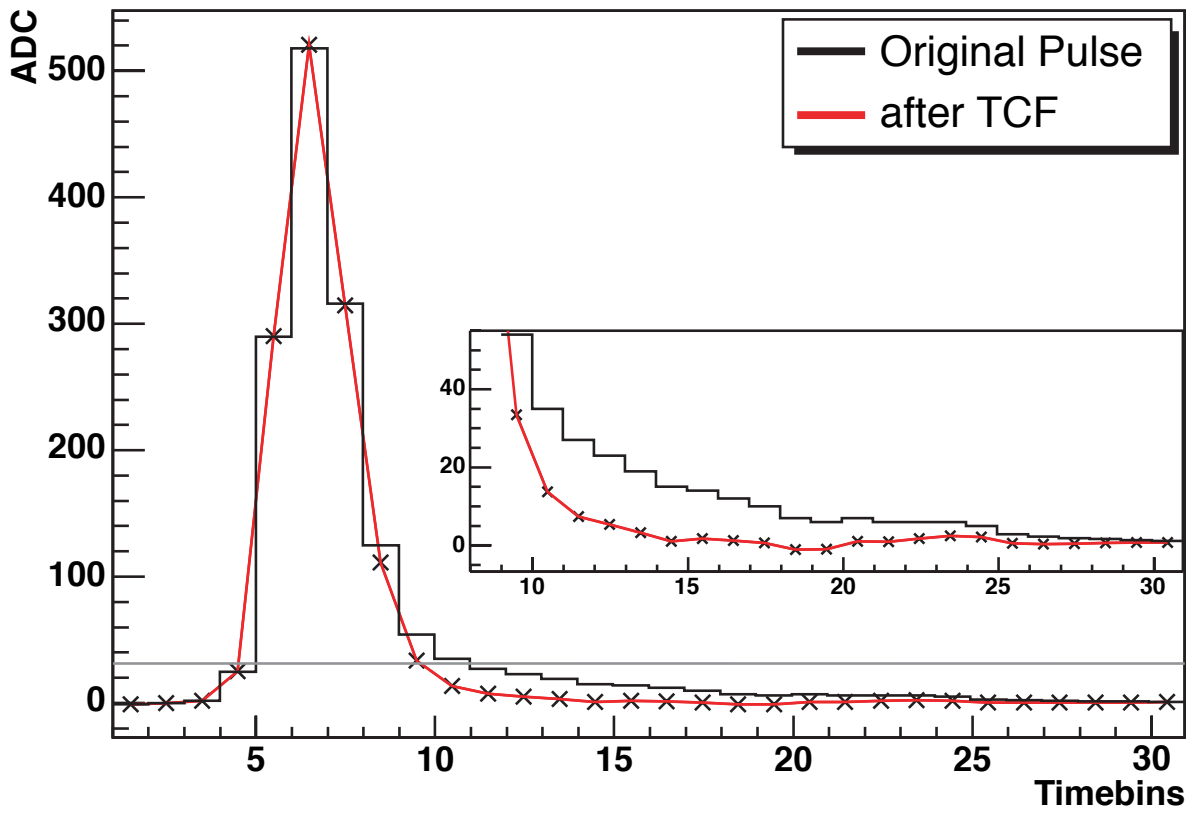

In this picture the performance of the TCF is shown to remove the signal tail without modifying the pulse and die amplitude.

### BCS2

The second baseline correction is only applied during the acquisition of an event and corrects non systematic signal perturbations, as shown in the plot below. It is realised as a moving average filter. The correction is calculated by using the average of eight presamples which were in the acceptance window. The acceptance window is defined by a configurable double threshold scheme.

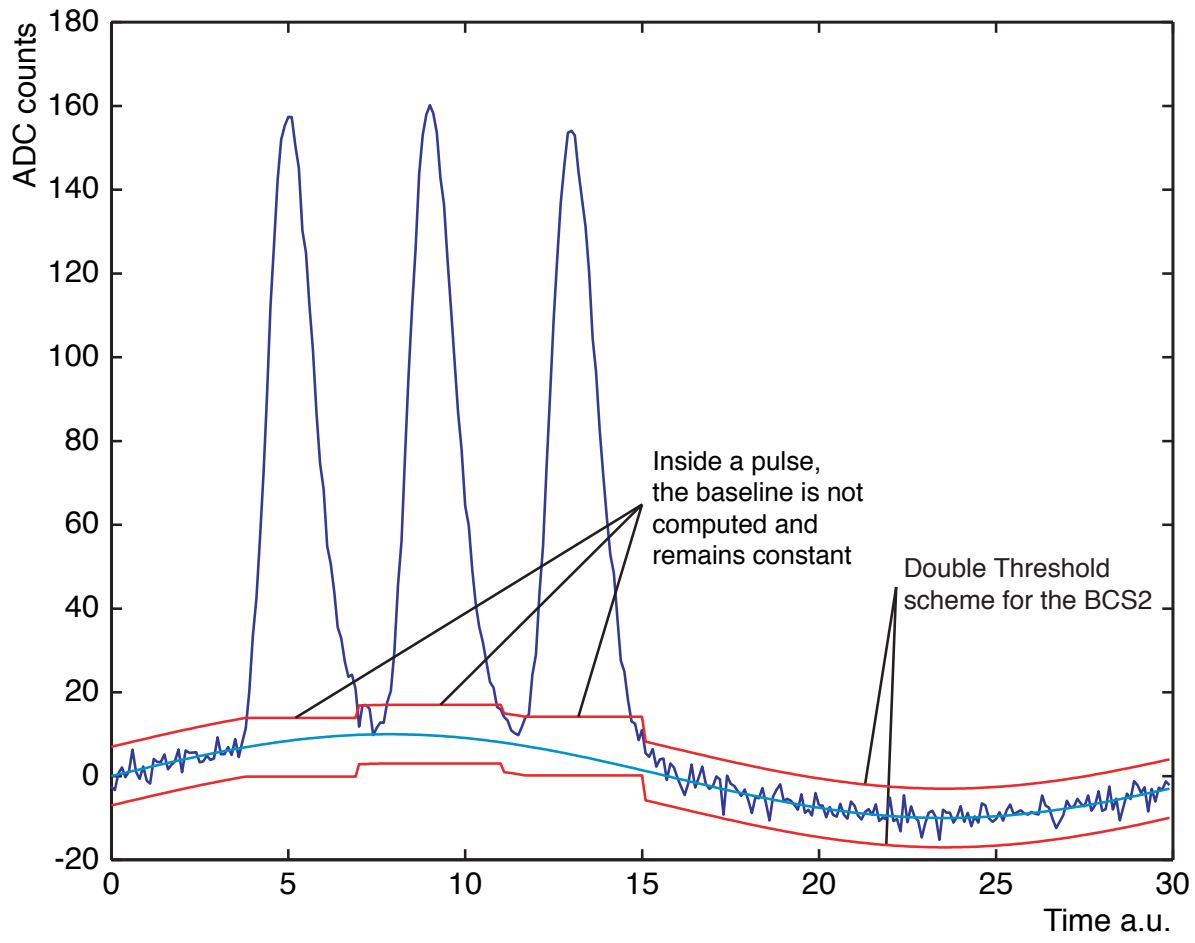

This picture shows the double threshold signal following scheme of the BCS2 (red lines). During a pulse the processing is frozen and the correction value is kept. The unsystematic perturbation (cyan line) is removed from the signal (dark blue line)

If the next sample is outside this window, it is not used to update the moving average value. This means, that if there is a big variation, which is normally induced by a pulse of a cluster, the correction value stays on the value which was previously calculated with the last sample in front of the pulse. After the pulse, the samples are again in the acceptance window and the correction is again calculated. In addition, to minimise the influence of the pulses, a configurable number of samples can be excluded pre and post the pulse from the calculation.

### ZSU

The last processing entity in the chain is the zero suppression unit. When compressing data, the most obvious way is to remove zeros, since they are not carrying information. In this case, these zeros are in between two pulses and they only carry noise. For this purpose, all samples which are above a threshold are marked. Glitches are removed by requiring more than one sample above the threshold. This limit is configurable, starting with more than three consecutive samples above the threshold down to one, which would not discard anything. To not loose any information of the pulse, additional samples can be marked as »to keep« by using the configurable pre- and postsamples. If there are two pulses closely together, they are merged when there is only one or two samples distance, because the DFU (see next section) adds for each found sequence two words. This merging increases the compression level.

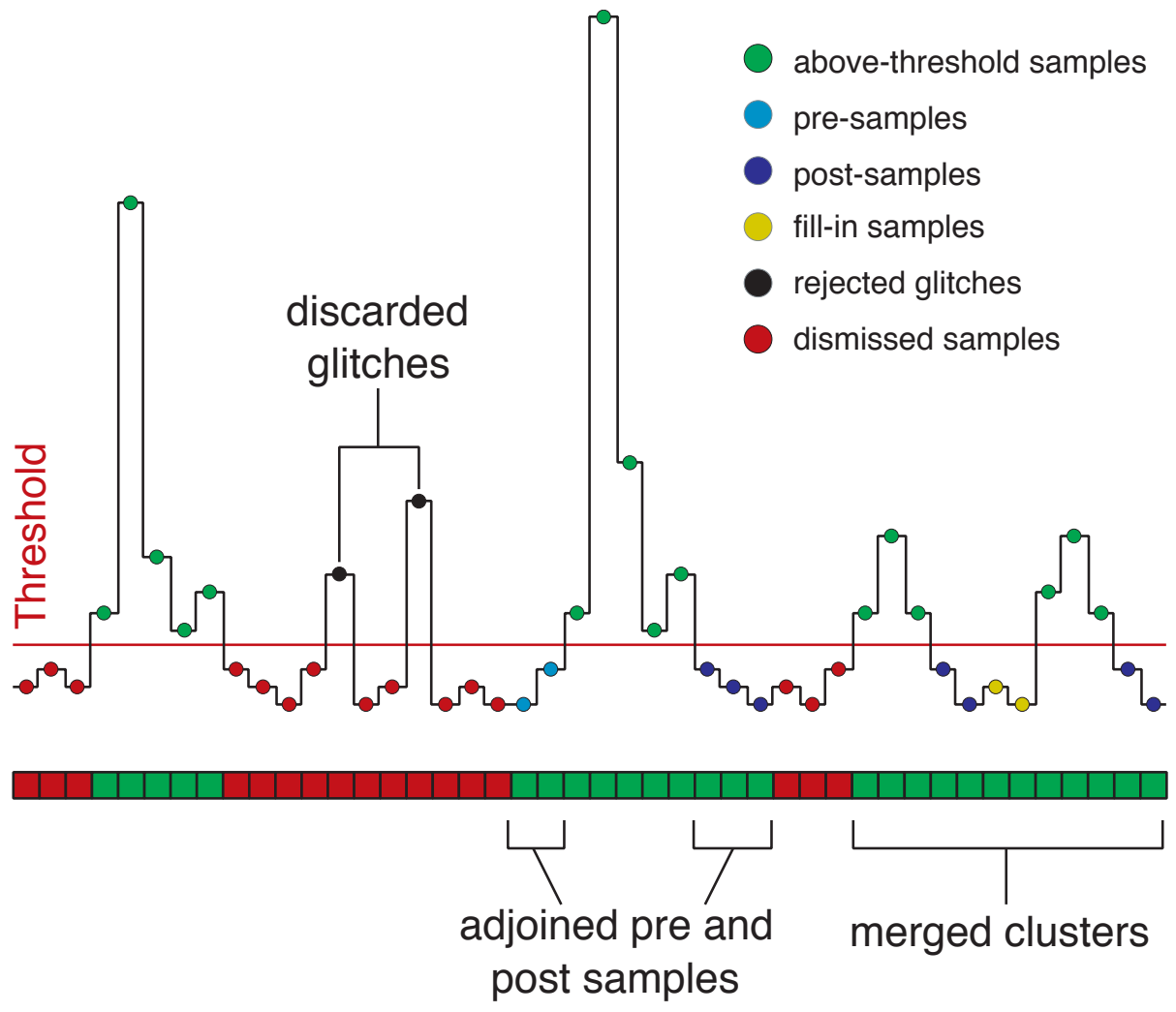

Here all features of the ZSU are shown, starting with the zero suppression threshold, the discarded glitches, pre- and postsamples and the cluster merging.

### DFU

When removing the samples in between the pulses one relevant information gets lost, the time information. As mentioned above, the DFU adds two words to each sequence, the first is the time information and the second is the total length of the sequence. The time information is the time distance in number of samples after the trigger. With this additional information a decompression is again possible.

In addition, this unit bundles the 10 bit words to 40 bit words since the ALTRO bus has a width of 40 bit. When the last 40 bit word is not completely filled the hexadecimal pattern 0x2AA will be added as often as needed to complete this word. Finally, the trailer word with a length of 40 bit is added. It consists of the total number of 10 bit words before and the »Hardware Address« which is unique for each ALTRO-channel in one readout partition (see »Monitoring« on page 34). The unused start is filled with the pattern 0x2AAA and in between the 10 bit word counter and the »Hardware Address« the number 0xA.

| | 39 30 | 29 20 | 19 10 | 9 0 |
|---|---|---|---|---|
| 40 bit Data Words | S 05 | S 04 | S 03 | S 02 |
| | S 10 | C 7 | T 06 | S 06 |
| | C 5 | T 12 | S 12 | S 11 |
| | ... | ... | ... | ... |
| | S 91 | S 90 | S 89 | S 88 |
| | 0x2AA | C 7 | T 92 | S 92 |
| Trailer Word | 0x2AAA | 10 bit word count | 0x A | Hardware Address |

This schematic shows the ALTRO format packing. »S« means Sample, »T« means the time position and »C« means the complete length of a sequence. The »Trailer Word« includes the total counter for this channel and the »Hardware Address«. The »0x« starting patterns are filling the empty positions.

### MEB

To reduce the dead time, the data transfer is decoupled from the data acquisition of the detector electronic. For this purpose, the ALTRO has a memory of 1024·40 bit and can be blocked in two, four and eight blocks. The MEB runs in both clock domains. It has to be interfaced from the DFU, which runs at the sampling clock speed, for data storage and it is accessed by the »Bus Interface«, which runs with the readout clock. On arrival of a level one trigger, the acquisition of an event is started and will be stored in the memory. If a level two trigger arrives, the event is frozen in memory and stays there until a CHRDO (**Ch**annel **Read**o**ut**) command is sent [9]. If after a level one trigger again a level one trigger occurs, the memory will just be overwritten by the next incoming event. With this scheme, the FEE can cope with bursts of events by filling up the buffers faster than the readout, which then catches up when the event rate is smaller.

## Backplane

The ALTRO bus from each front end card is firstly connected to the two backplane PCBs as shown in the picture on the next page. The backplane delivers the termination support for the bus. It also adds mechanical support in the fixation of the FEC. For each patch in one TPC sector there is a different backplane due to the fact, that the number of FECs is differing as well as the space in between the FECs. There are always two branches per RCU [1,15].

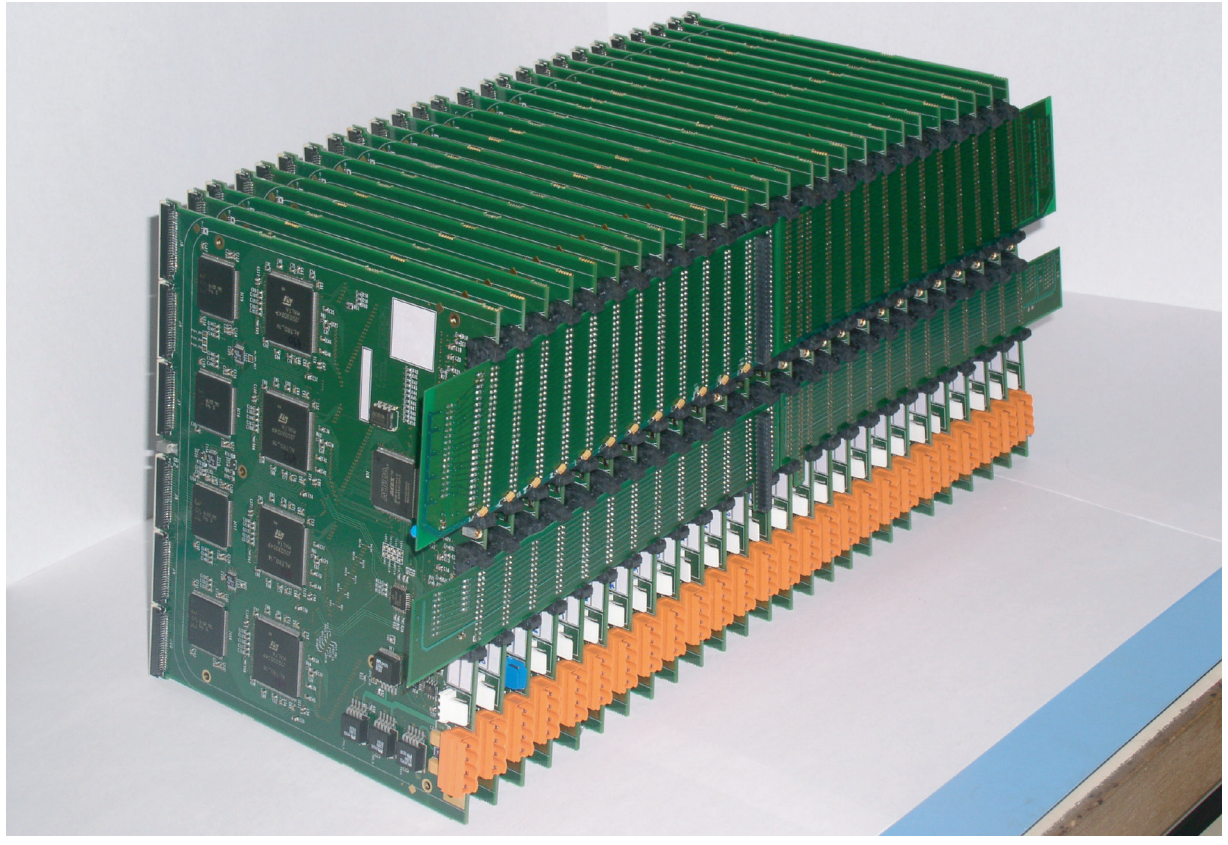

Both backplanes fully equipped with FECs are shown. The connector to the RCU is in the middle and the termination is on both ends.

## RCU

The RCU (**R**eadout **C**ontrol **U**nit) [2,5,16-18] is connected to the two branches of the FEC and has connectors for two daughter boards (SIU and DCS Board) as shown in the picture below. The RCU provides the bus termination. The purpose of the RCU is to be the interface between the FEE and the DAQ, DCS and Trigger. There is an ALTRO module for the communication via the ALTRO bus with the ALTROs. For the communication via FCB to the Board Controller there is the »Monitoring and Safety Module« via the FCB. The read out data is prepared by the »Data Link Interface« to cope with the needs of the SIU. The control over the RCU is handled by the DCS card which needs three interfaces. One is the configuration of the RCU FPGA itself, to change the firmware on updates and failures induced by single event upsets. There is the interface for the DCS system to control and configure the Front End and to the TTCRX (**TT**: Trigger and **C**: Control and **Rx**: Receiver) which delivers the different Trigger information [4]. Parts of the Trigger information have to be delivered to the SIU to build up the DATE (**D**ata **A**cquisition **T**est **E**nvironment) [4] event header.

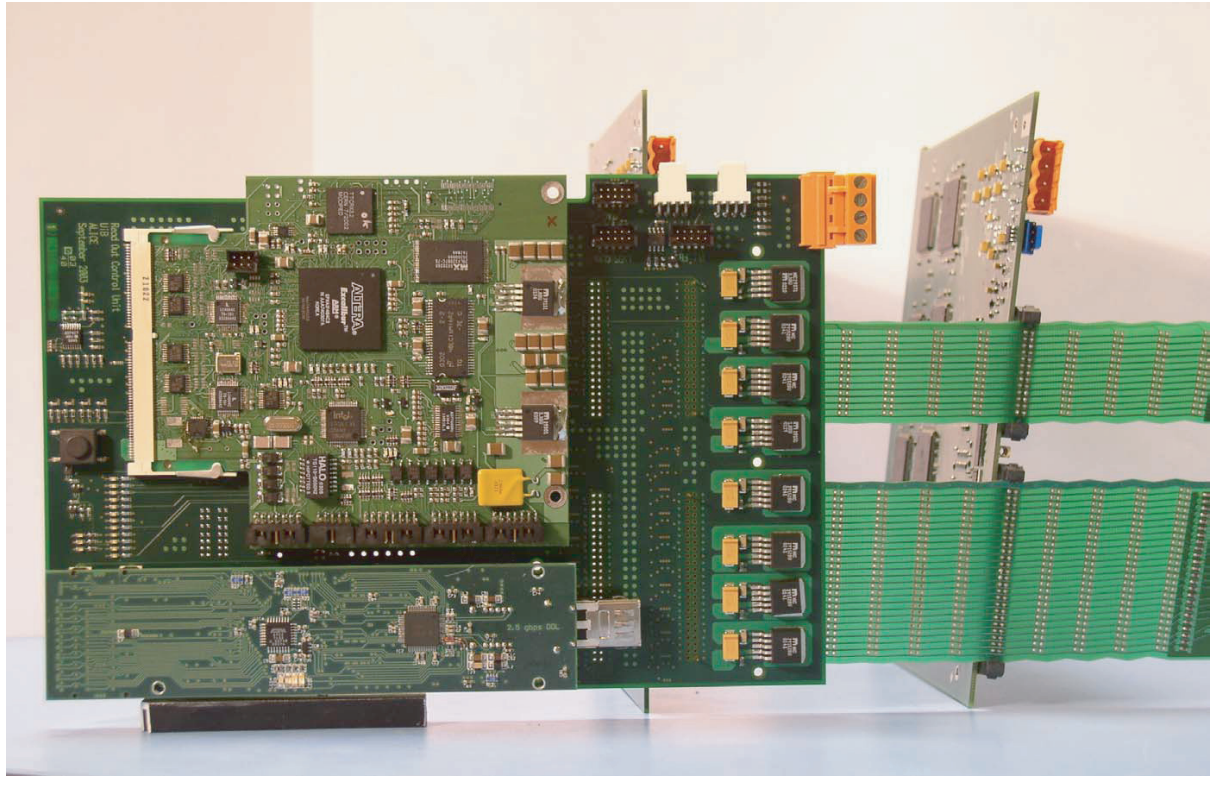

Here the RCU3 with both daughter boards (DCS as the upper board & SIU as the long lower board) and connected on the backside the backplane with two FEC is shown.

## SIU

The SIU (**S**ource **I**nterface **U**nit) is the detector end of the DDL (**D**etector **D**ata **L**ink) which is then connected to the ALICE DAQ system. The SIU uses a 32 bit half duplex data bus for the interface from the RCU and an optical transceiver to the DDL [4].

## DCS

The DCS (**D**etector **C**ontrol **S**ystem) daughter board is running a complete embedded Linux called μClinux [19] on an ARM 922T [20] hardwired logic on a FPGA. This is the end of the DCS system for the TPC, so this board handles the configuration of the FEC and the RCU and also the status control of the boards. The trigger receiver, the LHC TTCRX chip [21], is also located here. It delivers the L1 and L2 trigger information to the FEE [4].

# Jitter

## Jitter

Each clock has a finite jitter, which means that the time interval in between two clock pulses is not exactly constant. This is no problem or digital circuits, since all components are running synchronised with the clock, but when working with analog signals the inaccuracy in the knowledge of the exact time position leads to an inaccuracy in the measurement of the signal, because the time point when measuring is not exactly known. The clock accuracy is a compromise between the needed time accuracy of the measurement and the effort to build the clock. Since in this case the clock is needed on all 4356 FEC on the TPC, a complicated clock scheme would be complex and expensive. A simulation was done, to find out the needed accuracy. This simulation is described and the results are shown in this chapter.

## Simulation

The starting point is the signal generated by the PASA (described in chapter »The Front End Electronics« on page 15), which has the shape of a semi-Gaussian function of the fourth order.

$$f(t) = \begin{cases} k\left(\frac{t-t_0}{\tau}\right)^4 \cdot e^{-4\frac{t-t_0}{\tau}} & t > 0 \\ 0 & t \leq 0 \end{cases} \quad (1)$$

with the parameters $t_0$ as the starting time, $\tau$ as the relaxation time and k defined as:

$$k = Ae^4, \quad (2)$$

with A as the amplitude. The four is in both cases the order of the function which is defined by the PASA. This function, as shown below, is sampled without jitter and sampled at slightly different positions to simulate the jitter which results in an amplitude and timing error.

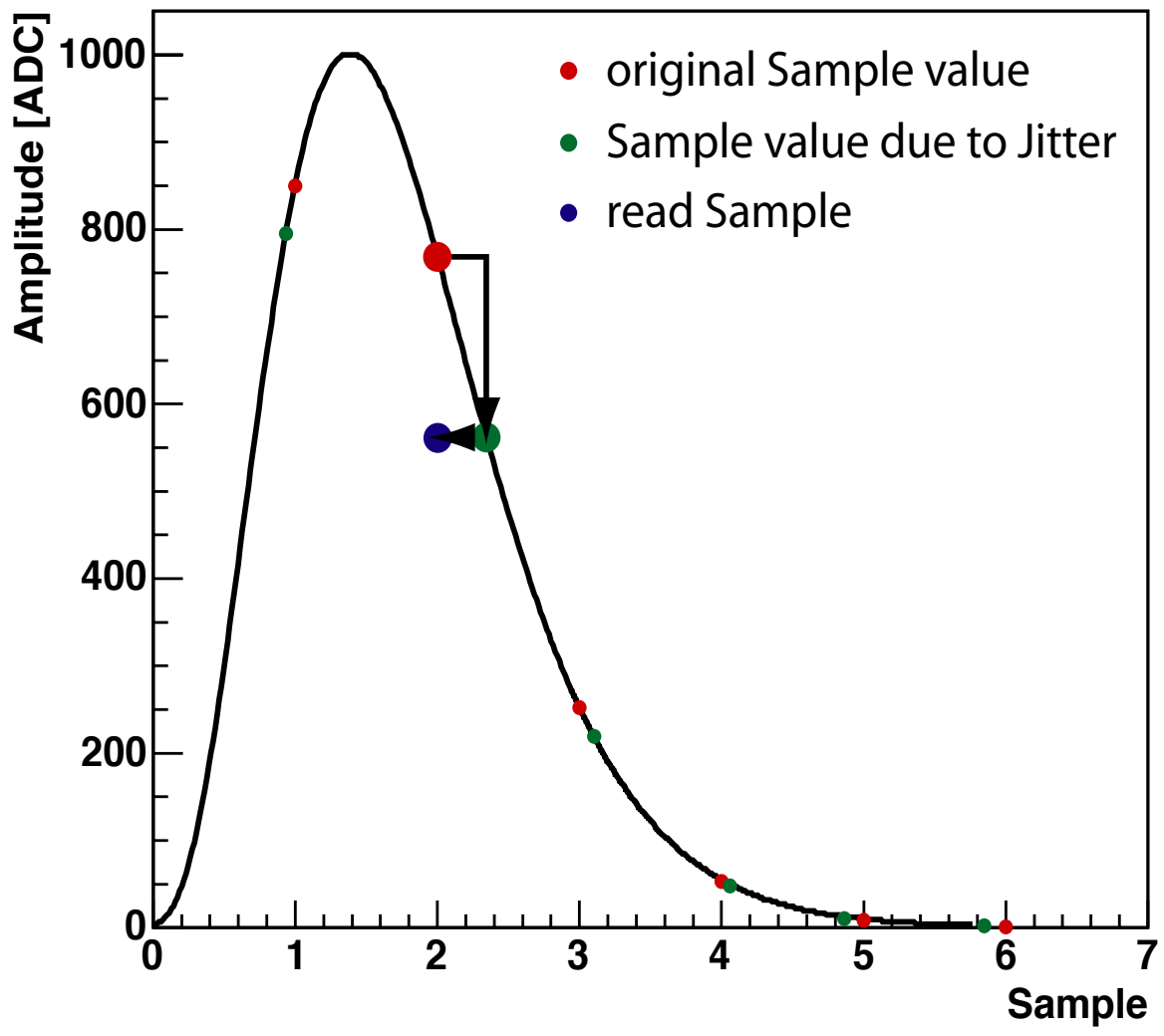

The generated signal (black line), the correct sample, the due to jitter disturbed time position and the read sample are shown.

The noise of the acquisition chain of PASA and ALTRO is added to these values. The noise was extracted from the data recorded for the pedestal calculation. For each channel the RMS was calculated and all are collected in the plot shown below.

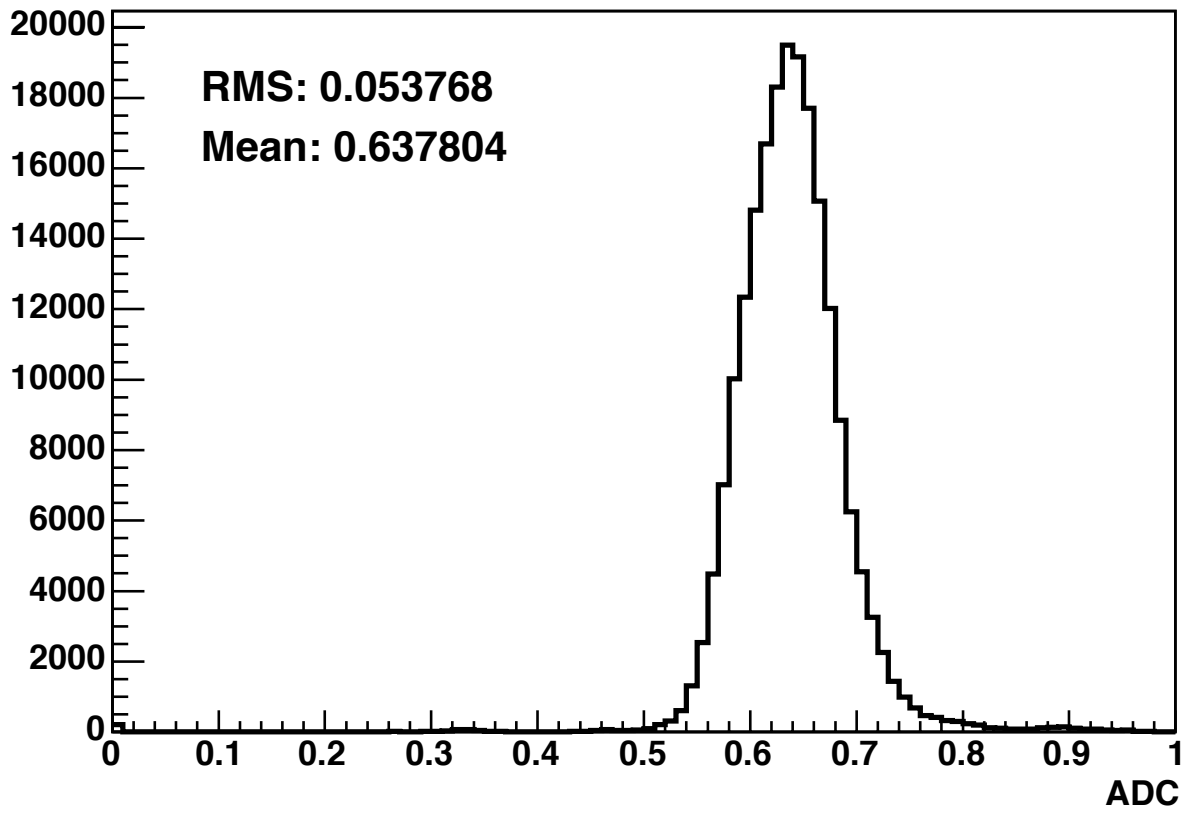

Noise spectrum of the acquisition chain, the mean RMS value is 0.64.

This leads to simulate the noise with a Gaussian probability distribution with a $\sigma$ of 0.6. The jitter is also simulated using a Gaussian distributed noise generator with a varying width $\sigma$ which represents the assumed clock accuracy. To circumvent systematic errors by always reusing the same starting position $t_0$, it is also randomly varied. All parameters are shown in the table at the end of this section.
These simulations result in three sets of samples: only noise, only jitter and noise and jitter together which are then rounded to integer values to add the quantisation noise, and finally fitted separately using the same function as fit function. The start parameters for the fit are the original values of the generated pulse. There are two important parameters of a cluster, the time position and the amplitude. To measure the impact of the jitter the difference of the cluster parameters of the original pulse and the disturbed one are calculated as shown below.

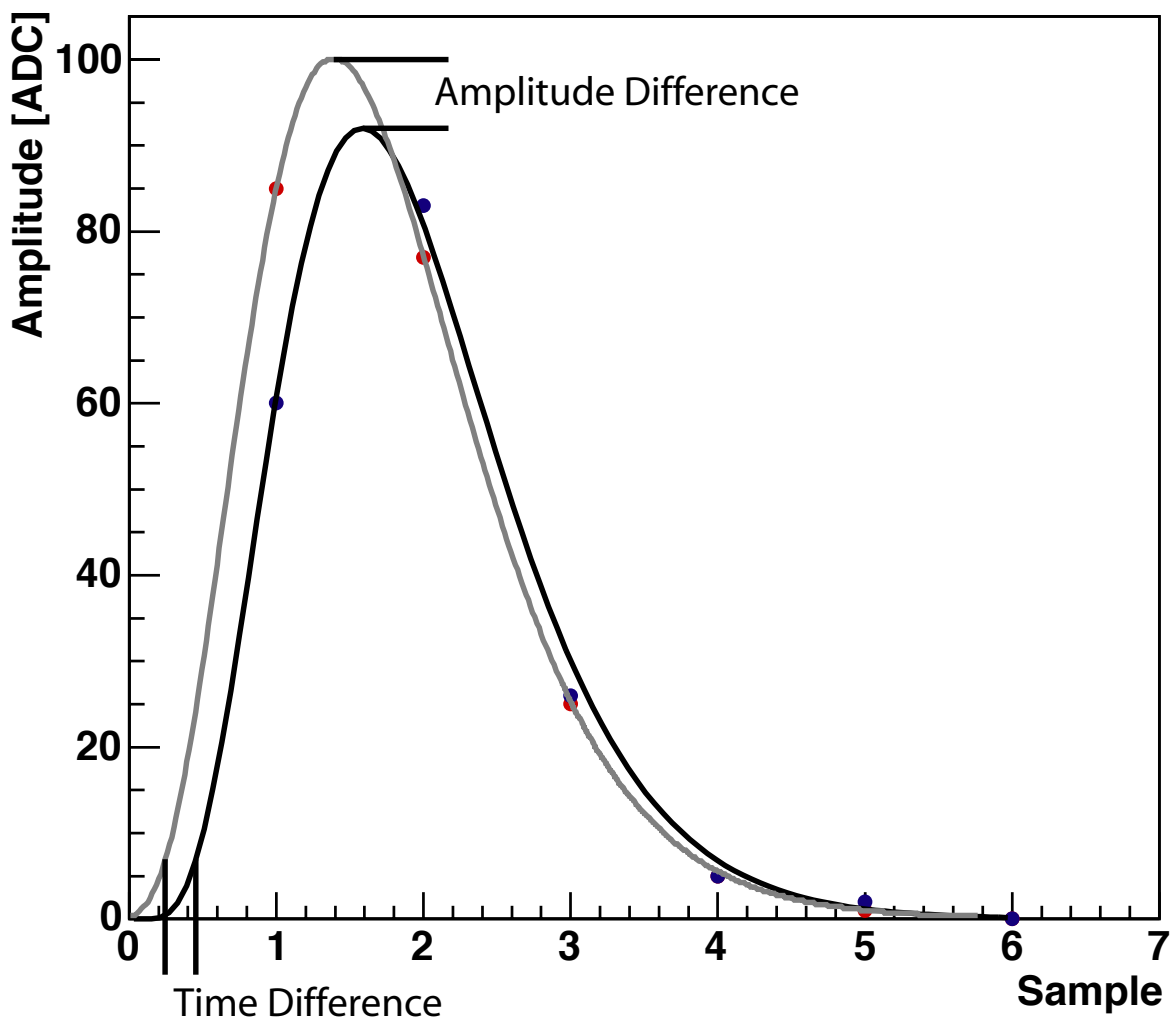

Here, the generated pulse and the fitted pulse is shown. The fit is based on the distorted data points.

| Parameter | Minimum | Maximum | Comment |
|---|---|---|---|
| A | 25 | 1000 | 18 steps |
| $t_0$ | -0.5 | 0.5 | Flat random |
| $\tau$ | 1.5 | 1.5 | Fixed |
| Jitter | $\sigma$ = 0 ns | $\sigma$ = 2 ns | Gaussian random |
| Noise | $\sigma$ = 0.6 | | Gaussian random |

Parameters of the simulation

## Results

These two differences, in the amplitude and in the time, are collected out of 2000 differing generated sets of pulses. From this distribution the RMS is calculated. For the amplitude difference the distribution is fitted by a Gaussian to get more stable results, as big amplitude differences are the consequence when the fit is not converging. The plot below shows the introduced error when increasing the amount of jitter by widening the $\sigma$ of the random generator and a fixed amplitude of 100 for the amplitude precision. The subsequent plot shows the time accuracy. In both plots the red line shows the influence of the noise alone, the green of the jitter alone and the black of both respectively. As expected, the noise introduces a constant error and the error of the jitter increases with the decreasing accuracy of the clock.

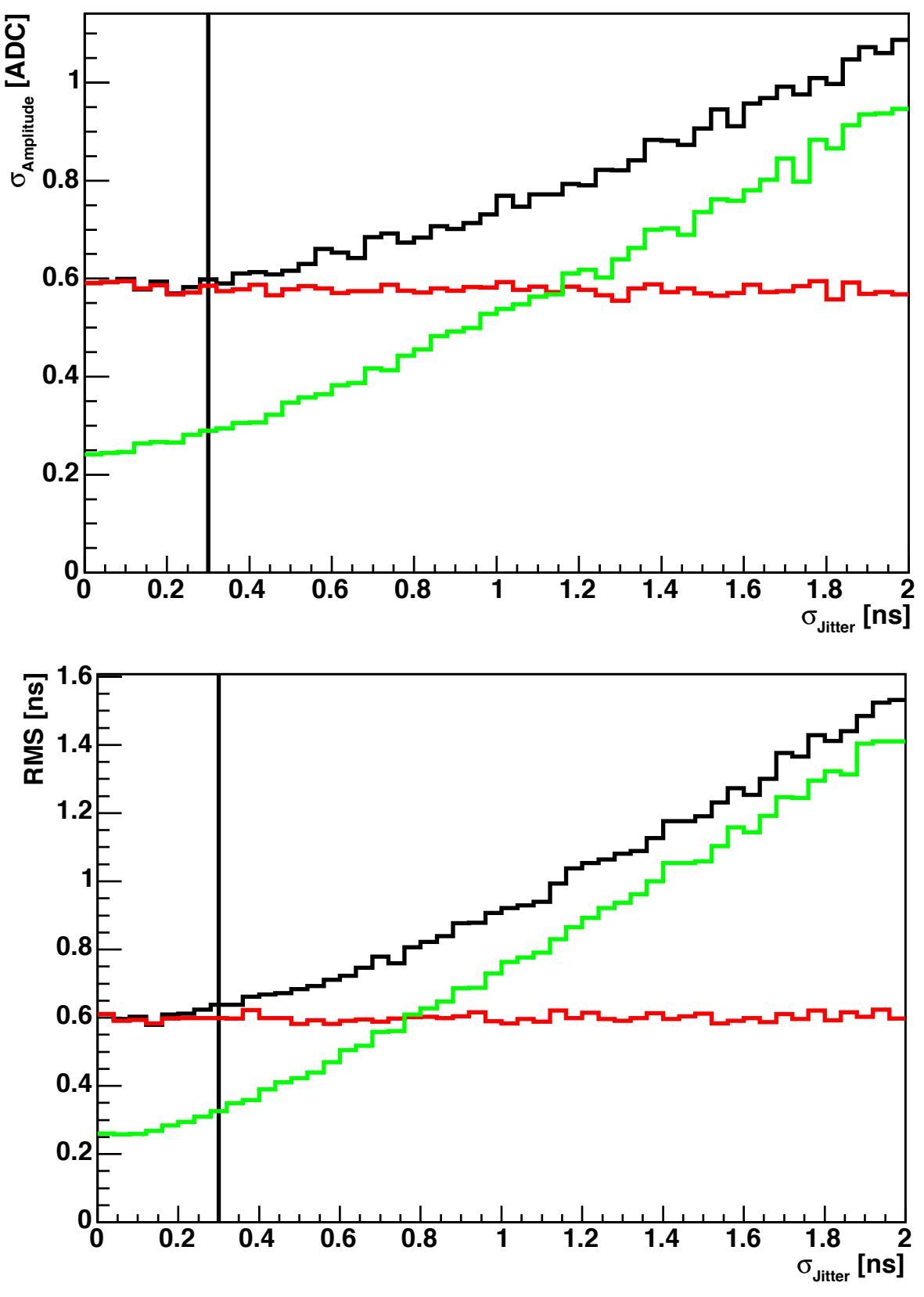

Error in the amplitude (top) and time (bottom) measurement with increasing amount of jitter. The red line is the influence of the noise alone, the green line is the influence of the jitter alone and the black line is the combination of both. The vertical black line indicates the expected clock accuracy.

The ADC value of a cluster in the TPC data is less probable, the higher the value is. This is shown in the plot below:

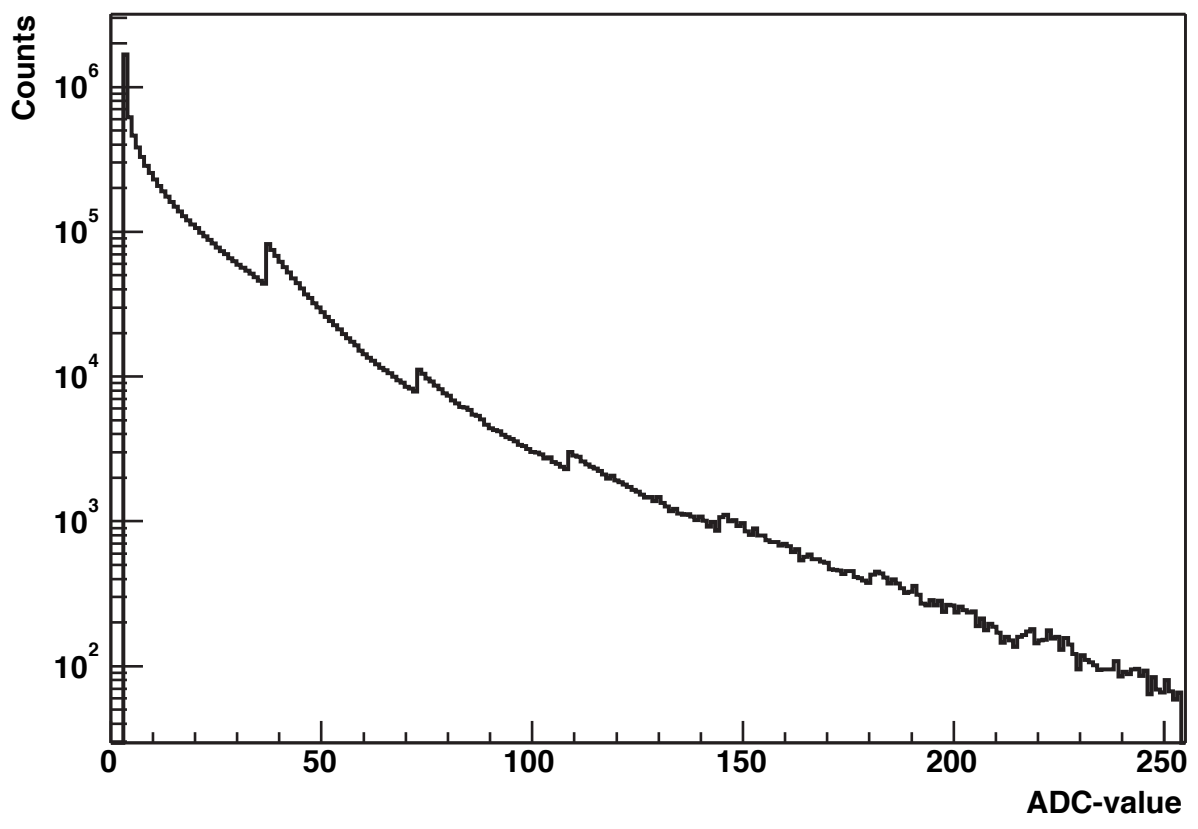

Distribution of the ADC values in a simulated event of the TPC [1]

The simulation was done for 18 different amplitudes starting from 25 up to 1000. The results are shown in the three following plots for the amplitude in absolute error and relative error and the time error. The red line shows the crossover in between the noise and the jitter as the main error source, the black horizontal line indicates the expected inaccuracy of the clock.

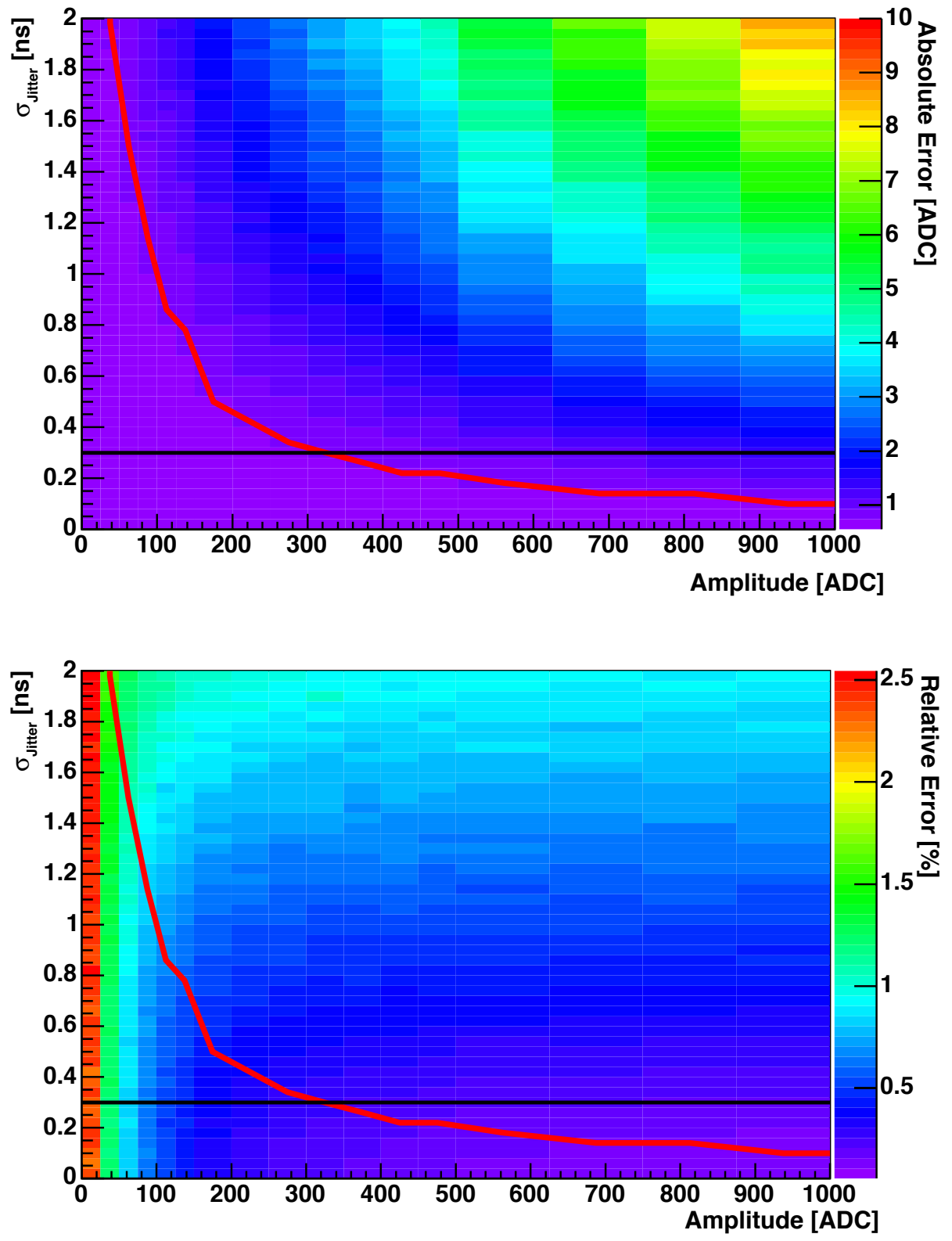

Absolute error (top) and relative error (bottom) in the amplitude measurement across all simulated amplitudes and jitter. The red line indicates the crossover in between the noise and the jitter as main error source. The horizontal, black line indicates the expected accuracy of the clock.

In the previous three plots the error of the absolute amplitude (top), the relative amplitude (middle) and the time (bottom) measurement at the foreseen clock inaccuracy of 0.3 ns is shown. It is clearly visible that in the domain of the most probable ADC values (≈ 30 ADC for a MIP) the effect of the jitter plays a minor role compared to the influence of the noise.

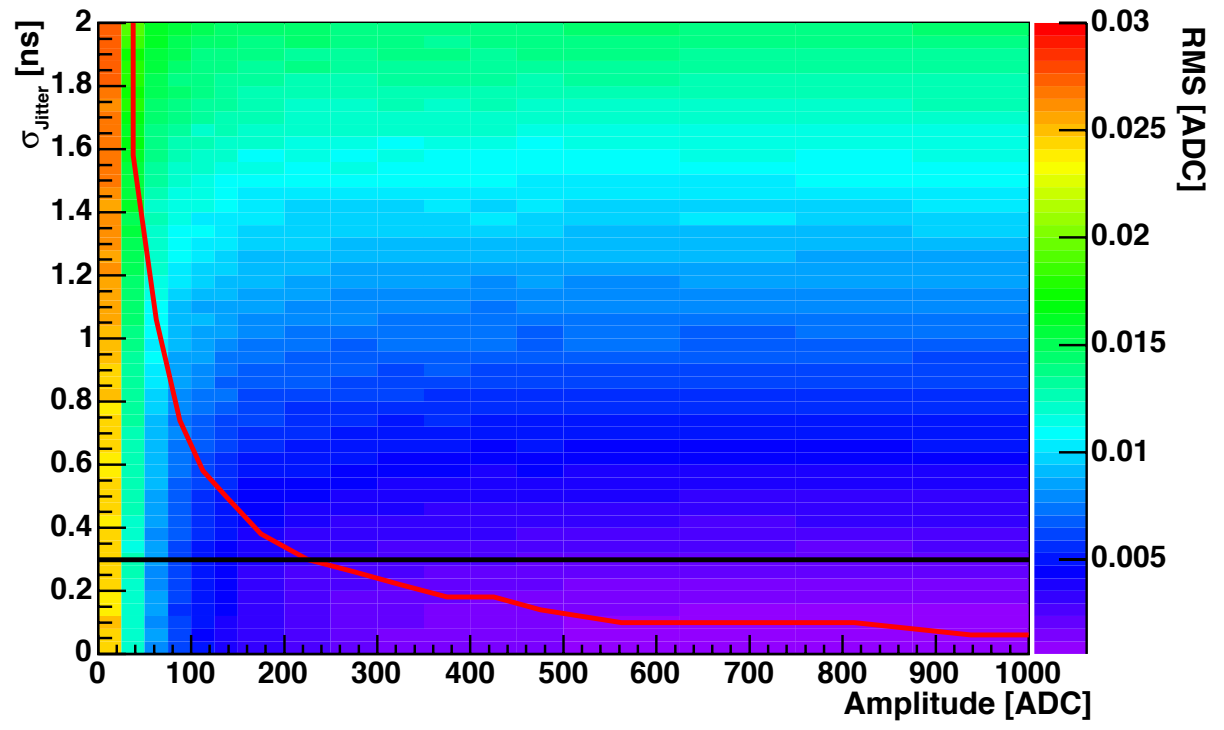

Error in the time measurement across all simulated time accuracies and jitter. The red line indicates the crossover in between the noise and the jitter as main error source. The horizontal, black line indicates the expected accuracy of the clock.

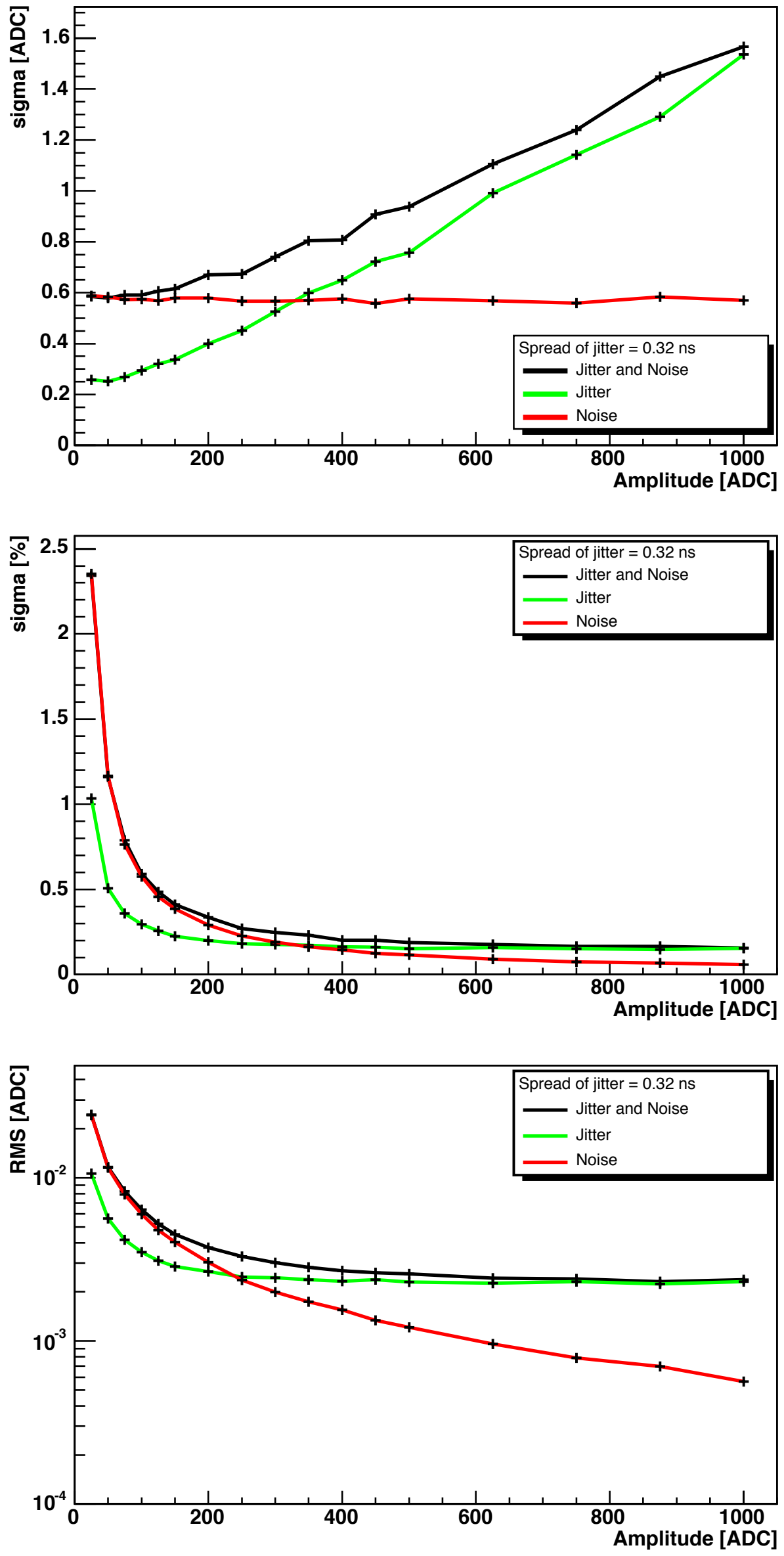

Error in the absolute amplitude (top), the relative amplitude (middle), and the time (bottom) measurement at the foreseen clock accuracy.

# Prototype Environment

# Prototype

A small prototype was built to do a TPC performance test. It consists of one IROC (**I**nner **R**ead**o**ut **C**hamber) module on one side and a complete field-cage with a central membrane in a gas-tight aluminium box. First tests were done to verify the electrostatic behaviour without a readout system. Later, for the complete TPC performance test, it was equipped with four FEC boards and a cosmic ray triggering setup using scintillators as partially visible in the picture below.

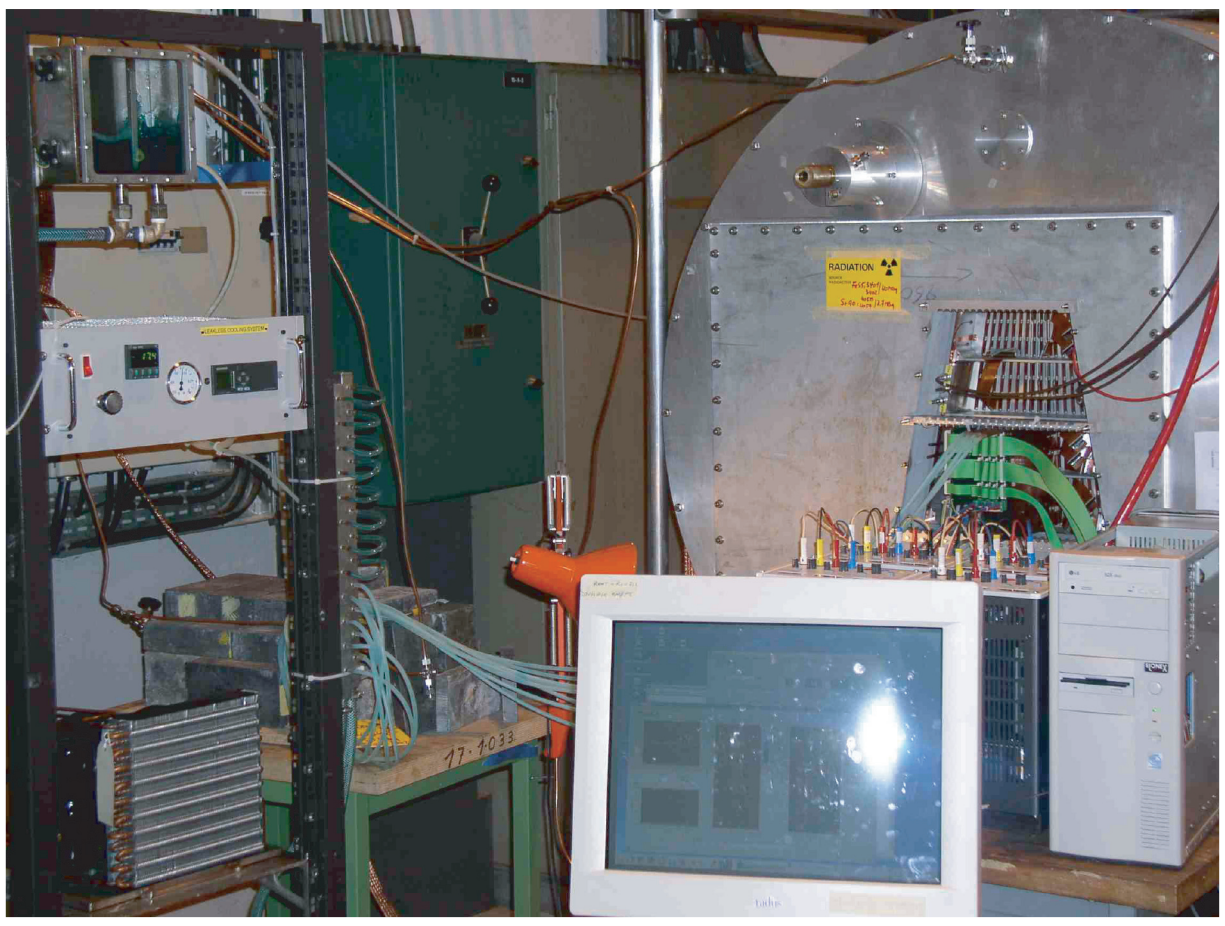

Picture of the prototype for cosmic ray measurement in hall 167 at CERN. Here, four FEC (behind the green flatband cable) and cooling (white silicon tubes) are equipped, in the front is the LabVIEW based DAQ PC.

At this time, the front end cards were not cooled and the test TPC was filled with Ar/$CO_2$ 90/10 as gas mixture. Later, a simple cooling setup and four additional cards were added and the gas mixture was changed to Ne/$CO_2$ 90/10. The cooling system is based on a underpressure liquid cooling and a copper shielding around the FEC as shown in the FEE chapter (page 16). This setup was used to gather some statistics of cosmic particles on MIPs as the most probable particle, cosmic showers as an estimate for the high multiplicity environment and particles with a big ionisation to see saturation effects, also two different gas mixtures were used to study differences in the signals. The data is available at [1].

| | |
|---|---|
| Length | 2.7 m |
| Diameter | 1.1 m |
| Drift Length | 1.35 m |
| Anode Voltage | 1245 V |
| Fieldcage Voltage | 55.8 kV (400 V/cm) |
| Readout Channels | 512 later 1024 |
| Gas Mixtures | Ar/$CO_2$ 90/10<br>Ne/$CO_2$ 90/10<br>Ne/$N_2$/$CO_2$ 90/5/5 |
| Oxygen Content | 50-65 ppm |

Detector parameters of the test TPC. The third gas mixture was used at the testbeam.

# IROC & Mapping

The endcaps of the ALICE TPC are circular and parted in 18 trapezoidal segments on each side. The pad plane follows this scheme and is subdivided in two parts, the IROC and OROC (**O**uter **R**ead**o**ut **C**hamber), as shown in the scheme below. Another subdivision is given by the order of the FECs with six rows of them, two on the IROC and four on the OROC, these subdivisions are called patches. Due to this trapezoidal form, different numbers of pads per row occur as well as three different pad sizes are used in the different patches and hinder a trivial direct mapping of the pad to one readout channel. Each pad has an unique number per ROC (**R**ead**o**ut **C**hamber), starting with 0 as the left pad in the innermost row and then counting every pad until the outermost row. A table of the pad index correlation to pad and row, FEC, cable, connector and pin for the IROC and OROC exists in [2].

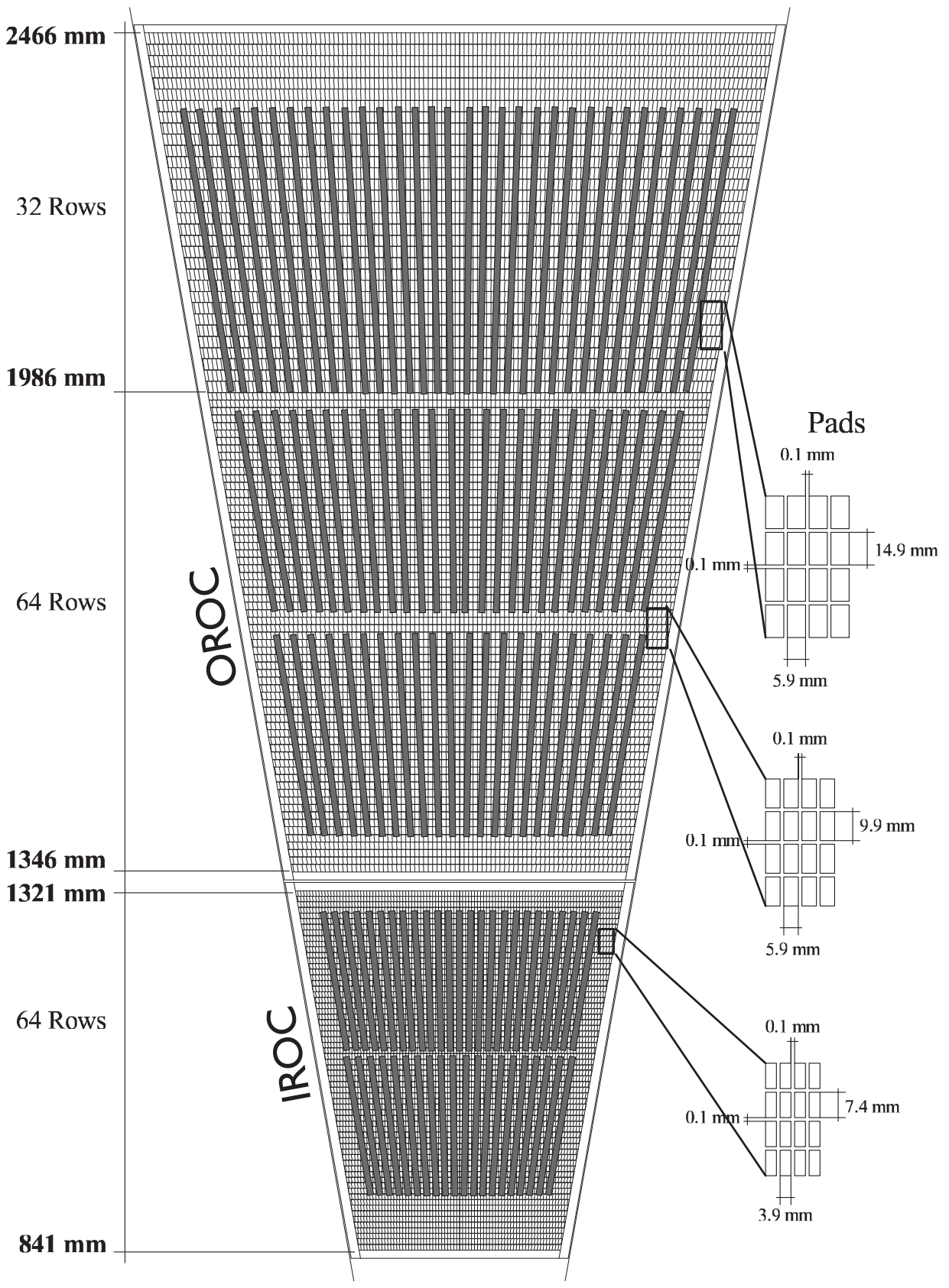

Distribution of the FEC on the Readout chamber. Both the IROC and OROC are shown. This is still the old layout with only 5 rows of FEC.

# Data Acquisition & Configuration

The data was read out via the ALTRO bus with an old version of the RCU often called RCU1 via a flat cable (the green cable in the picture of the previous column). This version was based an a commercial PCI (**P**eripheral **C**omponent **I**nterconnect) card, a board from PLDA [3] with a FPGA and a commercial PCI core also from PLDA [4]. Additionally there is a custom made mezzanine card on top of the RCU1 which implements the interface to the FECs via the ALTRO bus. The host operating system was Linux [5] with a low level PCI driver as interface to the RCU. The registers of the firmware in the FPGA were mapped in the address space of the computer. A set of small C [6] routines handles the communication to the driver as well as the coding and decoding of the readout memory where the data is stored in the ALTRO format as described in »Data Format« on page 33. These C routines are interfaced with a LabVIEW [7] based GUI (**G**raphical **U**ser **I**nterface). The LabVIEW software also implements the control, setup logic, graphic displays of the running status and storing of the data. The choice of LabVIEW determined the data format and the speed of the acquisition. The speed was limited to roughly storing 2 MByte/s with the equivalent of a 1 Hz event rate for 1024 channels. With the trigger rate for cosmics, especially when adopting the trigger to select high multiplicity or high ionisation events, this speed was more than sufficient. The data format as described in the two following »Data Format« sections is big endian encoded. There is the historical approach of building little endian systems, which means that they are working in low byte high byte order, as this approach needs less transistors. Outdated CPU architectures like x86 (for the Intel IA32 line and AMD Intel compatible line in 32 and 64 bit), the Intel Itanium and the Digital Alpha are using this scheme. There also are the big endian systems, which are working in order meaning high byte, low byte. This scheme is more efficient in handling integer data. Most CPU architectures, like the IBM and Motorola PowerPC platform, SPARC and MIPS are following this paradigm. Since LabVIEW has its origin on big endian systems they only use this format when writing binary data independent of the hosting platform. When reading these files with another program or programming language on a big endian system there is no problem, but when reading them on a little endian system the byte order has to be swapped as described in the following table.

| Length | C++ name | Big Endian | Little Endian |
|---|---|---|---|
| 1 Byte | char | $B_0$ | $B_0$ |
| 2 Byte | short | $B_1,B_0$ | $B_0,B_1$ |
| 4 Byte | long int | $B_3,B_2,B_1,B_0$ | $B_0,B_1,B_2,B_3$ |
| 8 Byte | long long | $B_7,B_6, \ldots ,B_1,B_0$ | $B_0,B_1, \ldots B_6,B_7$ |

Table showing the difference in the big endian and little endian coding.

### Data Format One

The first data format is in principle no format on its own, as the data is just written as a continuous stream of the ADC values in one channel as big endian coded short integer (16 bit) numbers over all channels in the increasing ALTRO address order. In other words, they are written in the readout order. Without the knowledge of the configuration as number of samples per channel and number of channels, which are not included in the file, the data is not decodable. A comparison with the successor format is shown in the table in the next column.

### Data Format Two

The second data format adds a header in front of the data, which is then saved in the first format. The header consists of the number of channels, the list of active channels, again the number of channels, the number of samples per channel and the data block. The double number of channels is only because of the way LabVIEW stores the data. In the header short and long ints (32 bit) are used. In this format now all needed data to decode the file is included.

| | Format 1 | Format 2 |
|---|---|---|
| #channels | - | long int |
| channellist | - | #channel * short int |
| #channels | - | long int |
| #samples | - | long int |
| data | n * short int | #channels * #samples *short int |

Data formats of the LabVIEW based readout software

## Monitoring

In this setup a quasi online monitoring system was integrated to offer a way for fast visual inspection of the data. Implementing a monitor by using LabVIEW would introduce a number of drawbacks like low speed, high complexity in debugging, complicated maintenance and expensiveness, since it would depend on a commercial product licence. As platform for the monitoring, ROOT [8], a well known data analysis framework, was chosen. A set of C++ [9] classes were developed to have a fast data decoding and analysis and were packed in one .so (**S**hared **O**bject). The interactive part is based on a set of CINT (**C Int**erpreter) [10] macros which call the functions and classes of the compiled .so. This adjudication of extending ROOT with the needed classes made the monitoring package independent of the running platform, as long as ROOT is available [11]. Since the online monitor is running in the interactive mode of ROOT all plot manipulation capabilities are available, as well as the save functionality. Additionally for experienced ROOT users, all C++ functions available in CINT are accessible, as well as the loading of private analysis macros to further process the data. The monitoring is working on a request base, so if the user requests a new event, the newest stored file is read and then displayed.

As stated before, the data is stored in the readout order which means starting from the first going on to the last FEC and in each FEC the channels are sorted in ALTRO addresses. Since there are four ALTROs on each side, the addressing on the top side is in FEC-channel order and the ones on the backside in the reverse order. The connection lines to the PASA are optimised to be short, so that the first eight channels of the ALTRO are in order, the last eight in reverse order. This internal ordering was added to the mapping table of the IROC. The pad index was replaced by the channel number in readout order. The readout order to pad mapping is different for each position of the FEC on the test TPC. This means that for every configuration in number and position of the FEC a different mapping table is needed. In the picture below the topview of the maximum ADC value of each channel of the eight FECs on the test TPC is shown.

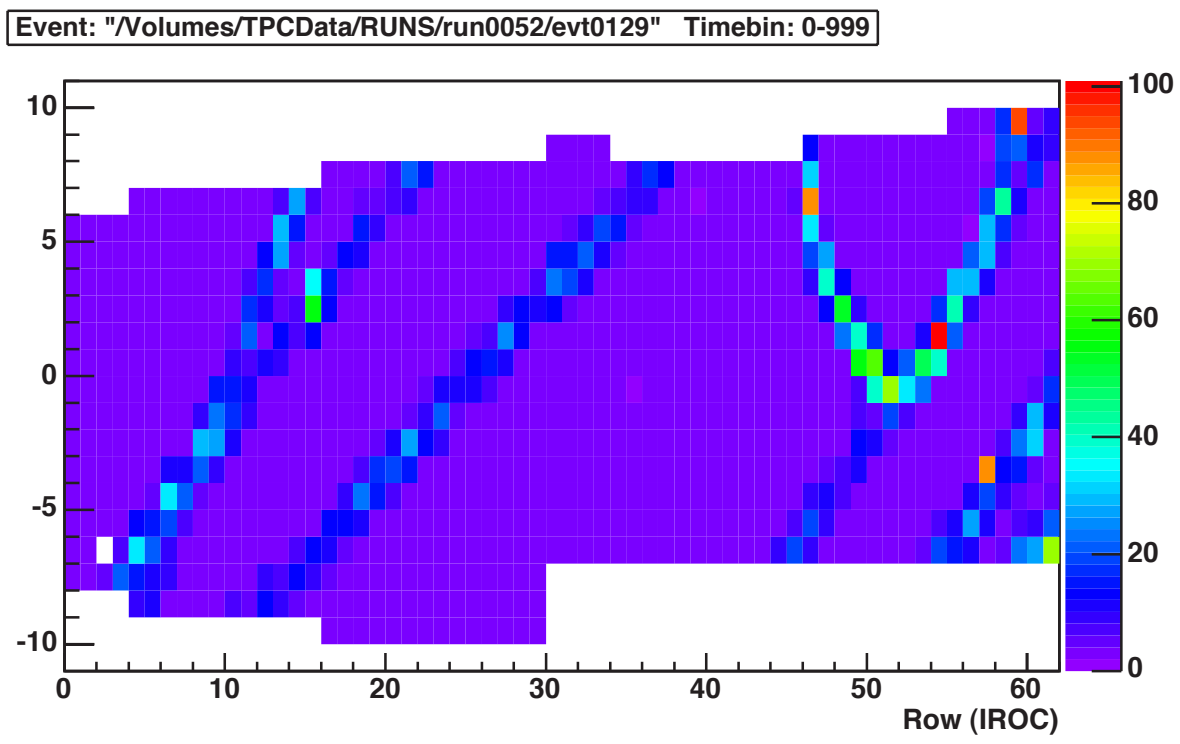

Topview of a cosmic event of the test TPC equipped with 8 FEC. Each bin shows the maximum ADC value of the corresponding channel.

To get a useful topview of the TPC the baseline has to be subtracted. There is either an online or an offline calculation function available. The online version is based on a double pass calculation, the first pass calculates the mean over all timebins regardless on the variance, the second pass then sets a double threshold scheme around the previously calculated baseline and recalculates it by only using the samples inside the window. For low occupancy events, this approach provides results with a negligible error compared to the offline version. The second method called offline favours either a special pedestal run or uses a normal data run. When having a pedestal run, the mean of each channel is calculated, then averaged over all events and stored. The scheme is slightly more complex when having no special run. Then only channels without a signal are used for the calculation what was possible, as most cosmic events have a very low occupancy. The pedestals are stored in the format readout number and pedestal.

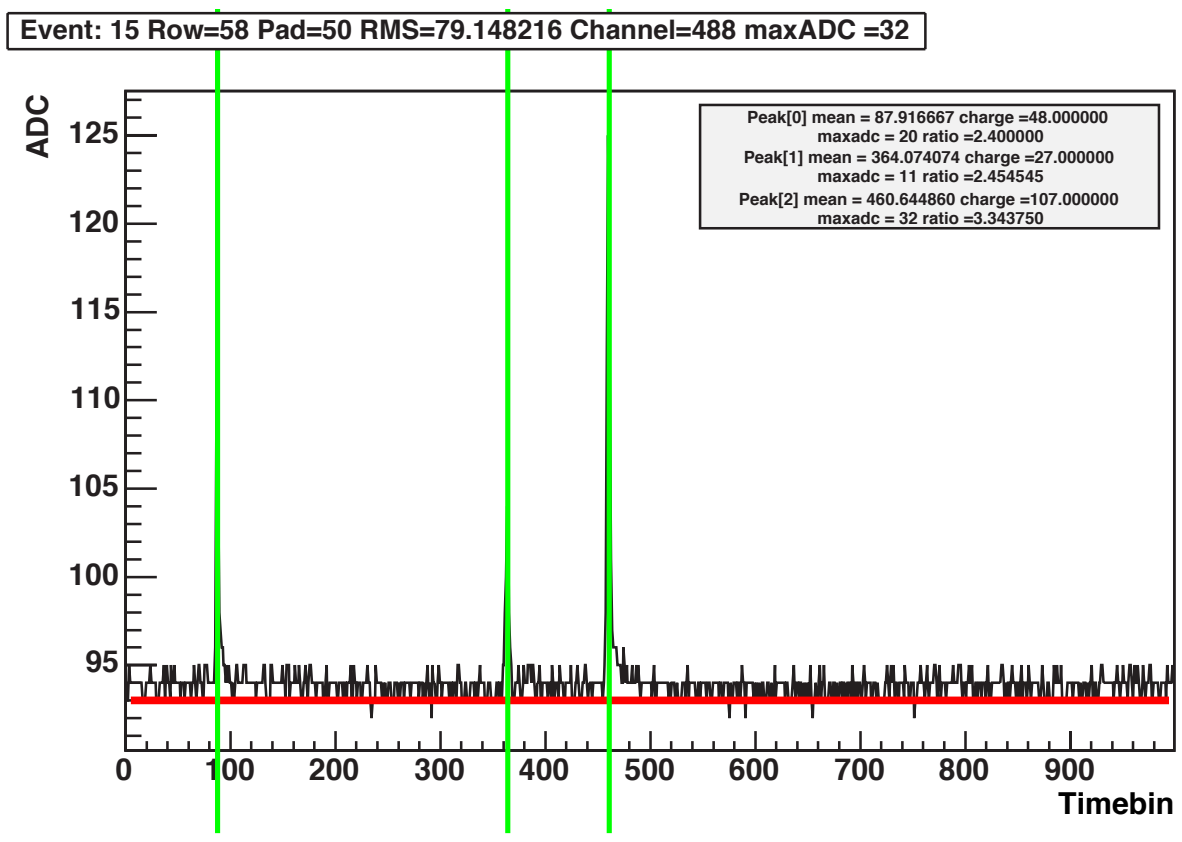

Pad view of a cosmic event. The black line is the signal, the red line is the baseline, the green lines are at the position of the weighted mean of the peak. In the grey box the parameters of each found peak are shown.

From the topview, the pad view is accessible by just moving the mouse over the pads. As soon as a new pad gets in the focus the channel view is updated with a maximum rate of more than 10 Hz. In the channel view, the baseline, a moving average calculation, the ALTRO++ calculation (see »ALTRO++« on page 43) and a pulse finder, as well as a zoomed view around the baseline to study the signal tail is available and configurable.

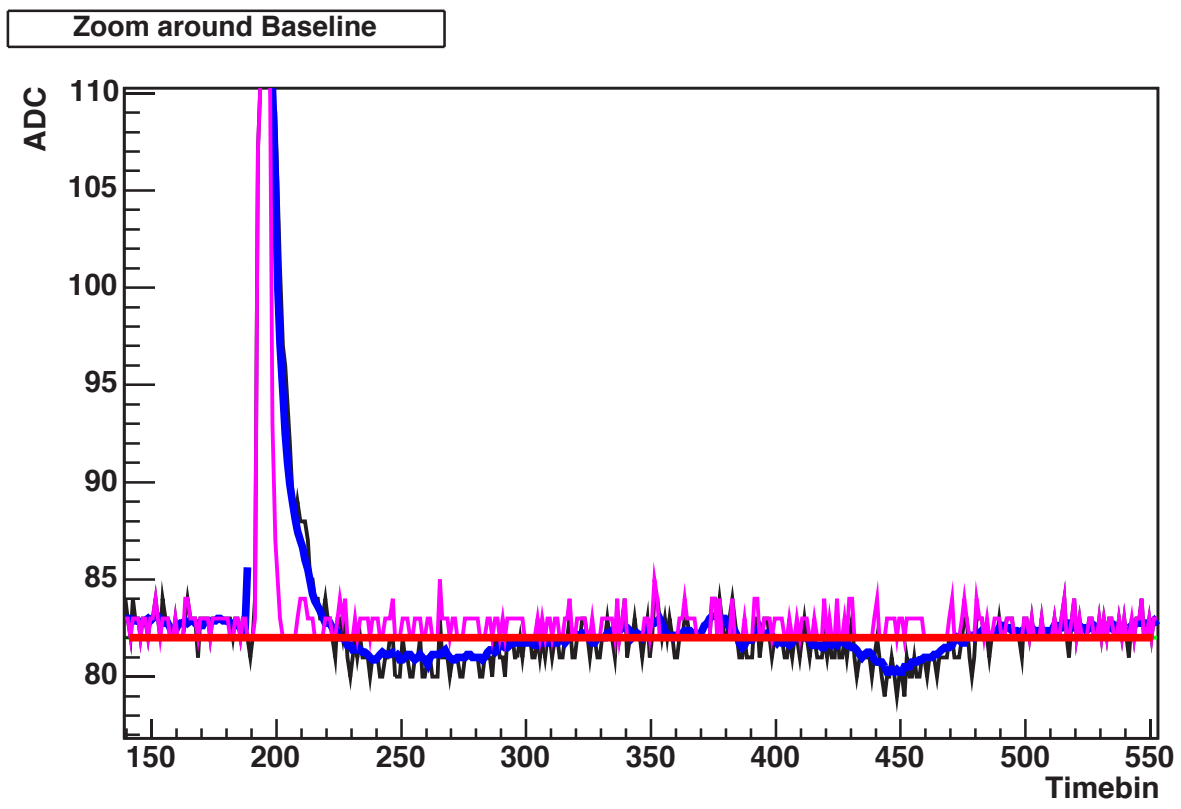

Zoomed view of a pad of an cosmic event. The black line is the signal, the red line is the baseline, the blue line is the moving average calculation and the pink line is the ALTRO++ emulation.

When clicking on a pad and holding the button the pad view is frozen to keep the wanted pad to allow to modify the plot like zooming in or fitting the signal or saving the plot as eps, ps, svg or gif (eps: **E**ncapsulated **P**ost**s**cript, ps: **P**ost**s**cript, svg: **S**calable **V**ector **G**raphics Format, gif: **G**raphics **I**nterchange **F**ormat).
On a double click the row view is opened as shown below. Here the ADC are displayed encoded in colour, a cluster finder is available, both the centroids of the pulse finder in each channel and then the merged cluster are displayable.

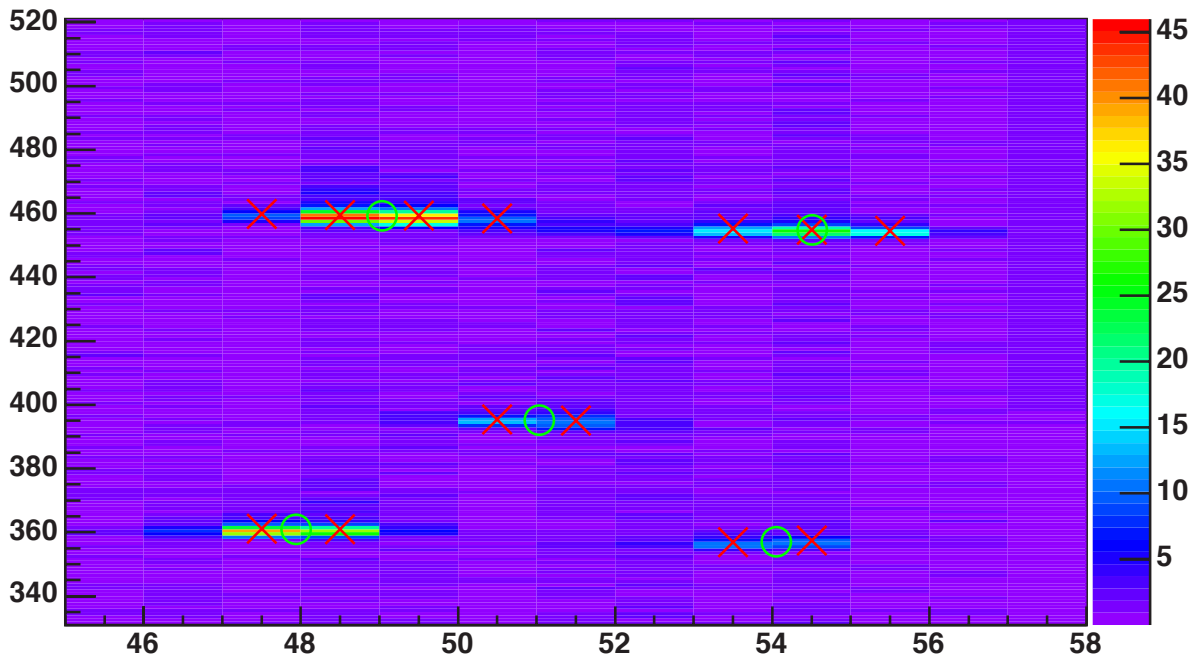

Zoomed row view of an cosmic event. Red crosses are the sequence mean values and the green circles are the cluster centroids.

The configuration was splitted in two files, one for the configuration of the monitor behaviour, like screen resolution or the configuration of the moving average display, and the other file for the run related parameters like the path to the run files or the pedestal files (see »Appendix« on page 65). As an independent entity there is a macro which could draw a 3D representation of the ADC values in the event.

This uses the OpenGL (**Open G**raphics **L**ibrary) [12] capabilities of ROOT which allows to zoom, move and rotate the display in real time.

## Testbeam

In spring of 2004 there was a beam test with the described setup in the PS (**P**roton **S**ynchrotron) testbeam, but fully equipped with FECs and cooling as well as the use of the gating grid during data acquisition. Since the testbeam was also seen as integration test of the complete data and trigger chain a preversion of the final RCU, the DCS Board a DATE based detector readout and a small trigger setup were used. The previously used flat cable was exchanged by the current version of the backplane. This large amount of changes had a big impact on the software needed to configure, control and monitor the TPC. The TPC gas was changed to the final ALICE choice of $Ne/N_2/CO_2$. In addition to the TPC, there was a silicon telescope and a TOF detector present and included in the trigger and data acquisition system.

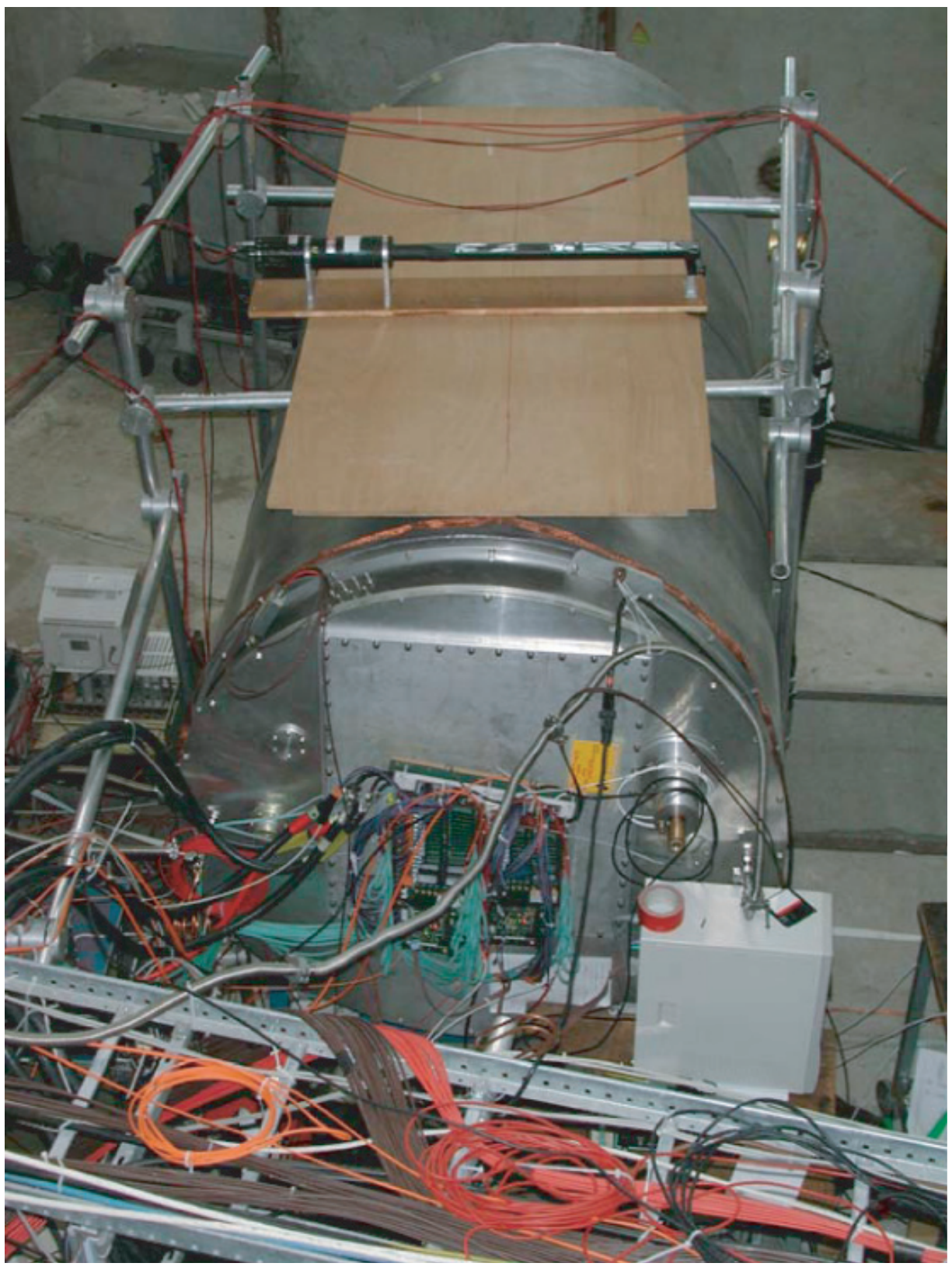

Picture of the TPC prototype at the PS testbeam. The beam enters from the right. On top of the aluminium barrel a cosmic veto trigger is located.

## Configuration

The configuration of the front end electronics is completely different compared to the previous setup. The prior setup with a PCI based RCU1 has changed, the PCI path is removed and there are now two paths to the RCU3, one is the DDL via the provided low level communication library and the FeC2 (**F**ront **E**nd **C**ontrol and **C**onfiguration) script language for easy development and debugging data transfer [13] and an Ethernet [14] connection to the DCS board. The internal communication layer of the DCS board and the steering host above it is based on the DIM (**D**istributed **I**nformation **M**anagement) client server system and is called InterCom Layer [15]. It is foreseen to implement the InterCom layer also over the DDL to get the same access via the different physical layers. This will replace the FeC2 script language or the low level DDL communication library which have no user C/C++ interface.

## Readout

The readout is now done via the DDL and the DATE system [16], what made a new online monitoring scheme necessary. When the electronic is set up and the trigger is started the data acquisition is started. There are two patches in one IROC, as well as two RCU cards and two DDLs for the acquisition system. The data of these two links as well as the data from the silicon telescope and the TOF are then merged in one DATE file.

## Data Storage

During this testbeam roughly 0.6 TB of data were taken. These data were first stored on the local discs of the DATE computers at the experiment and then transferred to CASTOR (**C**ERN **A**dvanced **Stor**age Manager), a CERN central taping system for the LHC data [17,18]. Irritatingly, off site access to the data in CASTOR is complicated and insecure.

## Data Format

The usage of DATE as readout system also introduced a complete new data format as well as the usage of the new RCU, which introduced the final ALTRO format. Each RCU reads out both branches of one complete patch and converts these 40 bit data into 32 bit data, as the DDL only supports these. This is done as shown in the scheme below.

| | 31 24 | 23 16 | 15 8 | 7 0 |
|---|---|---|---|---|
| 1 | w(1) [31...0] | | | |
| 2 | w(2) [23...0] | | | w(1) [39...32] |
| 3 | w(3) [15...0] | | w(2) [39...24] | |
| 4 | w(4) [7...0] | w(3) [39...16] | | |
| 5 | w(4) [39...8] | | | |

Translation scheme to convert four 40 bit ALTRO words (w(n)) into five 32 bit words by the RCU data sampler.

By this scheme, four 40 bit ALTRO words (w(n)) are converted in five 32 bit words. Ahead of this data block a header assembled by the RCU is added which is similar to the standard DATE event header. This header consists of seven 32 bit words and is shown below. At the end, the total number of 32 bit words is added what is extremely important, because it is needed to retranslate the 32 bit data back into 40 bit. This shown block is called »payload« in the DATE language.

| | 31 … 24 | 23 … 16 | 15 … 8 | 7 … 0 |
|---|---|---|---|---|
| 1 | Format Version | L1 Trigger Type | Res. MBZ | Event ID 1 (Bunch Crossing) |
| 2 | Res. MBZ | Event ID 2 (Orbit Number) | | |
| 3 | Block Att | Participating Subdetectors | | |
| 4 | Res.Bz | Status & Error Bits | | Mini Event ID |
| 5 | Trigger Class Low | | | |
| 6 | ROI | Res. MBZ | | Trigger Class High |
| 7 | ROI High | | | |
| 8 | ALTRO channel 1 data | | | |
| 9 | ALTRO channel 2 data | | | |
| | ...... | | | |
| | ...... | | | |
| | ALTRO channel n-1 data | | | |
| n-1 | ALTRO channel n data | | | |
| n | Event Length (n) | | | |

Complete data block generated by one RCU at the testbeam. The grey entries are set, the white ones are fixed to zero. The Event length is the total length in 32 bit words [19]

From the RCU these data are transferred via the SIU into the DATE system which collects the data from all different data sources and merges these to one event. DATE collects several events in a quite unhandy file format by just concatenating them in the event number order. A scheme of the DATE event structure is shown in the next column. Each event starts with an event header composed of the GDC (**G**lobal **D**ata **C**oncentrator) which consists of the total event size the event id, the event type, the run id and other informations used by DATE. The event id is important, since there are certain types of events without physical relevance which have to be excluded from any analysis or monitoring. After the event header, a subevent header follows which includes the information coming from the first LDC (**L**ocal **D**ata **C**oncentrator) which finishes the data transfer. Now the equipment header follows, it consists of the equipment id and additional data of each DDL, since it is possible to have several DDLs with one LDC as target. Now finally, the previously described data block is concatenated. In the case of the testbeam, both DDLs were plugged in one LDC, so that after the first data payload the next equipment header followed with the data block. The equipment id is used to identify the different RCUs, or taking the detector layout into account, it is used to identify the different patches of the TPC. It is an integer number and unfortunately the meaning of each number is not included anywhere in the files as well as no fixed definition exists, so to decode the data it is mandatory to also know the exact setup of DATE. If there are additional LDCs in the setup, a new subevent header is concatenated with the previously described equipment header and data payload. The number of equipments following each subevent header is not stored anywhere, so it is impossible to cross-check the event structure. At the testbeam, two additional LDCs, one for the silicon telescope and one for the TOF, were installed. The order of the LDCs in the files is mixed, since, as described above the first finished is the first stored. Irritatingly, also in the event header, the number of following subevents is not stored. The only way to get back on track, if an error is in the event structure, would be by scanning for the event magic number in the event header which is fixed and in principle used for identification if a endiannes swap happened. This sequence can also occur in the data, so this method is insecure. In the DATE files acquired during the testbeam some files do not have a correct structure. All readable events up to the point of the incorrect structure are used and the remaining events are discarded. At last, nowhere in an event file a pointer to the included events or even the number of the events is stored, the complete event has to be parsed to get this information. It would have been a trivial task to concatenate this information at the end of the file.

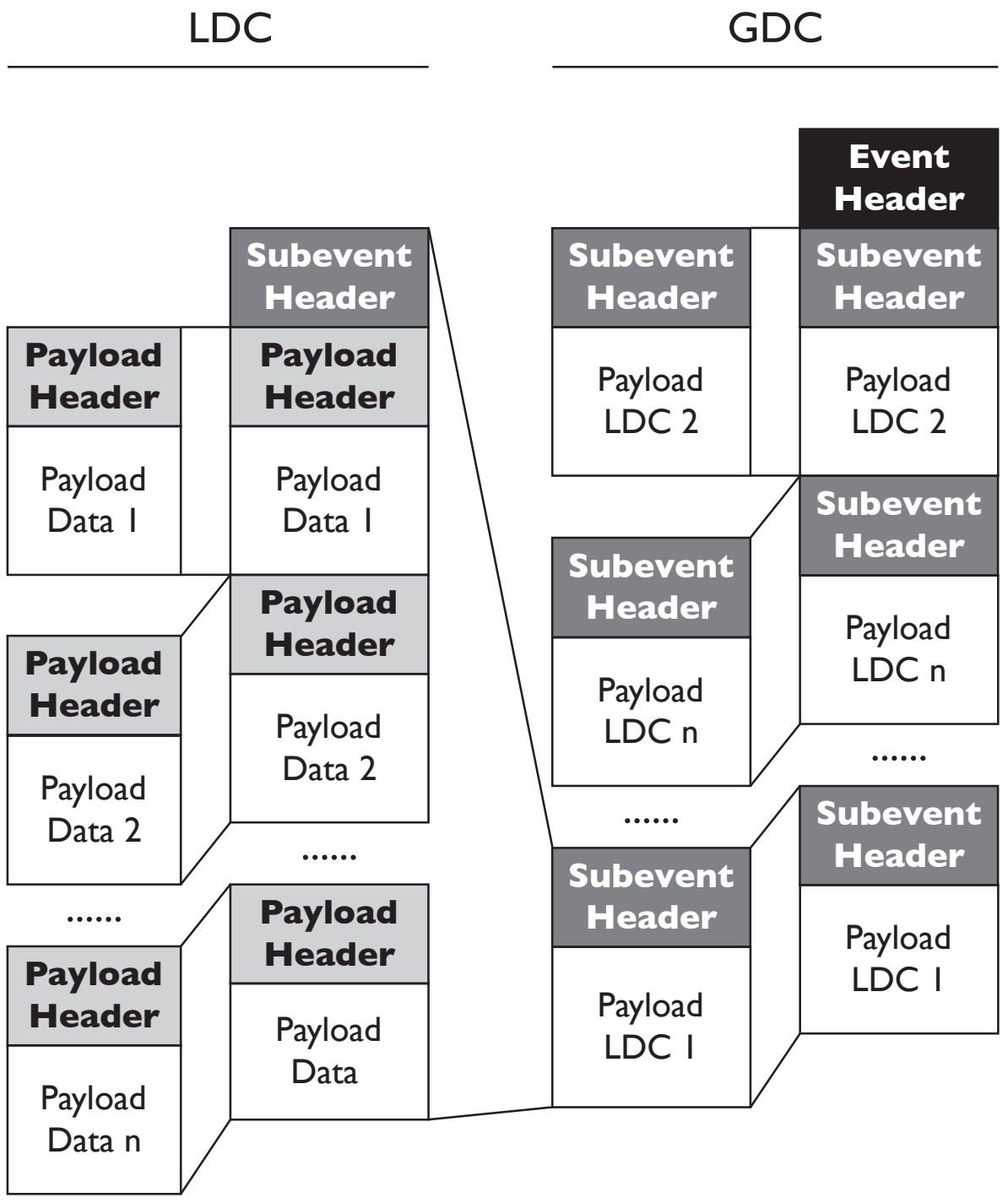

DATE format scheme. Per LDC N payloads are concatenated to a Subevent. Per GDC M subevents are concatenated to one event. Several events are concatenated into a file.

## Monitoring

Plenty of changes and new implementations were needed, e.g. backporting to an completely outdated operating system (Red Hat 7.3) and a non standard compiler (gcc 2.96). To begin with, the idea of having the written event files as interface was not possible anymore, since the DATE group did not want this simple approach. Now there are two run-ning modes in the online monitor, one uses a DATE library (libmonitorstdalone.a) to access the data online and the other mode reads stored DATE files. Irritatingly, there was neither a library nor a class which encapsulates the internal DATE format which is roughly explained above and the available documentation [20,21] was outdated and incomplete as well. Several C++ classes were implemented to get an encapsulation of the monitoring library and of the file access. Based on this, a class to encapsulate the DATE format was implemented, it uses a slightly modified version of the provided DATE event description header file (event.h) to decode the header and gives access to the data payload and the included information in the header. The change in the header file was necessary due to an incompatibility with the dictionary generator of CINT.
The online monitor extracts the two data blocks of each RCU and first translates the 32 bit data into 40 bit data, which are stored in 64 bit integers and then decodes the internal ALTRO data. At the present moment, the RCU firmware does not include the branch number into the ALTRO address, so that every address there is doubled. This is cured via the AltroFormat class which searches for a falling edge to find the crossover point in the branch and then sets the twelfth bit to code the branch. Both encoded data blocks are then merged and the double addresses are eliminated, setting the patchnumber at the thirteenth bit. This can also be extended for more patches by the usage of higher bits. In the mapping, the readout count number was exchanged with the patched ALTRO address, which is now unique for each channel.
During the testbeam several format errors in the ALTRO format encoded data occurred. It happened repeatedly that the total number of 10 bit words in the ALTRO Address was wrong, as well as the two last 10 bit words, carrying the time and pulse length information as shown in the section »DFU« on page 19 was set with wrong numbers. This would render the data decoding impossible, so this was circumvented by setting these numbers to the correct values in the decoding routine. This was possible since all data taken at the testbeam had the same acquisition window.

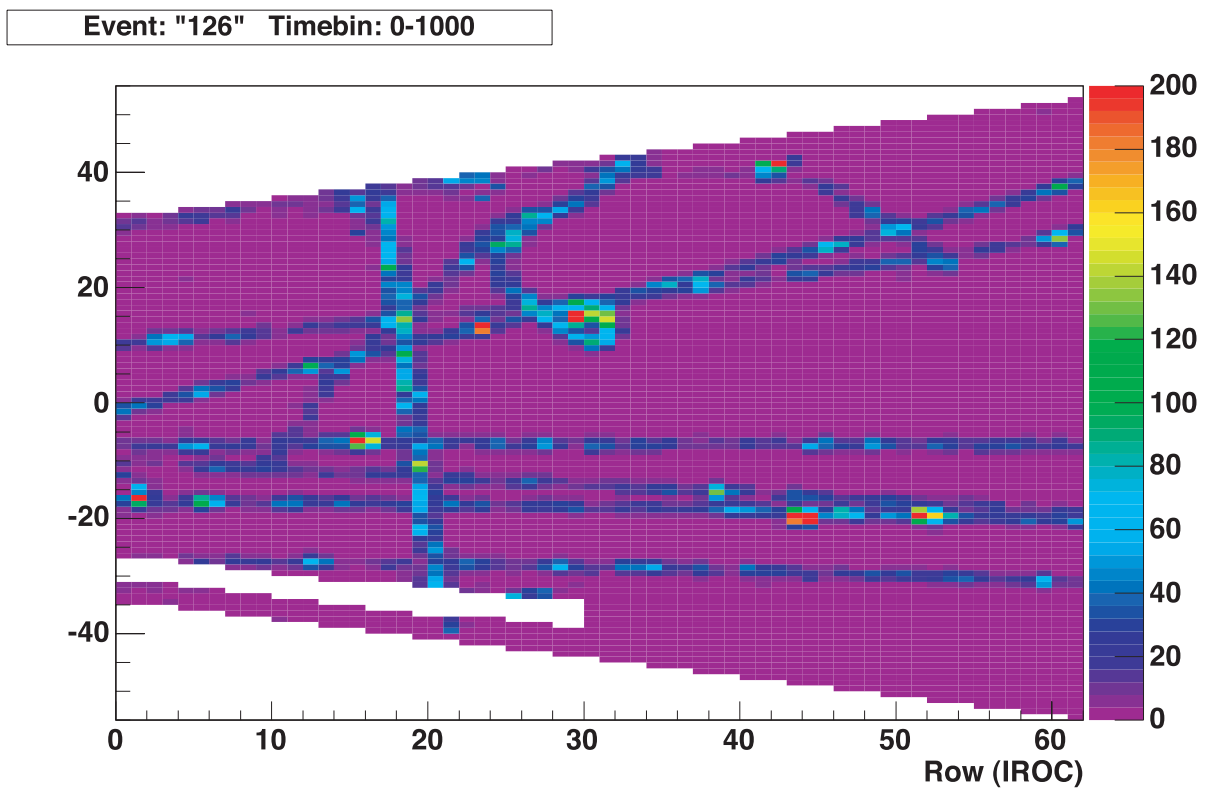

Topview of an testbeam event of the fully equipped test TPC. The beam is entering the topview of the TPC from the left.

Another change is visible in the pad view. There are always twelve samples at zero in the start, this is due to the length of the processing pipeline of the ALTRO [22]. Additionally, the length of each channel is not known, this is set at the time when the ALTRO is configured. Since this is done manually and independent of the data acquisition this information is stored nowhere. The structure in the beginning of the channel is induced by the switching of the gating grid.

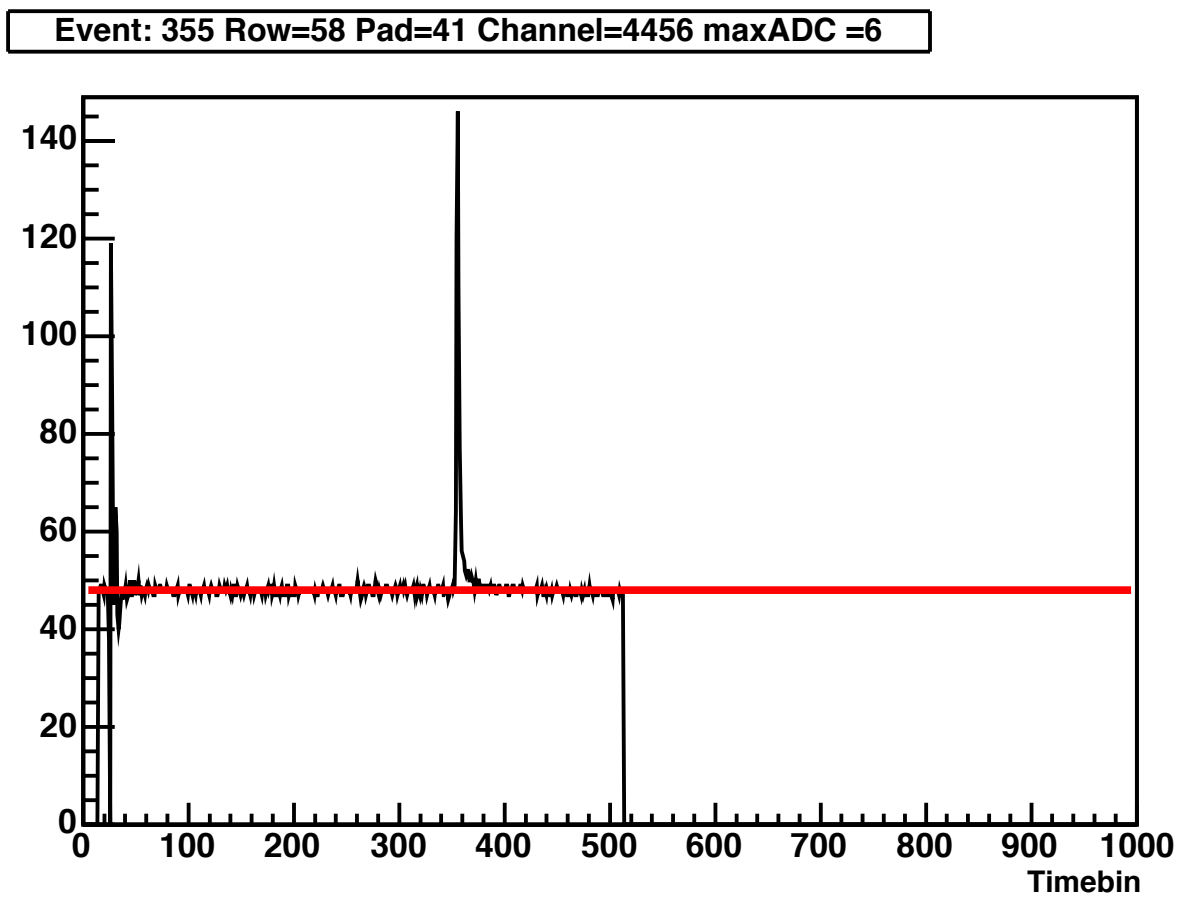

Pad view of a channel from a testbeam event. The first 12 samples are zero due to the length of the processing pipeline of the ALTRO. The acquisition length was configured to 500 timebins. The structure at the start is induced by the switching of the gating grid.

This information could have been extracted from each ALTRO coded channel as long as the zero suppression was turned off. It was foreseen to take data also including the zero suppression, but there were problems in the configuration of all channels with the correct baseline, which is mandatory.
There is a monitoring subsystem from the DATE group, but it was not available during the testbeam, it could not handle the TPC data and the scheme to implement the monitoring of the data in the data acquisition system is also questionable.

## HLT

The HLT was also included in the testbeam as a data receiver as well as data producer. The HLT gets the data payload of both RCUs. In the HLT publisher subscriber [16,23,24] system for the data processing the AltroFormat class was included to decode the ALTRO data. Unfortunately, during the testbeam there was not enough time to also complete the monitoring to run as a HLT client.

# ALTRO Parameter Optimisation

# Parameter Optimisation

The digital processor of the ALTRO has to be accommodated to the detector response of the ALICE TPC by configuring the different processing units in the digital processor. A different scheme to extract the parameters has to be used for each of these units. Since the effects of the processing in the ALTRO are not completely reversible it, is important to have a good crosscheck of the influence of the parameters on the data and the wanted impact during the extraction.

## BCS1 Parameters

Depending on the working mode [1] the BCS1 processing part of the ALTRO needs to be configured with a correct pedestal pattern (f(t): LuT data) and an overall fixed baseline (fpd: fixed pedestal data) for the channels. The extraction of the fixed baseline is described in »Monitoring« on page 31. To extract the pedestal pattern a similar approach was chosen: all channels which have a signal at a certain time after the gating grid effect are discarded. All accepted channels are divided into their timebins and each ADC value is stored in a data array of the dimension channel and timebin. In a different array of the same dimension the counting of found valid ADCs per channel and timebin is saved. After processing all events the mean of these values is calculated and stored with the extended hardware address as unique identifier, as described in the chapter »Prototype Environment« (page 29). The configuration process to finally send this look-up table to the ALTRO is described later in this chapter in »Computing«.

### Extraction

Based on data of a run taken at the TPC testbeam the look-up table was computed. At this step, several problems occurred with the integrity of the stored data. A few data files were not completely written by DATE, so that the event structure was inconsistent, these files were discarded. The more severe problem in many runs was that the hardware addresses of the read out ALTRO channels had errors. Three check criteria were implemented. The first check just verifies that the read hardware address is smaller than the biggest allowed one. This check is more important for the implementation of the mapping table since this is a boundary unchecked array, as usual in C/C++, and accessing a non existing position causes a crash of the program due to memory protection of the OS (**O**perating **S**ystem). The second check verifies that the read hardware address is valid, so that it is in the predefined set of addresses of the IROC module. The third check verifies that each address is unique. When a doublet is found the event is checked again in reverse order to find the second address of the doublet. The error log of run 820 is attached in the »Appendix« (page 66) as an example. To create a pedestal pattern a correct mapping is mandatory what results in discarding all events which have at least one of the previously described errors. The reason of these hardware address problems is most likely located at the RCU firmware level, since there was only a small amount of time to implement the firmware on the new RCU hardware before the start of the testbeam. Timing problems in the FPGA programming could easily generate errors like these. These problems are currently investigated at the FEE group at CERN [2]. The used default values to extract the pedestals are attached in the »Appendix« on page 67.

### Result

An interesting part of the signal is the influence of the gating grid switching which should be a systematic effect constant over time. The plot below shows the signal, the corresponding baseline pattern and the result of the correction.

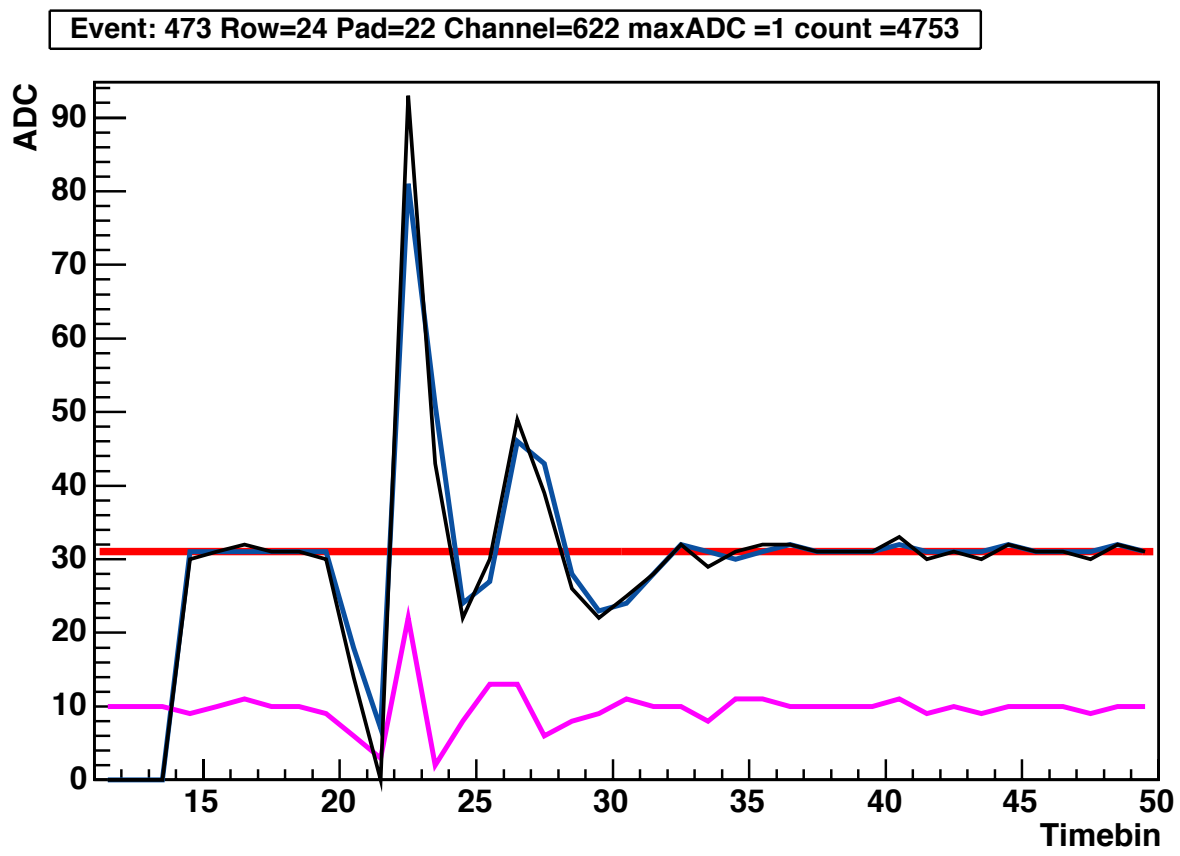

Signal of one channel. The black line is the ADC data, the blue line is the corresponding pedestal LuT, the red line is the fixed pedestal and the pink line is the signal after LuT subtraction. For a better visibility an offset of 10 was added.

The influence of the gating grid switching is substantially decreased by this correction, but not completely removed, as shown in the plot above.

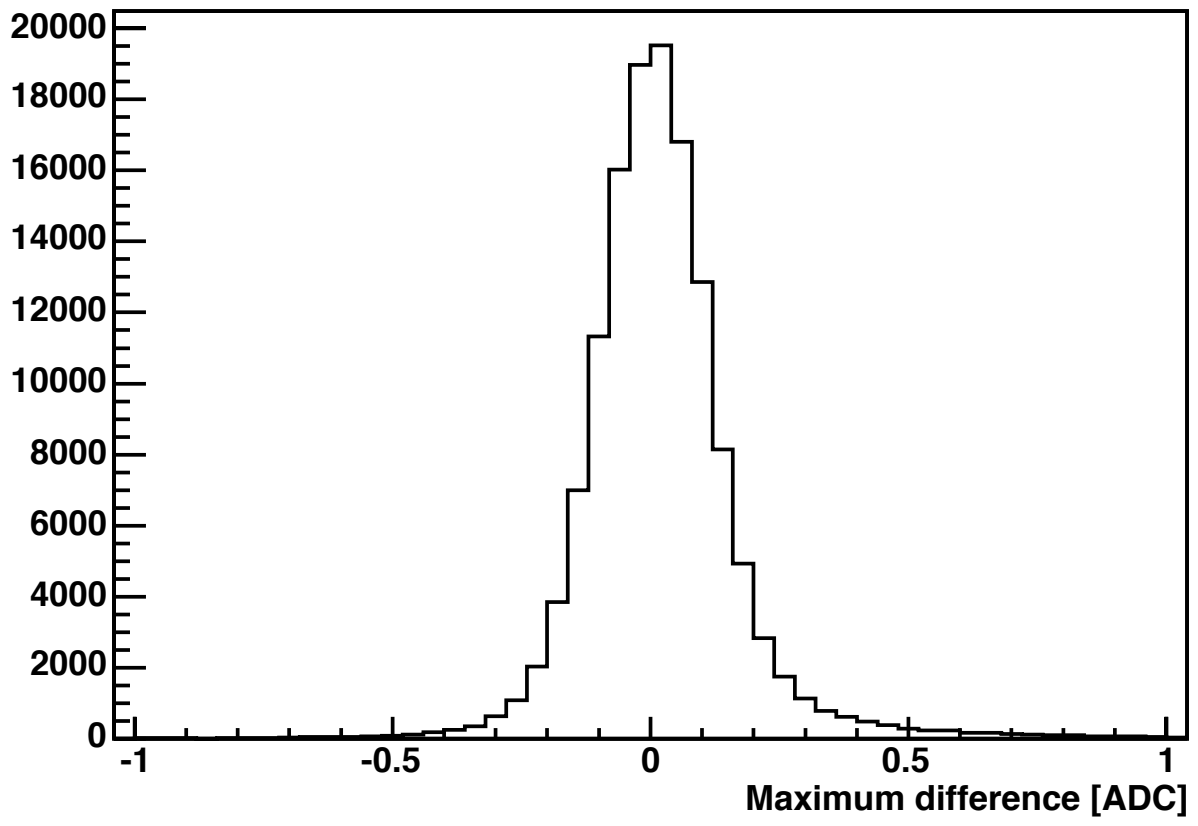

Maximum dispersion of the baseline over the total acquisition time of this particular run (run 820).

To get a general overview of the correction capability on this data, the instability of the data was analysed. At first,

each timebin of each channel over all events was fitted with a line to observe possible baseline variations over time. In the plot before the maximum time dependence of the timebins past the influence of the gating grid is shown, via an extrapolation of the line over the time. It leads to no relevant difference since mostly every slope leads to a maximum difference below the quantisation noise level. For this data the baseline correction capabilities of the ALTRO were turned off.

### Baseline Dispersion

Secondary, the dispersion of each channel and timebin over all events was calculated using the RMS. Only channel-timebin doublets with a small inner noise can be sufficiently corrected by the use of the LuT in the pedestal memory of the ALTRO. For the time domain of the signals past the gating grid influence, this inner noise is sufficiently small and therefore the removal of the baseline is applicable. In the region influenced by the gating grid the effect is not completely correctable since the inner noise is beyond the correctable limits. In the plot below the RMS values of all channels and the specified timebin is shown. The peaks are artefacts of signals since this calculation was done on data with signals, not like the final approach to use a dedicated run with the complete detector and trigger, only without signals inherited by tracks, called pedestal runs. The fit with a line shows that the weight of these glitches is small.

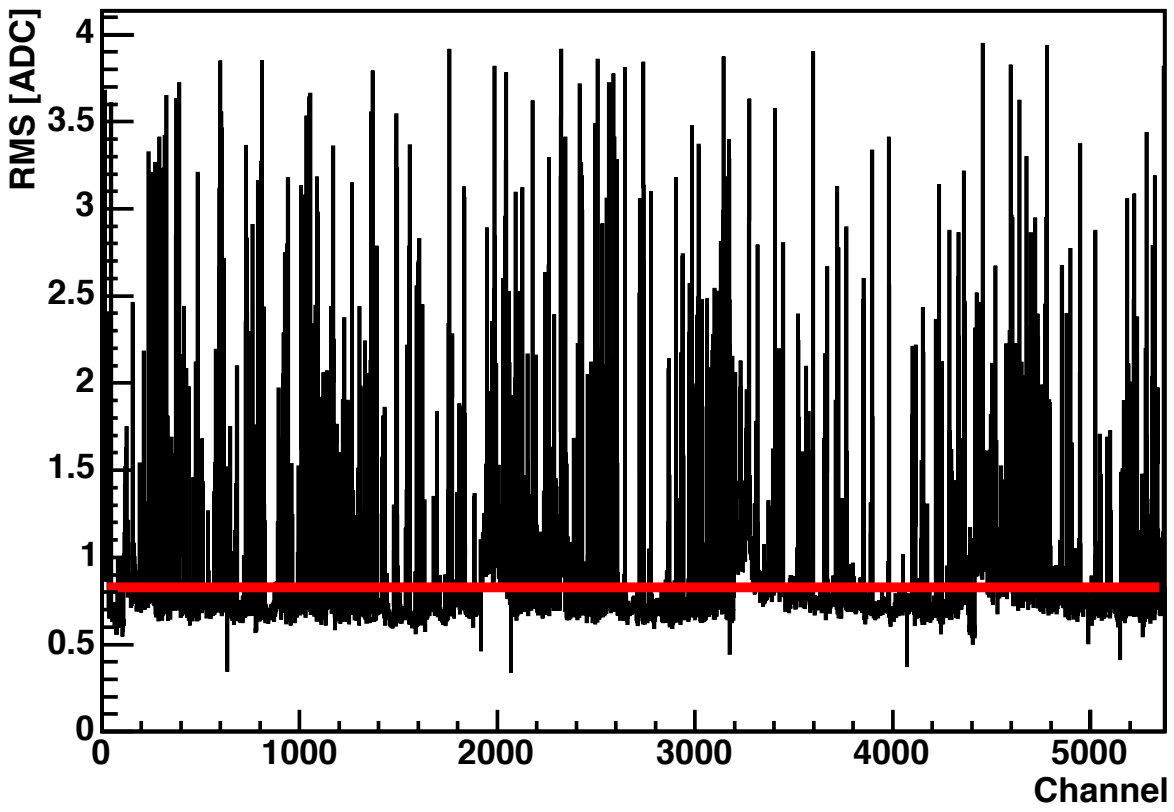

RMS values of timebin 78 over all channels. This timebin is located past the influence of the gating grid switching. The mean value is 0.83 ADC (red line) including the visible glitches. It is expected that a pure pedestal run would have a mean value of around 0.7 ADC.

In the plot in the next column, the same inner noise is shown, but calculated at the timebin 22 which is the most influenced one by the switching of the gating grid, as shown in the channel plot on the previous page. Clearly, a wide spread is visible so that the influence cannot be fully corrected. In the testbeam the electrical version of the gating grid pulser was not the final one as well as the switching of the gate was not synchronised with the ALTRO sampling clock. These are the two main reasons for the big visible variance. Additionally, the grounding of the two patches had a different quality as clearly visible at the change at channel 3328, as well as the different branches of the RCUs at 1664 and 4352.

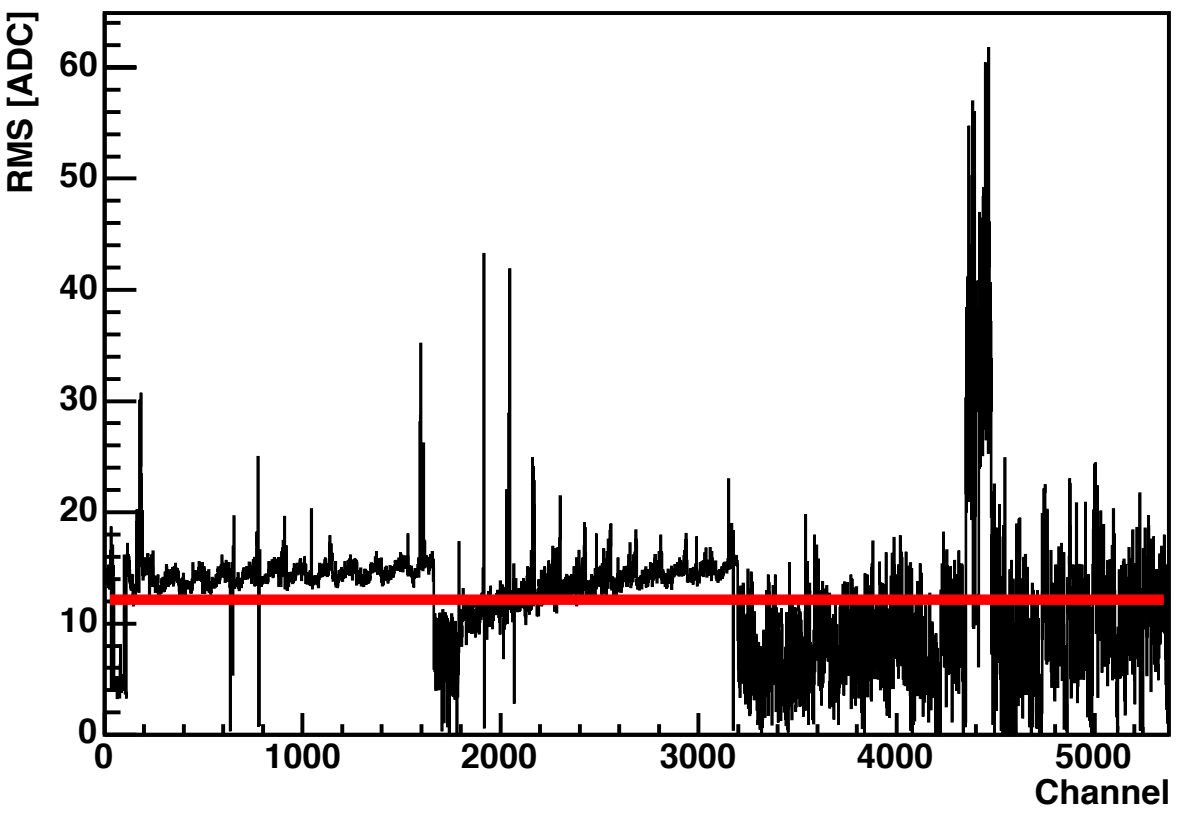

RMS values of timebin 22 over all channels. At this time the gating grid influence is maximal. The trend changes at channel 1664 and 4352 are the different branches and at 3328 the RCUs change.

As shown in the first plot, the correction of the baseline using the LuT in the BCS1 is working on real data, as the influence of the switching of the gating grid is also up to a certain extent removable. This analysis was only based on normal data instead on a dedicated pedestal run as well as the detector electronics was not final the conclusion is, that the baseline correction of the complete data is completely possible.

## TCF Parameters

The TCF has six parameters to be accommodated to the real signal shape which means that these parameters are extracted from the real detector response on tracks of charged particles. In general, the accommodation has to follow the working principle to shorten the signal, but not change the amplitude or create over- or undershoots after the pulse.
The basic fact that the parameters are extracted from the data opens up two schemes to find a set of parameters. At first, to create a universal pulse by overlaying selected pulses from the data by normalising and positioning them. From this universal pulse the parameter set would be derived. Secondary, to search for the best set by extracting the optimum parameters for each pulse individually and then choosing the best overall working set. This latter path has been implemented and tested.

### Pulse Finder

The first step is to find a good set of pulses. There are certain requirements for each of these pulses. They should have a sufficient amplitude, should not be disturbed by glitches or other pulses and the tail should be inside of the acquisition window. Technically, the chain to find these pulses starts with discarding data influenced by the gating grid (discard the first 12 timebins of the pipeline delay of the ALTRO and 26 of the gating grid influence), followed by the amplitude criteria. Only pulses which have an amplitude inside the set band (minimum: 600, maximum: 800) are accepted. The band should be narrow to get similar pulses, but wide enough to collect enough statistics. It should accept high amplitudes to maximise the signal-to-noise ratio, but it should be smaller than the maximum amplitude (1024 ADC) to avoid overflow and saturation effects. Only one pulse per channel is allowed to avoid crossover effects. The position of the pulse should be at small times (maximum position timebin 200) to have a sufficient large time window in the acquisition window left to also include the tail of the signal. Additionally, the time position also defines the allowed amount of multiple scattering in the signal, which increases with increasing time. Some pulses have an extremely big integral, compared to their amplitude, what can be created by several detector effects. The ratio of amplitude and integral can be limited to remove this type of pulses. All remaining signals are then saved with a smoothed tail to avoid influences by the noise. The smoothing is done by a moving average calculation which starts after the pulse. All parameters and their default values used in this analysis are briefly described in the »Appendix« on page 67.

To crosscheck the set parameters there is an additional program which reads the stored pulses and plots them into a postscript file, including the pulse information. Below an example pulse is shown.

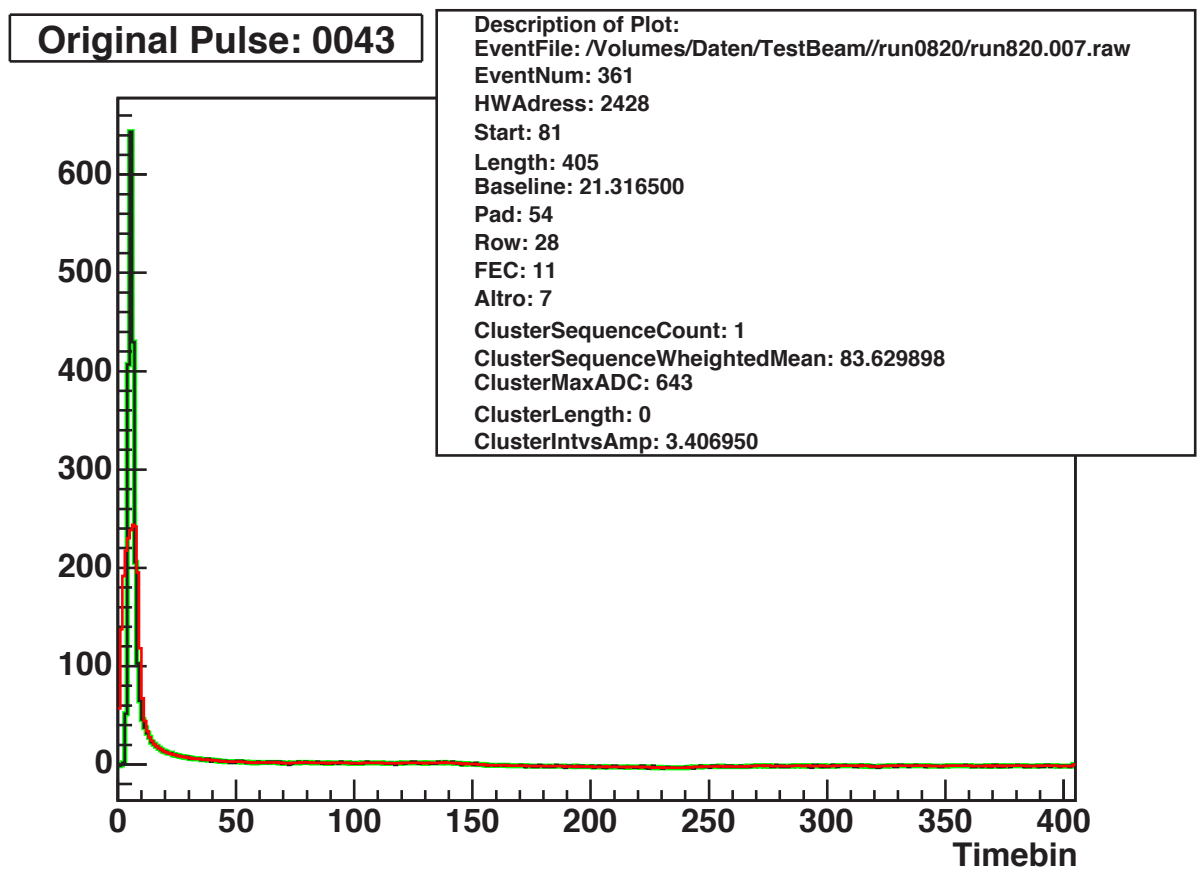

Example of an extracted pulse found by the pulse finder. The black line is the signal, the red line is the moving average and green is the resulting smoothed pulse. All pulse parameters are shown in the parameter box.

### Parameter Set Finder

To derive a parameter set from a given signal an algorithm was developed and implemented in MATLAB [3] and is described in [4]. This algorithm was reimplemented in C++ and optimised to reduce the calculation time by roughly two orders of magnitude and to increase the portability, since it is now not depending on the commercial MATLAB licence anymore. Each pulse will be processed and the optimum parameter set will be stored. Two stages of the TCF are used to remove the tail, the third stage is used for equalisation to keep the amplitude. Each stage can be individually configured in the parameter set finder. All parameters and their default values used in this analysis are briefly described in the »Appendix« on page 68. At this stage any other algorithm to optimise a parameter set to a pulse can be injected, the interface to the data is defined by the found pulses, and the output of the parameters is defined by the six parameters of the TCF in the ALTRO. The access to the input and output file is encapsulated in a C++ class.

### Correlator

The correlator applies all parameter sets on one given pulse by a floating point version of the TCF algorithm of the ALTRO. The idea is to compare the optimal parameter set, the set which was created by the parameter set finder for this pulse (optimal set), with the result of all other parameter sets of the other pulses (correlated set). To determine the difference in the resulting pulses three quality measures were defined, as shown in the plot below. The first one is the amplitude difference in between the optimal and the correlated set. The second one is the difference in the shortening of both sets. At a configurable level the length in between the two crossing points of the application of the optimum set and the correlated set is calculated. The third one is the difference in the undershoot integral after the pulse. All these differences are stored. All parameters and their default values used in this analysis are briefly described in the »Appendix« on page 68.

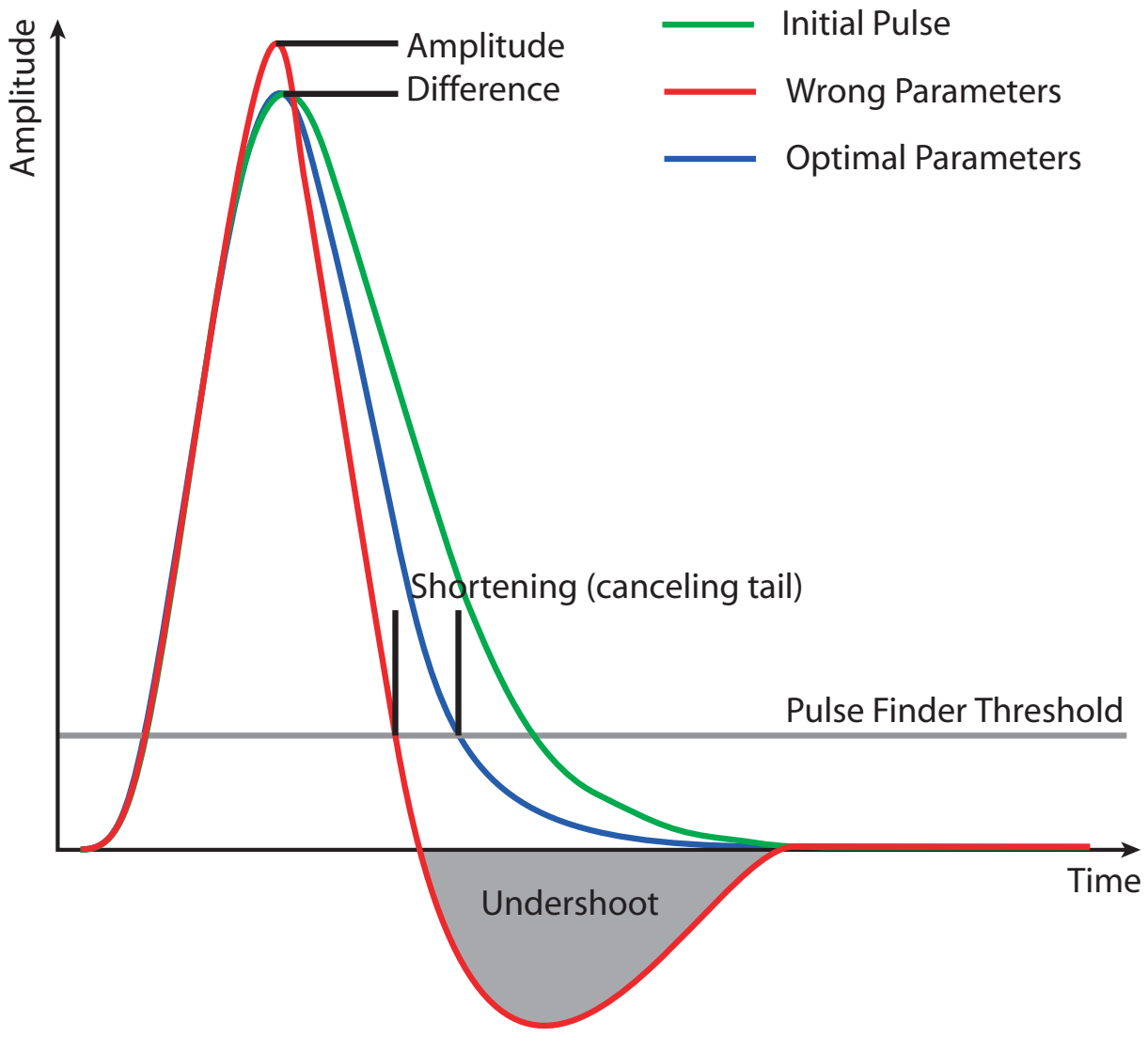

Schematic of the original pulse and the application of the optimal and the wrong TCF parameters on the original pulse and the corresponding quality measures.

### Best Set Finder

The final task is to find the best set of coefficients. The first step is to define the wanted criteria, which, for the TCF, is the sustainment of the amplitude, minimising an undershoot and maximising the shortening of the tail as previously described. This is preserved by the parameter set finder. For the best set finder the criteria is that the differences in between the optimal set and all correlated sets should be minimised. In other words, the search is done to find the set which works best for one pulse and still works best on all other pulses.
For this purpose, two schemes were implemented which start from the same data base as shown in the two scheme flowcharts on the next page. All parameters and their default values used in this analysis are briefly described in the »Appendix« on page 68.

### »Weighted Quality«

At first all differences in each of the quality measures of all correlations in between the result of the pulse with its optimum set and all results of this set with the other pulses are summed. This sum is then normalised and the RMS of all differences is calculated.
The first scheme »Weighted Quality« tries to combine these three quality measures to one quality measure and then to find the best one. Each quality measure has a weighting parameter to vary the importance. Additionally, each measure can be weighted by the RMS value to reward the set with the most steady performance over all pulses. All weighted values are then added and sorted afterwards. The result is a sorted list starting with the set with the lowest value as the best one. This scheme has the advantage to be simple but the disadvantage that the addition of the different measures is not explicit, since the normalised and weighted distributions can have big differences in their shape.

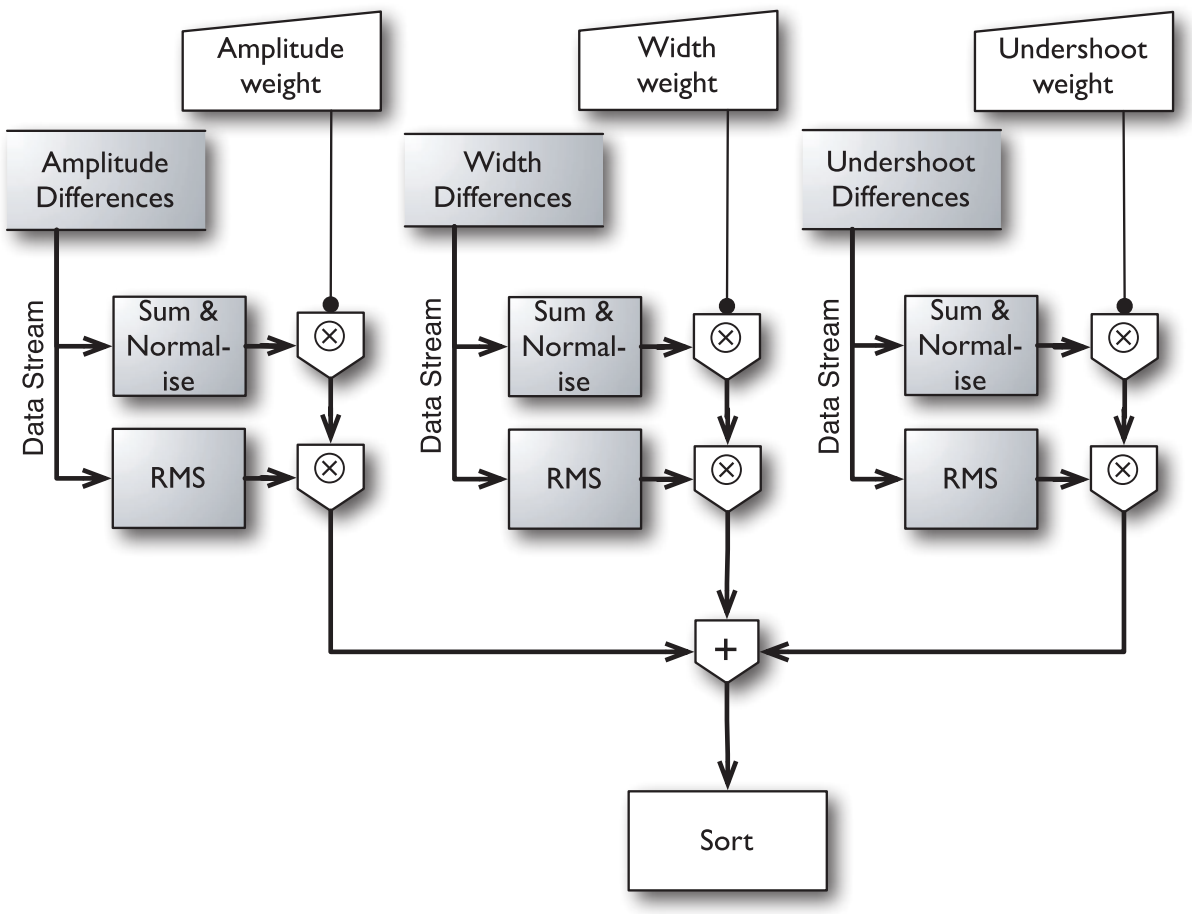

Flowchart of the "Weighted Quality" scheme. The greyish boxes mark the common block of the schemes.

### »Weighted Vote«

The second scheme »Weighted Vote« uses the idea of an election. Each quality measure can be weighted by the RMS value to reward the set with the most steady performance over all pulses and is then sorted individually. There are now three uncorrelated elections of the best set for each quality measure separately. These elections are combined by adding the slot number of the individual elections. The set with the lowest sum of the slots is the best one. There are additional weights to vary the importance of the separate elections, in other words, the importance of the different quality measures can be weighted.

This scheme is more complex than the previous one but has the advantage that differences in the separate quality measures are replaced by a position.

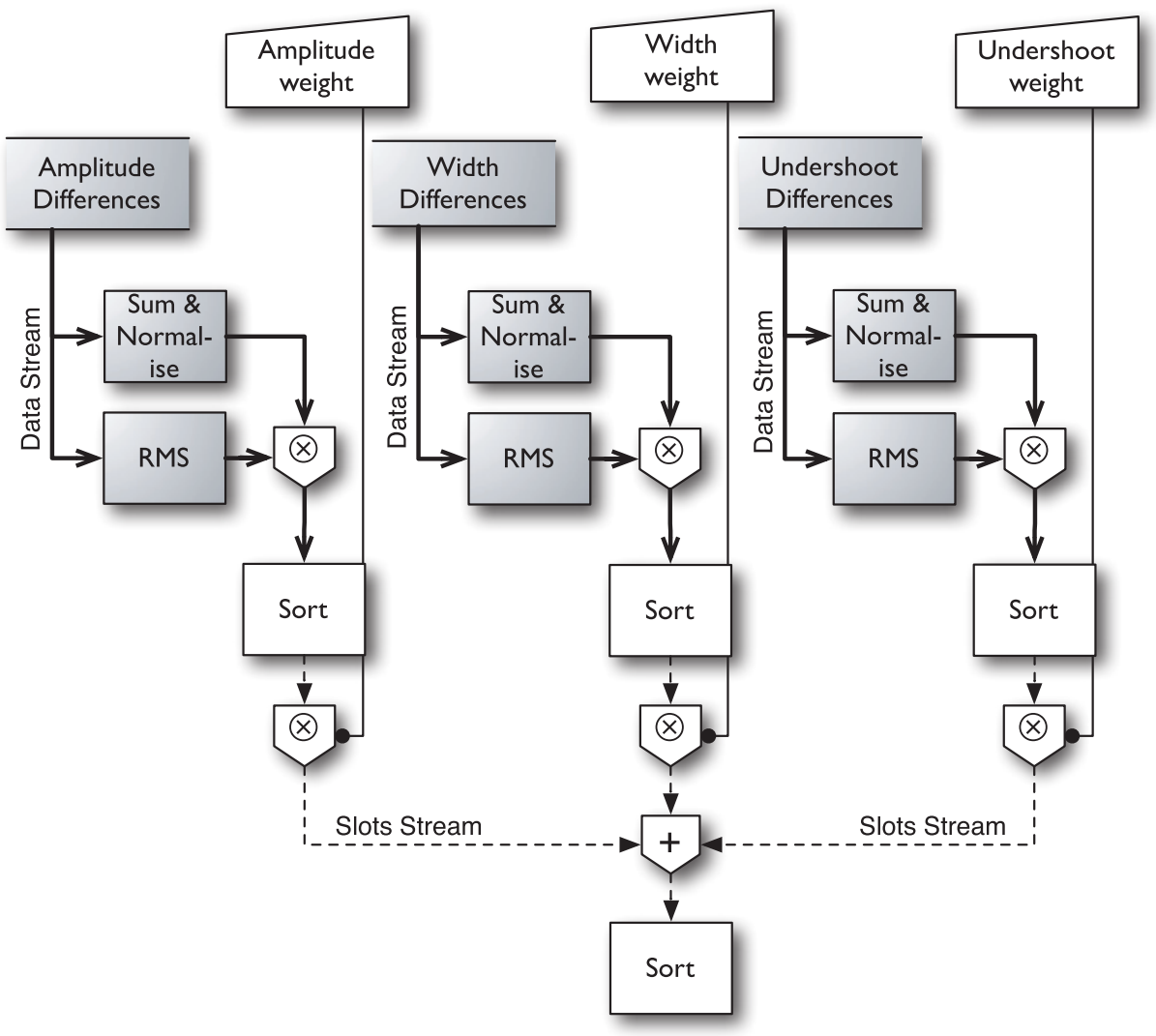

Flowchart of the "Weighted Vote" scheme. The greyish boxes mark the common block of the schemes. The black lines indicate a data stream, the dashed line a slots stream.

### Set Performance Check

To crosscheck the performance several plots are produced. As a summary for each quality measure, the results are sorted and plotted. Problematic pairs of sets and pulses are then on the left and right due to their big difference to the optimal set. An indication that the best set is not working well, is when the extreme ends are strongly populated. Translated, this means that the set works on many pulses quite well but does a lot of harm to the rest. The next step is to inspect the pulses which are the base for the not proper working coefficient sets. The reason of this behaviour can also be that there are strange pulses found. These pulses should be removed and then the calculation has to be started again with the parameter set finder.

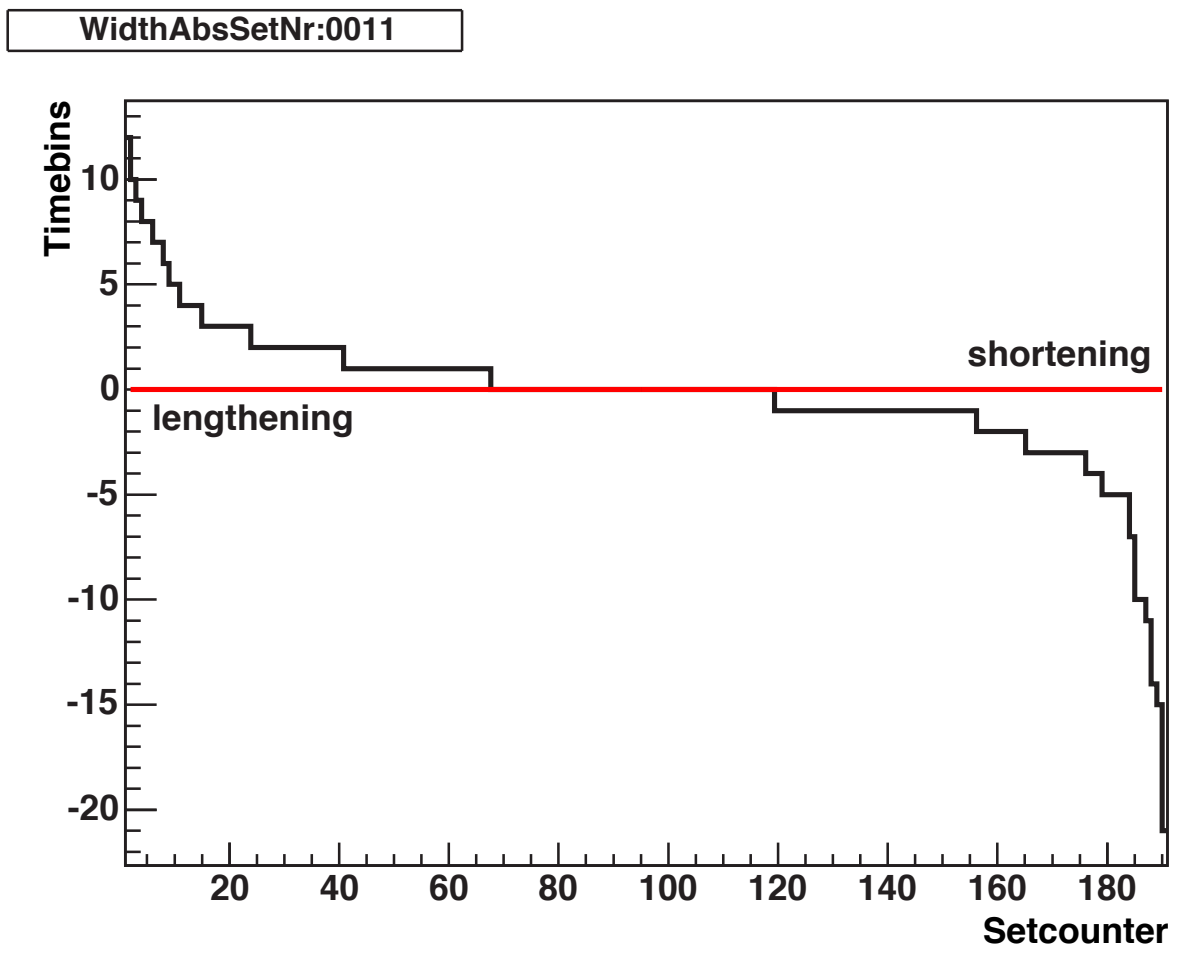

Example quality plot for the shortening amount. Negative values indicate a lengthening of the pulse, positive vice versa. Overpopulated ends indicate a non optimal parameter set.

To have access to the original pulses and the results of the TCF using the optimum set of the pulse and the one which was found as the best set, a set of four plots per pulse is created, starting with an overview of the three pulses and followed by a zoom on each of the three quality measures.

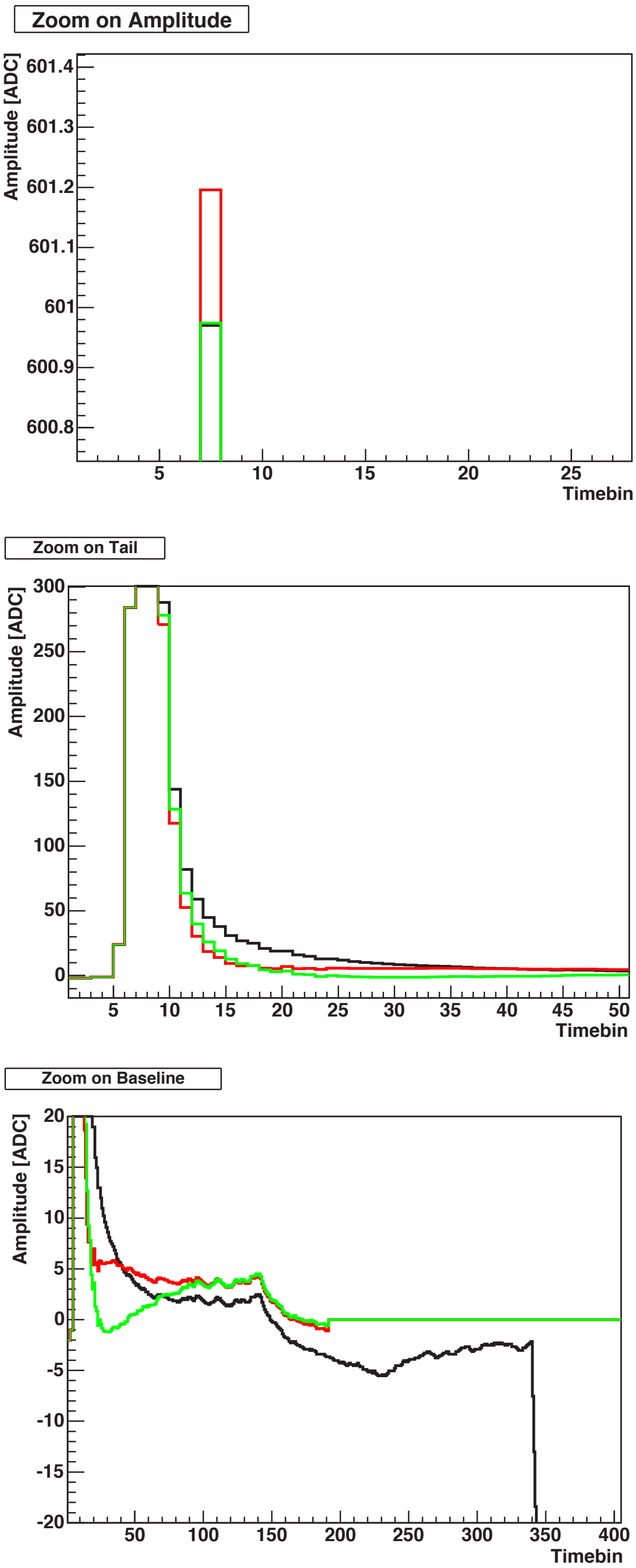

Zoom on the amplitude difference (top), the shortening difference (middle) and undershoot difference (bottom). Black is the original set, green is the signal processed using the optimum set and red is the signal using the correlated set.

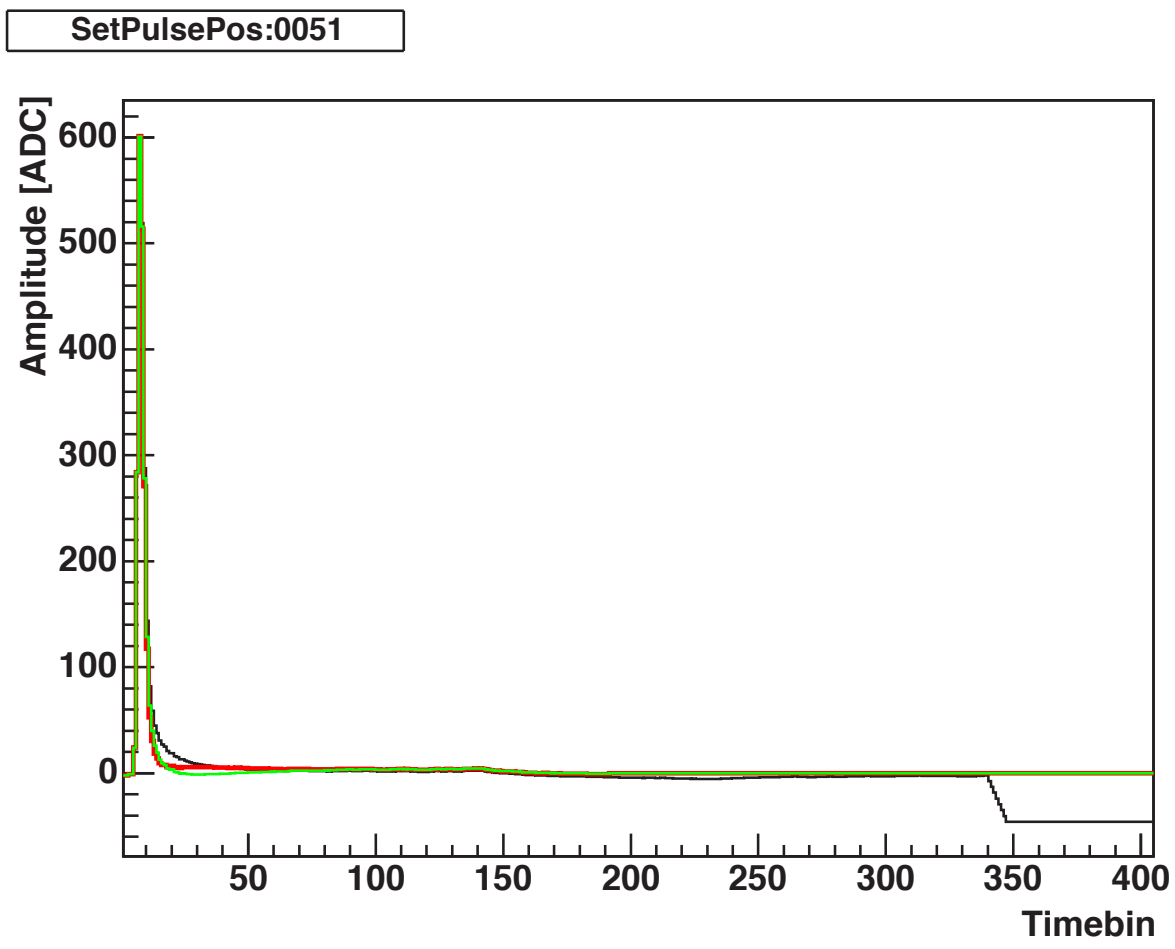

Overview over the three pulses. Black is the original signal, green is the signal processed with the optimum set and red is the signal with the correlated set.

# BCS2 & ZSU Parameters

The parameter set of these two units is correlated, since the goal is to keep all signal information, but also get a good compression. As described in the ALTRO chapter, the ZSU removes all samples below a threshold. It can happen that, due to pile up effects, a small signal would fall below the ZSU threshold. This can be cured by the BCS2 unit, as long as the BCS2 can track the baseline. When allowing the BCS2 a large acceptance window it can happen that it will start following a signal and then never get back again on track of the real baseline. With the use of the pre- and postsamples this overreaction can be removed, but the amount of baseline ADC values is decreased so that, again, it can happen that the BCS2 looses track.

The ZSU can only work and keep all signals as considered, if the BCS2 stays on track or the baseline variation of the channel is smaller than the threshold. Clusters with a big charge have a visible, long living and significant tail as discussed in chapter »Ion Tail analysis« on page 47. These could shift down the baseline far enough, so that small clusters get lost. Even low occupancy events can have baseline distortions harming small signals. When moving to high occupancy channels the distortion increases, so that the basic assumption is that the optimising goal is to keep all clusters, but maximise the compression without loosing cluster information.

The result of this optimisation gives a set of possible parameters for each occupancy tested. This scheme is not implemented, but parts are existing, like the BCS2 and ZSU units as described in the following section.

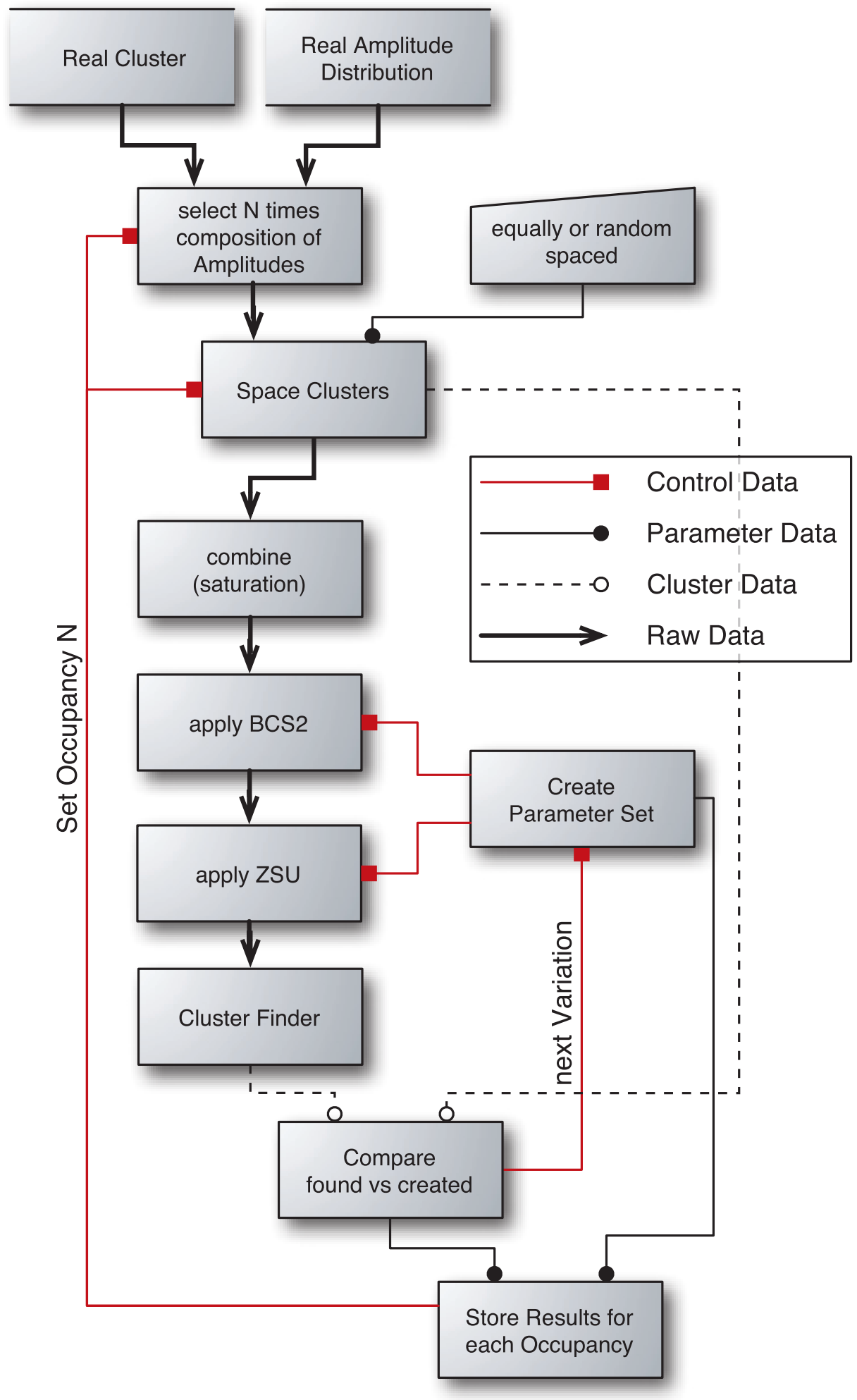

Flowchart of the BCS2 & ZSU finder

## ALTRO++

As clearly indicated, the ALTRO parameters will be optimised by using the digital chain of the ALTRO. To reduce the overhead in complexity, price and speed when introducing real hardware in this process, a ALTRO emulator was developed. The aim was to build up a software module which reproduces exactly the results of a real chip on the cost of speed or possibly higher precision, due to the increased capabilities of the CPU of a normal computer. This includes a bit-precise fixed point implementation of the TCF. The ALTRO++ can be configured to switch individual units of the digital chain on or off. Additionally, as an inconsistency compared with the real ALTRO, the clipping can be turned off. This feature is extremely helpful when analysing data which is dropping below zero, which would be forced to zero by the real ALTRO.
There is a limitation, the AUTOCAL circuit, as described in the chapter »Front End Electronic« page 15, cannot be implemented, as the data in between two events is not existing in the stored data. The ZSU in the ALTRO++ only calculates the compression factor. The DFU is not implemented at the moment, so the ALTRO++ does not create the ALTRO data format as described in [1]. This feature will be implemented. The ALTRO++ will then produce a correctly sized array of 64 bit numbers to store the 40 bit ALTRO formatted data. Not implemented is the MEB since this is not needed in software.

## Computing

The final ALICE TPC consists of roughly 560000 channels, assuming the worst case that each channel needs a different configuration for the digital processor additionally to the pedestals, this puts the attention on the computing time to extract these parameters. The stability of the parameters will define the update rate to revise or recalculate the parameters. Also, the pure data volume which has to be transferred to the detector electronics before a start of run requires a clear scheme. Finally, this configuration data has to be archived for the offline data reconstruction [5,6]. This leads to the questions of computing time, computing frequency, data volume and storage frequency, which will not be known completely before the ALICE physics program starts.

### Pedestals

The parameters of the pedestal configuration are easily extractable, but a problem arises in the pure data volume. The computing time is negligible, since only a few hundred events have to be parsed to get the mean pedestal value. Additionally, this computation is completely independent for each channel and extremely simple, so this task can already be done in the RCU or later in the LDCs of the DAQ or the HLT nodes. The data for the pedestal calculation will not be archived. The problem of the pedestals is more the pure data volume, since for all channels nearly 700 MByte is needed. The distribution of the data should as consequence be done in parallel. As already partially existing and implemented two data paths, the DDL or the DCS are usable as described in the chapter »Prototype Environment« on page 29. The data volume for each RCU is 5 to 10 MByte depending on its position on the TPC, since pedestal data could be highly compressed by a entropy coder [7] like the huffman coding [8], its volume could be reduced by a factor of five. As long as the data will not be sent from one source to the detector on both data paths, the 200 MByte/s DDL or the 10 MByte/s DCS are sufficient. For archiving these data it can be slowly collected and centrally stored.
The calculation and storage frequency will not be known before the complete ALICE setup is completed, detector wise, as well as cooling and electronics wise. The upper limit can be defined by the experience of NA49 [9] of three pedestal runs per running day (24h).

### TCF

The determination of the TCF parameters shows a completely different picture of problems. The previously described scheme consists of several computing steps with different computing prerequisites. The pulse finder has at the moment an inspection rate of roughly 26000 channels/s of 500 timebins on an Opteron 246 [10] system using xfs [11] as file system and gcc 3.3.4 [12] as compiler. To extract the coefficients for one found pulse the same system needs 0.2 s. The running time of the correlator is increasing quadratically from 1.2 s for 100 to 30 s for 1000 correlations respectively without storing the correlated pulses and from 14 s for 100 to 21 min. for 1000 correlations respectively when storing the correlated pulses what is only needed for a debugging purpose. The best set finder needs 50 s to find the best set. The most time (90%) is spent by reading the current implementation of the data set, so that a speed-up below 20 s is easily possible.
The first problem is the uncertainty, if each ALTRO channel needs its own optimised TCF parameter set or if each ALTRO reduces the needed effort by a factor 16 or if bigger structures like TPC rows or patches reduce again the needed effort by a factor of 6 to 30 resulting in 96 to 480 in total. There is not enough data of pulses with a sufficiently high amplitude to answer the question, if there are differences in the ALTRO channels or in the different ALTROs which are big enough that a channel-wise TCF configuration is needed. Another question is the stability of the sets, which should be quite high, since only changes in the signal shape affect the TCF. The complete chain can run in parallel, since no communication in between the different set finding blocks is needed, so the that calculation time scales with the number of CPUs in a cluster.
Assuming the worst case that every one of the 557568 channels need its own configuration and that 1000 channels, fulfilling the pulse finder requirements, are needed to extract the optimum set. I assume that one million events should be sufficient to get enough statistics on each channel leading to an inspection time of 6000 h for one CPU. The extraction of the coefficients would take 30000 h for one CPU. Without the major speed-up due to the reimplementation this scheme would be completely impossible, since the calculation time would be still half a year on a 2000 CPU cluster. The correlator needs 4500 h without writing and finally, to find the best set, 8000 h are needed. In total, this leads to a quite big amount of data and computing time, but is easily manageable with a cluster.

| Program | 1 CPU | | | | Cluster |
|---|---|---|---|---|---|
| | | On Set per | | | 2000 |
| | | Chip | Row | Patch | CPUs |
| Pulse Finder | 6000 | 375 | 38 | 2.4 | 3 |
| Coefficient Maker | 30000 | 1900 | 200 | 12 | 15 |
| Correlator | 4500 | 290 | 30 | 1.8 | 2.3 |
| Best Set Finder | 8000 | 490 | 49 | 3 | 4 |
| Total | 48500 | 3055 | 317 | 19.2 | 24.3 |

Running time in hours of the different steps to extract the TCF Parameters

### BCS2 & ZSU

For optimising this parameter set, it is not expected that these parameters differ for each channel, since they are mostly dependent on the occupancy. The problem here arises when scanning the complete parameter space since there are eight parameters with 6.6 trillion possible combinations per occupancy. Fortunately, many combinations can be excluded, since they are quite senseless.

### Configuration

When all parameters are extracted they have to be stored and prepared for sending them to the FEE. For this purpose a set of classes were implemented. There are encapsulations for the different hardware components:

» **ALTRO**COMMANDCODER
» **RCU**COMMANDCODER
» **BOARDCONTROLLER**COMMANDCODER

By the use of the component encapsulation classes all commands can be translated to their correct bit pattern for the hardware. The extracted parameters are set via the use of the matching commands. Additionally, to the ALTRO digital chain parameters, the BC needs adjustments in the configuration of the controlling thresholds. Functions to read and parse the various parameters and error registers are also included in the classes.
The layer above is parted in the different interface encapsulations.

» **INSTRUCTIONBLOCKCODER**
» **CONFIGIO**
» **FEC2WRITER**

The RCU has a memory where command sequences can be stored. The coding of these command blocks is encapsulated in the InstructionBlockCoder. It supports both configuration modes of the FEE, the individual channel configuration as well as the broadcast mode, if the configuration parameters will be automatically sent by the RCU to all channels on one FEC as long as the commands support a broadcast, otherwise it automatically creates the sequence for the individual channels. These sequences can be translated into the FeC2 language [13] to use the DDL by the intermediate step to write out these sequences as FeC2 script. This is included in the FeC2Writer. For the configuration using the InterCom Layer [14] no intermediate step is needed, since the function creating the code sequences is called by this layer and returns the command block which is then handled by the DIM client server system. Later the communication via the DDL will be also included into the DIM system.
The data source for the InstructionBlockCoder is a binary file encapsulated into a class (ConfigIO). It can be easily extended to communicate with a database without any need of changing the other parts of the system. The actual partitioning of the data is derived from the transport granularity, so that all configuration data from one RCU is collected in one file.

# Ion Tail Analysis

# Ion Tail Analysis

During the analysis of the cosmics data of the test TPC an additional effect got visible following the normal signal tail. Every avalanche signal is the result of the contribution of a large number of positive ions leaving the anode wire in various angles, and, by following different paths which can last for several tens of microseconds, induces a long ion tail as described in the section »Signal Creation« on page 11. At first, the effect of the ion tail was visually found in cosmic data at the test TPC in pulses with extremely big amplitudes (>700 ADC) during the usage of the online monitor. After applying a moving average calculation in order to smooth the data the shape of the ion tail was also found in clusters with smaller amplitudes (>200 ADC).
Neither the spread of the avalanche around the anode wire nor the variation from avalanche to avalanche has been accurately understood and quantified yet. Since this effect was visible in normal data, a data based analysis was developed. Due to the big influence of the gas mixture on the TPC properties as well as on the signal shape, the analysis for the ion tail was repeated for each gas mixture.

# Pulse Extraction

To characterise the signal tail and its variation, adequate pulses were extracted from the data. Like in the pulse finder of the TCF parameter extraction (on page 39), single pulses of at least a maximum amplitude of 200 ADC at an early time position are needed, since these criteria are fulfilling the prerequisite of an complete, undisturbed and visible tail. The end of pulse position of the simple clusterfinder is defined as the time position of the last sample above the threshold.

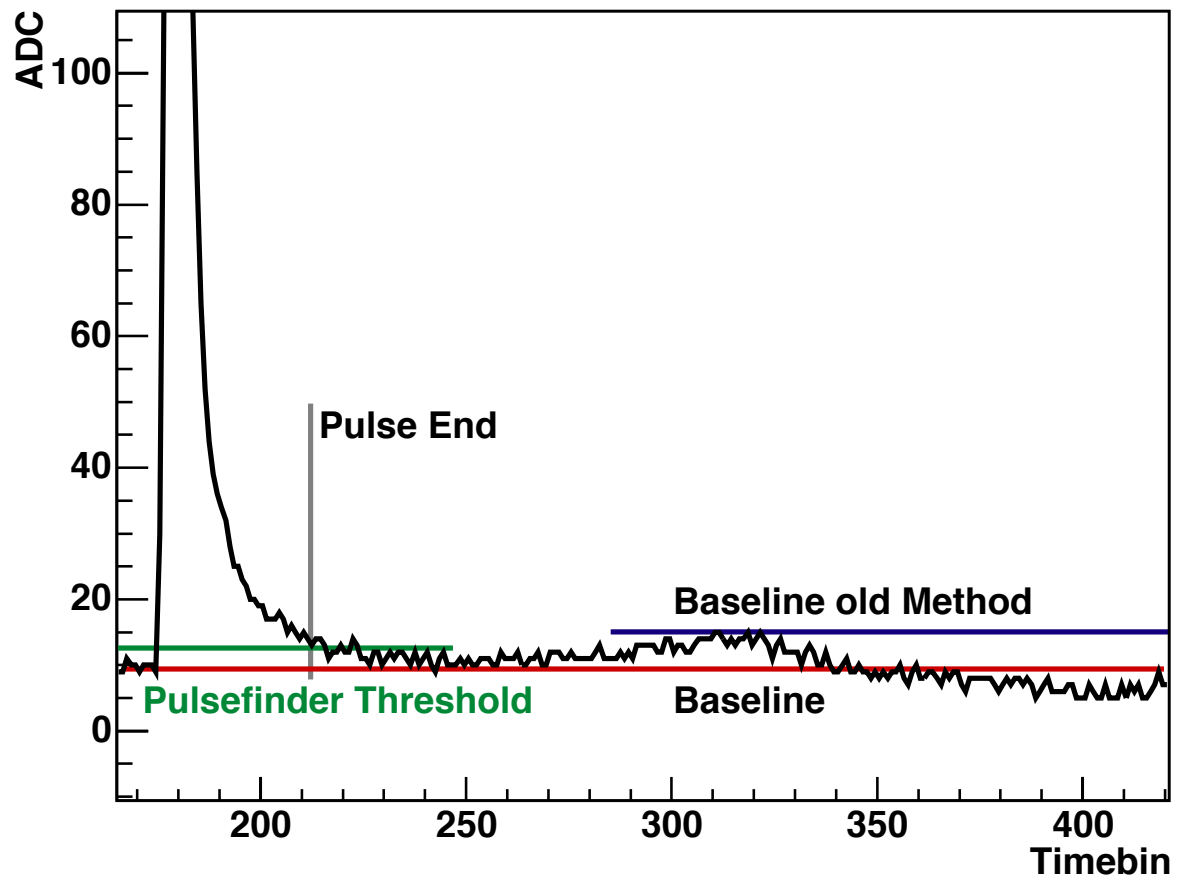

Part of extracted signal. The red line shows the new baseline calculation method, blue is the failing old one. The grey line shows the end of the pulse, which is also the start of the tail, the green line shows the pulse finder threshold.

The endpoint of the pulses also defines the starting point of the ion tail. From all pulses 500 timebins (50 μs: 500 timebins with 10 MHz sampling frequency) starting with the ion tail start point are then saved. The determination of this point requires a proper knowledge of the baseline which is no problem during analysis of the cosmics data as described in the section »Monitoring« on page 31, as in this case a proper baseline table and a correct mapping procedure exists. In the testbeam data this is not the case. The extraction of a correct baseline table is possible with additional data reject and check algorithms as described in »BCS1 Parameters« on page 38, but for the tail analysis the mapping of the ALTRO address of the baseline pattern or value to the correct ALTRO Address in the data is problematic, since the address can be incorrect in the data. If applying the filter used by the baseline pattern extraction, the statistics for sufficient pulses is dramatically reduced and renders a tail feature extraction at high maximum ADC values or high pulse charges impossible. Additionally, the filter is not completely correct, since switched address errors are not found, so that wrongly reshifted ion tails would spoil the analysis, what is quite sensitive to small errors due to the small signal itself. So the baseline for each pulse has to be extracted from the channel hosting the pulse, which faces the problem, that there is always a big pulse with ion tail, so that there are not many ADC values at the baseline level. The first approach using the double threshold scheme as described in the section »Monitoring« on page 31, leads to an extreme signal drop as shown in the following plot, which was not discovered in different data sets.

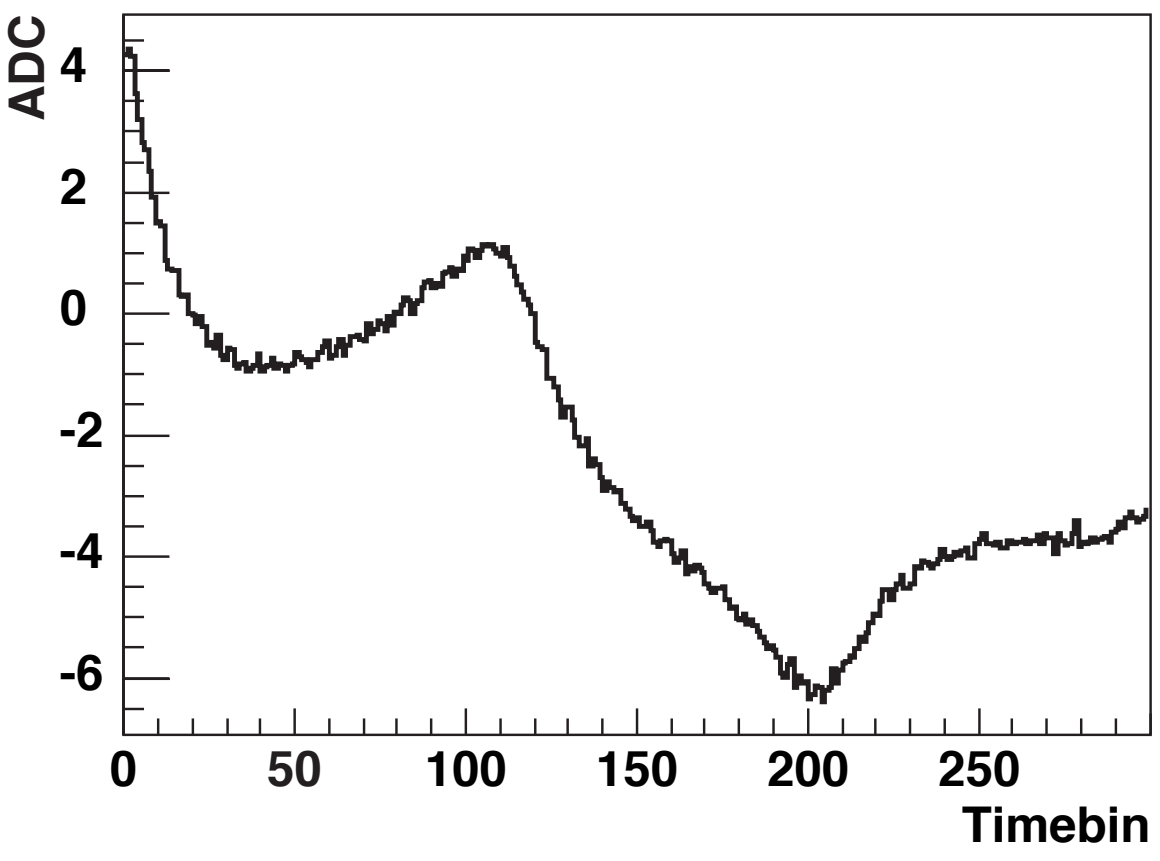

Mean of the ion tail of all pulses of an maximum amplitude of 600 < maximum ADC < 700 using the old baseline calculation. The signal drops by 4 ADC.

This is caused by the limitation of the double threshold baseline calculation, which can fail in following the baseline if the mean of all samples lifts the threshold away from the signal baseline, as shown in the plot to the left.

### New Dynamic Baseline Method

This old method was replaced by a more precise and also stable one, which extracts an ADC histogram of all ADC values after the gating grid pulse of the channel and then calculates the mean of the channel by using the most probable ADC value bin with a configurable amount of bins next to it. Since this analysis searches for a signal drop after the pulse, an asymmetric window with only one lower bin and three higher bins was chosen too minimise the ion tail influence on the baseline calculation. To check the performance of this method, the pedestal table and the wrong address filter are used. The remaining mapping failures are discarded by comparing the file baseline with the dynamic one. If the difference is bigger than two ADC the plot was displayed and checked via the eye. After removing the ambiguous addresses by the eye scan, nearly no outlayers of the new method are left, as shown in the plot below, which also includes the double threshold calculation result on the same eye scanned data.

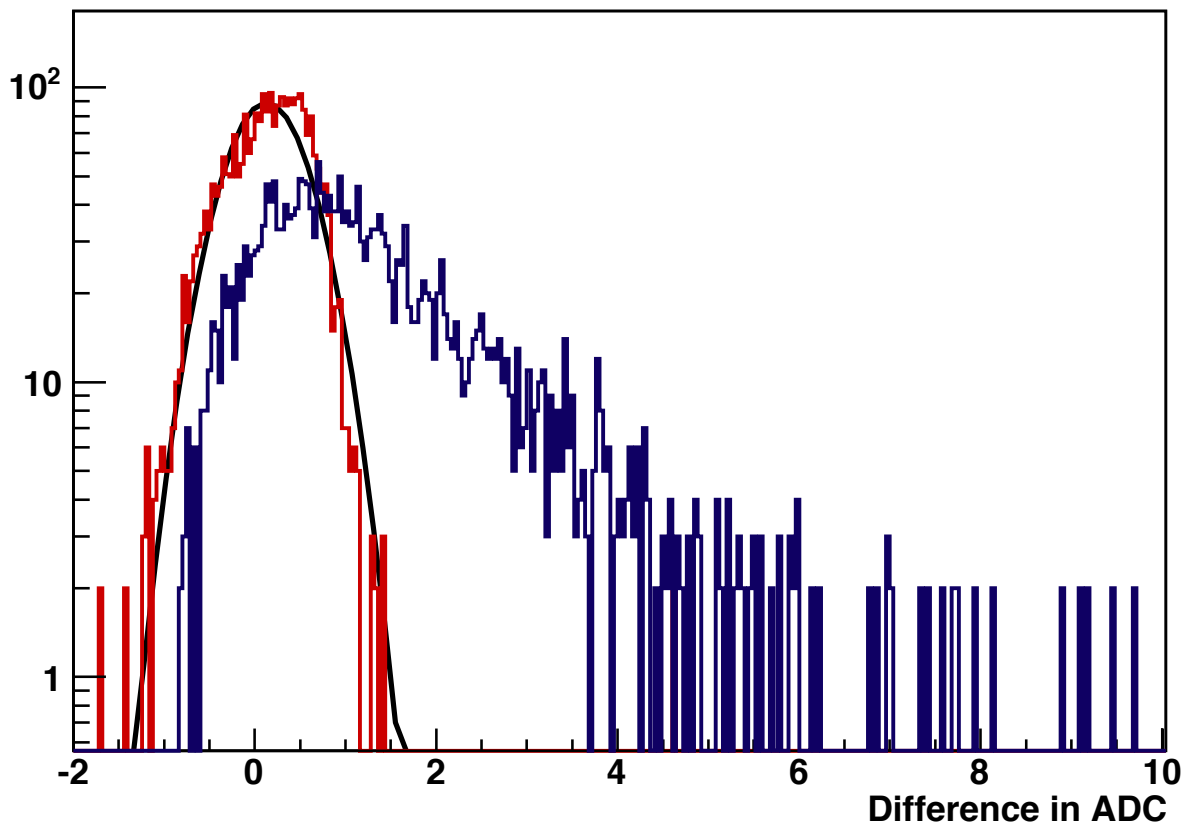

Difference (dynamic - file) in between the file based baseline and the two dynamic baseline calculation methods. The red line is the new method, which is quite symmetric (mean = 0.13, $\sigma$ = 0.46). The blue line is the double threshold scheme which clearly shows a trend to overestimate the baseline (mean = 1.7, RMS = 2.6).

This method works for all pulses in a low occupancy environment, which is a prerequisite of this analysis as long as there are no slow developing effects.
Additionally, the ALTRO++ class (described on page 43) is integrated here to check the performance of the ALTRO digital circuit to remove disturbances like this ion tail.
All extracted signals are stored in a ROOT file as histograms. These are the time normalised tails as pure ADC values and as smoothed values by the moving average calculation. To check the analysis additional data is stored, these are complete pulses, histograms on all applied cuts and histograms to crosscheck the baseline calculation and validity of the cluster end refinement. Additionally, all cut parameters and cluster informations are stored.

## Cosmics

At the cosmic ray site of the TPC a few special runs to acquire statistics on pulses with an high amplitude were taken. At this time, the TPC was flooded with Ne/$CO_2$ as described in »Prototype« on page 30. Out of these runs the sufficient pulses were extracted and then collected to increase the statistics. As clearly visible, in the plot in the next column, the statistics is quite poor when going on to pulses with a high pulsecharge. A ADC threshold of 200 was set for the data acquisition, which reduces the trigger rate to roughly one in a minute. This set represents a running time of more than a week, so increasing the statistics considerably would take a very long time.

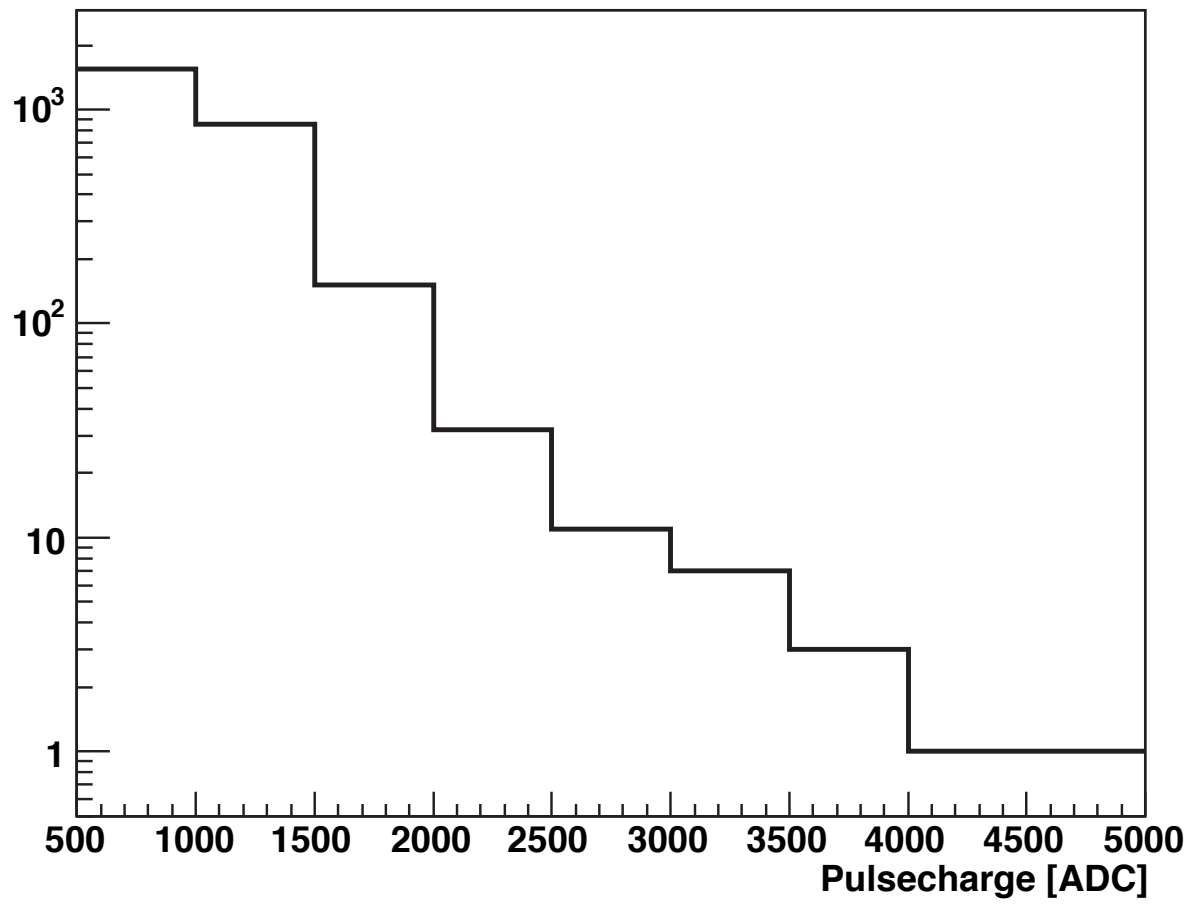

Total collected statistics of 2620 sufficient pulses in the cosmics data subdivided in the later used pulsecharge binning.

The mean of all pulses in one class is calculated to reduce the noise influence as well as the effect of the extremely limited statistics. Eight timebins are averaged and collected in the plot shown below.

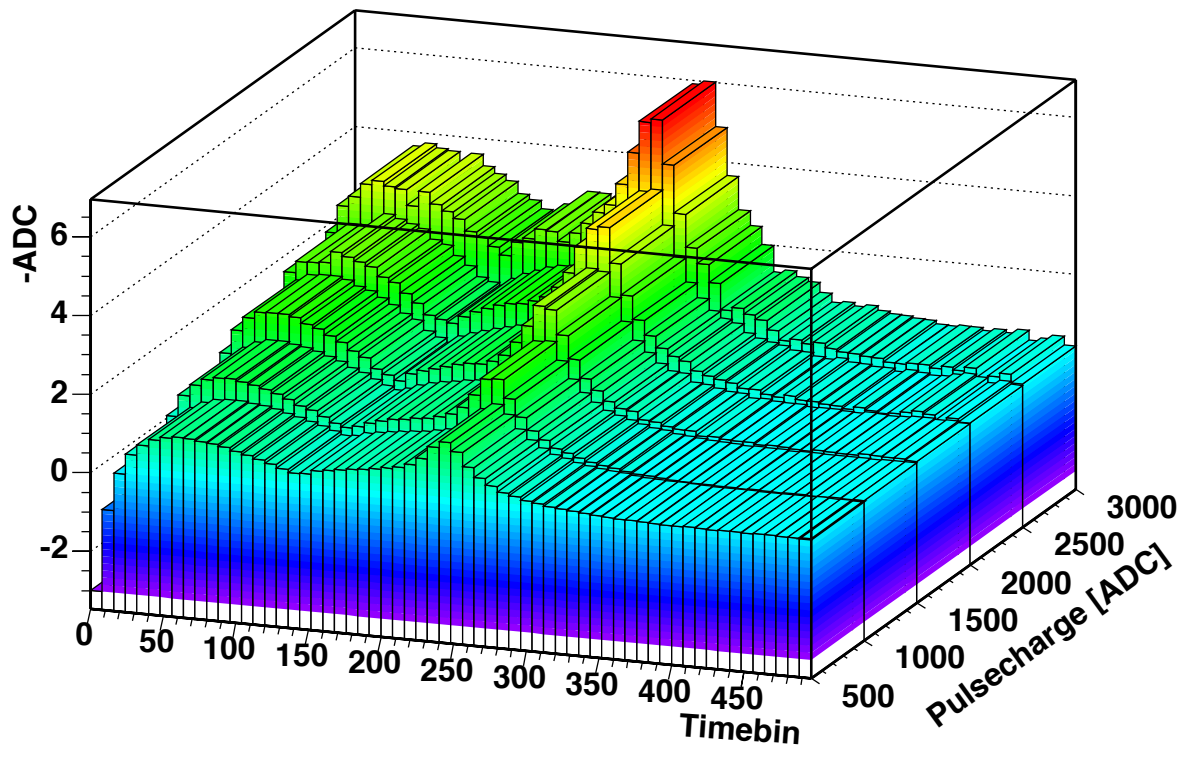

Shape of the inverted ion tail, separated in pulsecharge bins. Eight timebins are averaged to reduce the effect of the limited statistics and the noise.

The plot above shows two minima (the view is inverted for better visibility), a slow changing undershoot, a local maximum, at the beginning a slow falling signal which then gets a fast deep drop and finally recuperates to the baseline. The second drop is located 25 µs (250 timebins) later than the end of cluster point, which is consistent with the simulation introduced in »Signal Creation« on page 11 and as shown in the plot on the next page. It is impossible to determine the relative portion of ions drifting in the direction of the different targets from the achieved result, but it indicates that a significant portion is drifting towards the cathode wires. The portion drifting towards the pads is not visible, since no overshoot is present in the data. Additionally, as a first approximation, the time profile of the ion tail scales linearly with the cluster charge and no time position change is visible.

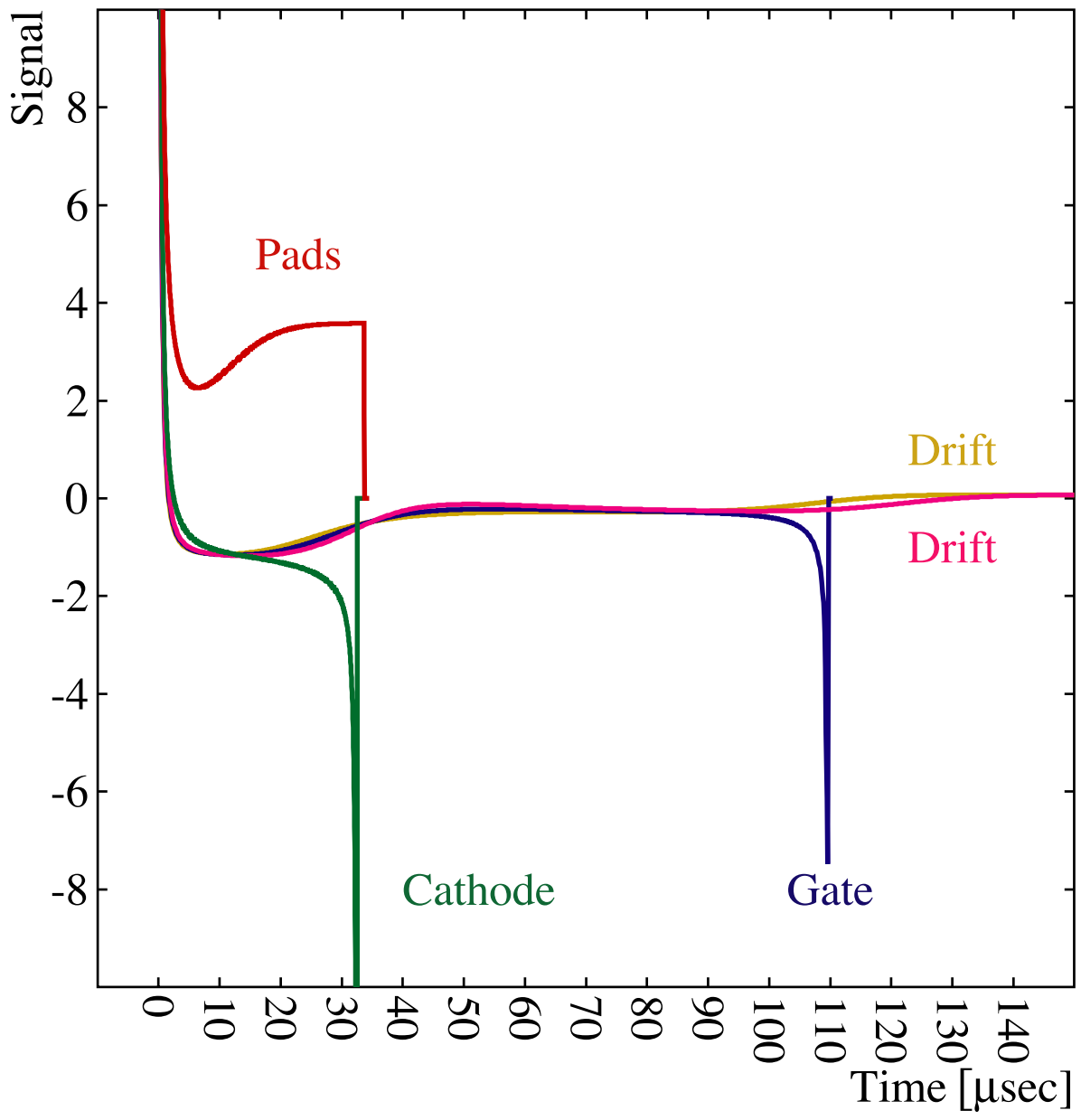

Signal induced in the pads for individual ions moving in different directions in the readout chamber: Drift Region at 95°, Gate wire at 110°, Cathode wire at 140° and directly to the pads at 190° [1]

Another interesting aspect is the spread in the development of the ion tail. All ADC values of all pulses of their class are collected in a hit graph as shown below.

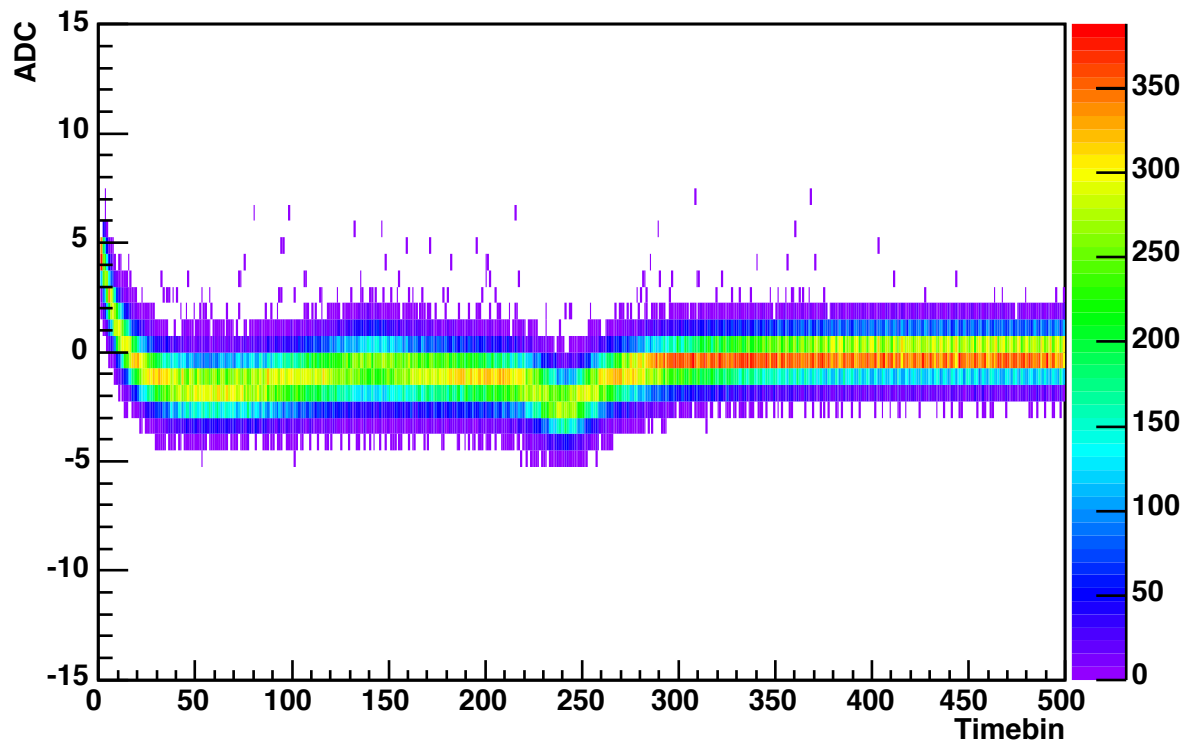

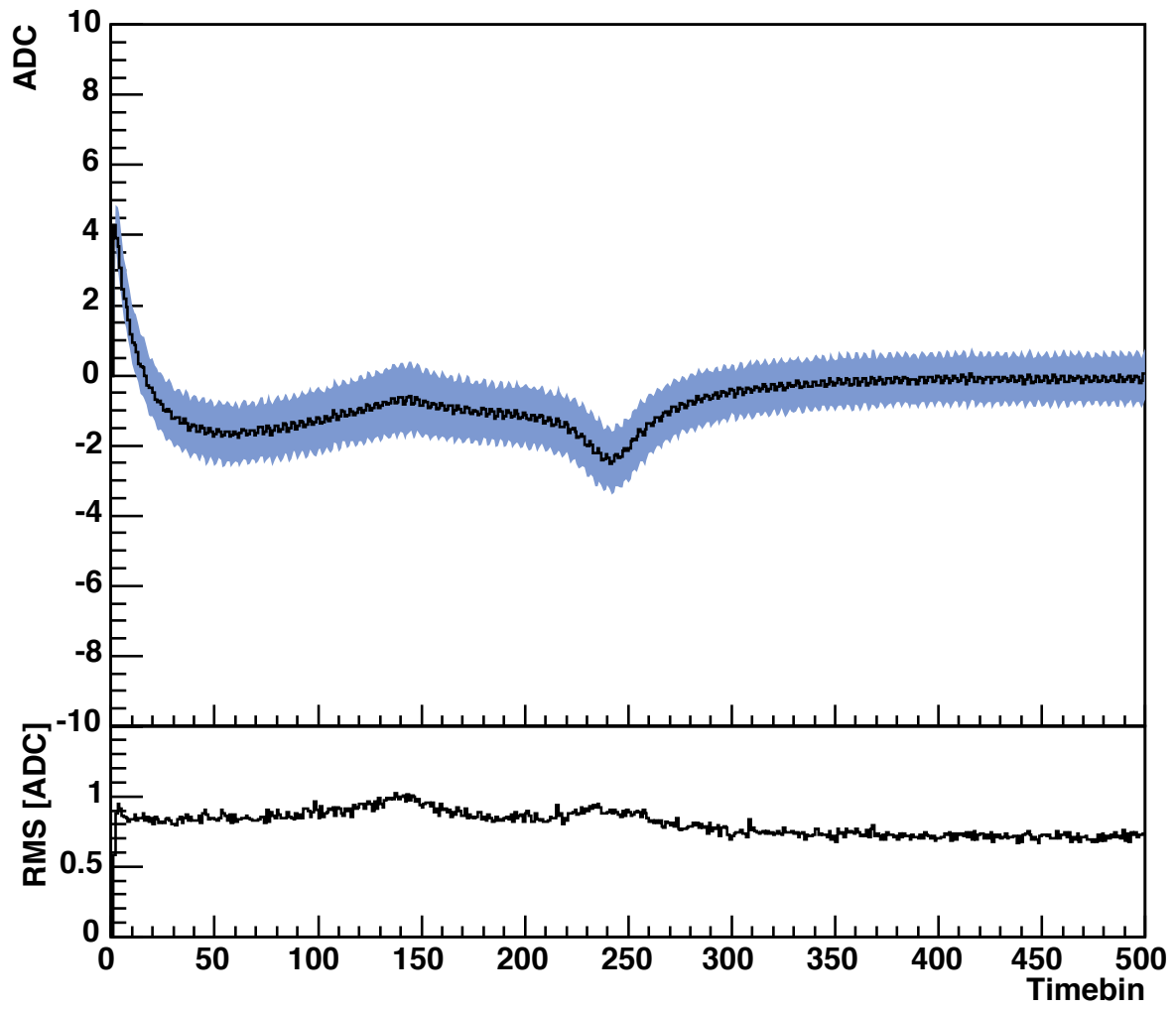

Hit graph (top) and extracted mean and RMS values (bottom) of all pulses with a clustercharge in between 1000 ADC and 1500 ADC. The blue area shows the amount of variation on the signal.

The previously described shape is still visible but the accuracy of the ADC and the noise distorts the picture. To quantify the spread, the mean value and the RMS value of each timebin of the hit graph is calculated, what leads to a better visibility of the shape and the spread is diluted by the noise. To get a clearer representation the noise was removed by using a moving average calculation before collecting the pulses in the hit graph, as shown below. The moving average calculation uses three samples to the left and four to the right, so that the influence of the noise is reduced by a factor of three (from RMS = 0.75 to RMS = 0.25).

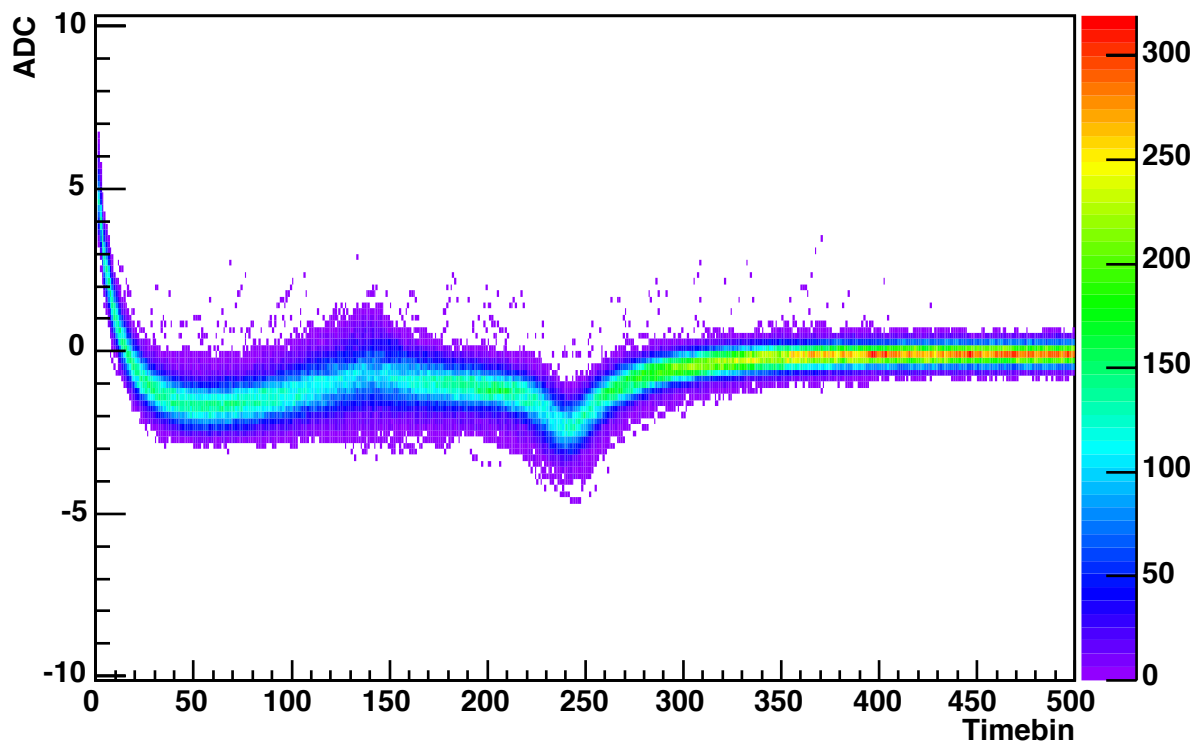

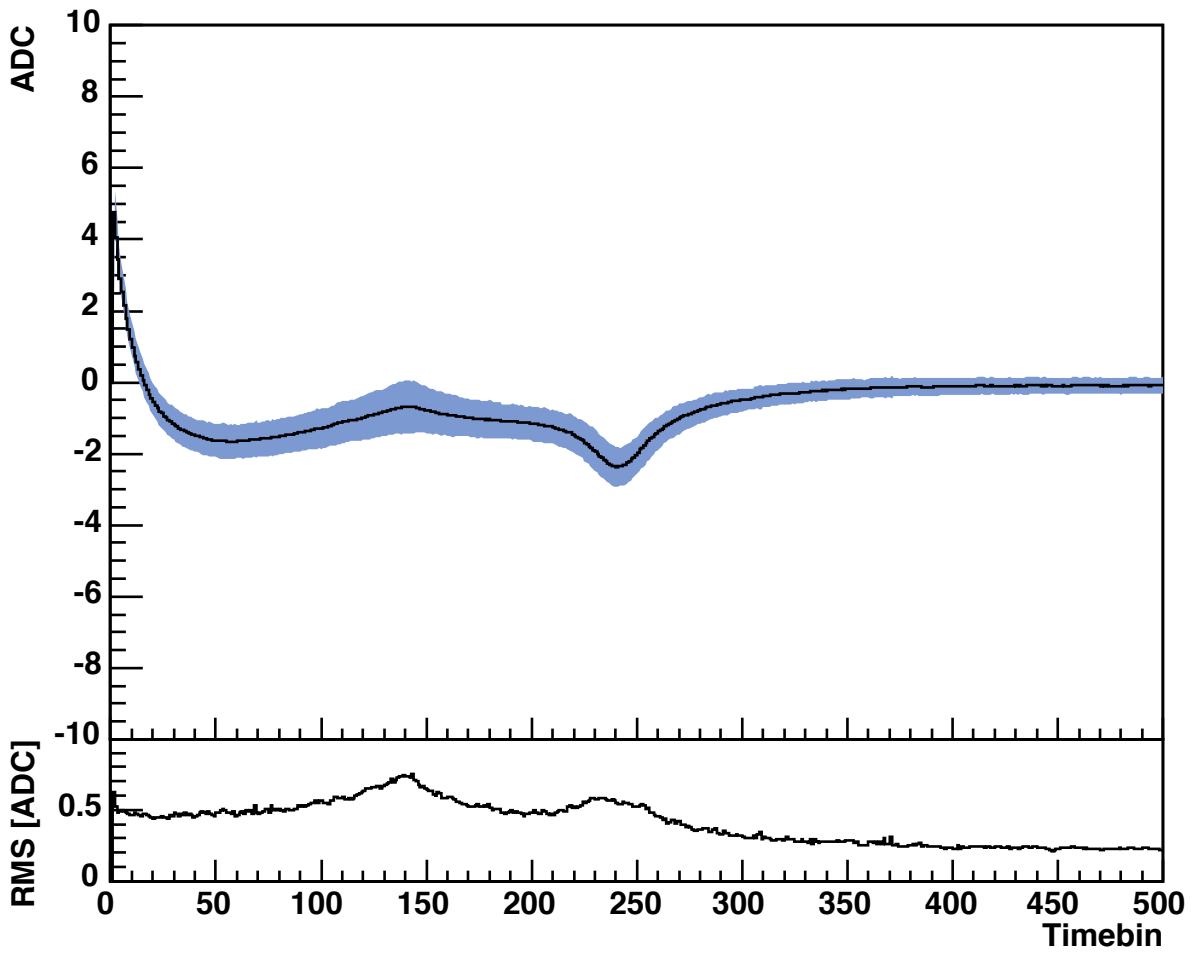

Hit graph (top) and extracted mean and RMS values (bottom) of all pulses after the moving average calculation with a pulsecharge in between 1000 ADC and 1500 ADC. The blue area shows the amount of variation on the signal.

The spread of the tail shape is clearly visible as well as the variations along the time axis. To quantify this, the mean value and the RMS value of each timebin of the hit graph is calculated. The spread now shows also a time dependence that has its maxima at the local maximum and the second minimum. The plot also shows that only a very few samples are lying outside an ADC window with a width of one, so that the tail can be considered as constant. This result implies that the angle of incidence of individual primary electrons plays a minor role in the overall signal shape.
This ion tail only appears if a pulse with a big amount of charge exists. These pulses are quite seldom as shown in the statistics in the beginning of this section and as shown in the chapter »Jitter« on page 23, but small pulses can get lost during the zero suppression stage if they sit on the ion tail of a big pulse. The presence of the ion tail nevertheless will produce a baseline shift in a high occupancy environment due to their pile-up.

### BCS2 Performance

The BCS2 unit in the ALTRO as described in Chapter »Front End Electronics« on page 15, is built to remove disturbances like these. The same data used before to quantify the ion tail is piped through the ALTRO++ (page 43) by the use of the BCS2 to check the correcting capability. The result is shown in the plot below.

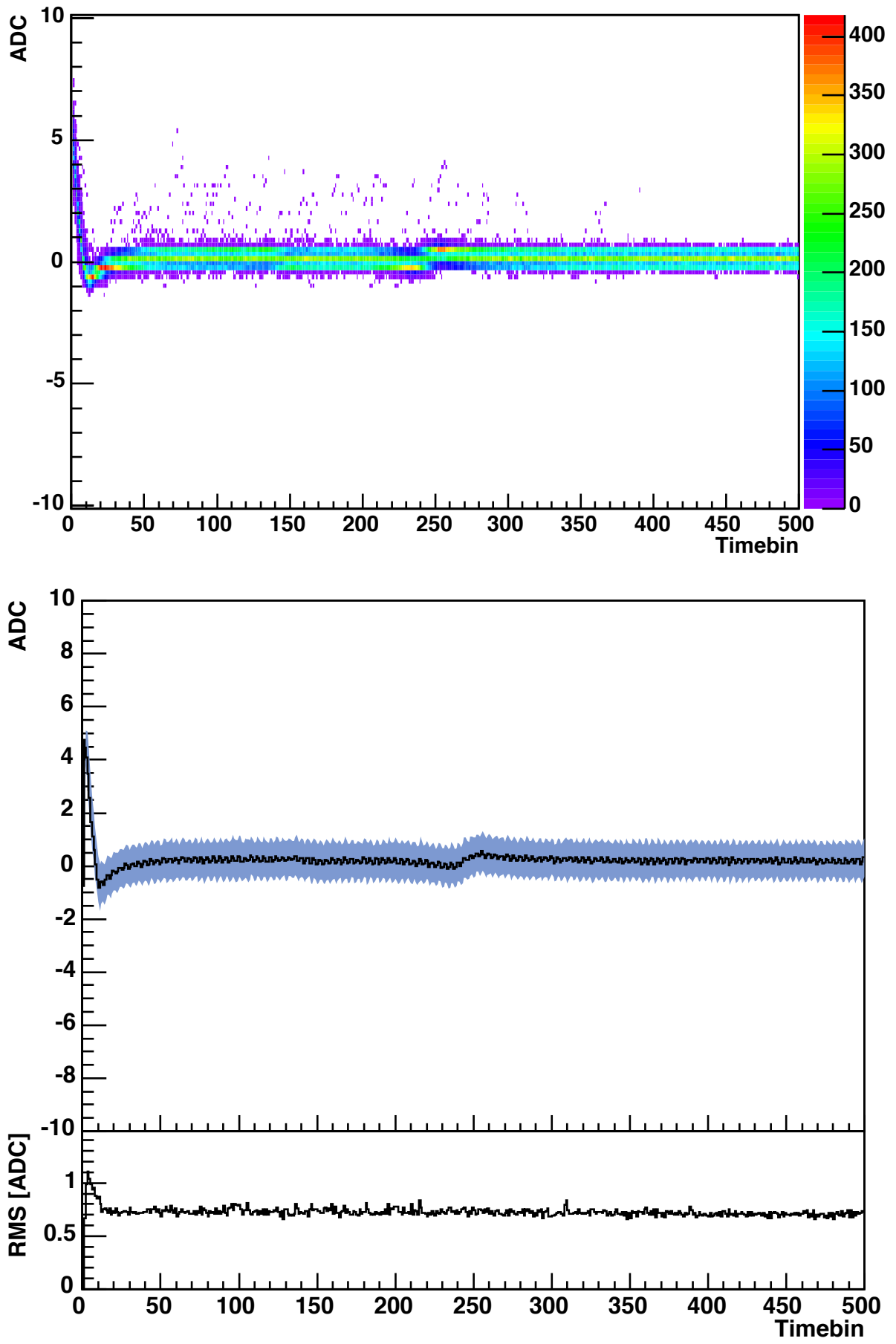

Hit graph (top) and extracted mean and RMS (bottom) of all pulses after the ALTRO++ module and the moving average calculation. The ion tail structure is removed.

The parameter optimisation scheme as described in »BCS2 & ZSU Parameters« on page 42 is not implemented, so that here a few parameters are tested and visually inspected. It is nevertheless unlikely that these parameters give the best performance. This result is achieved by using a low and high threshold of five and two pre- and post-samples.

The remaining structure is far below the noise level of the ALTRO, so that the ion tail influence can be cured by the BCS2 unit of the ALTRO.

## Testbeam

Plenty of data was archived during the testbeam time, but no special runs were taken to study the effect of the ion tail. Here a different gas mixture of $Ne/N_2/CO_2$ was used. It was not possible to include a filter in front of the data acquisition system to remove events without or with small signals, but the capabilities to acquire a huge amount of statistics removes the need of such a filter. The available statistics is shown in the plot below subdivided in their pulse charge classes.

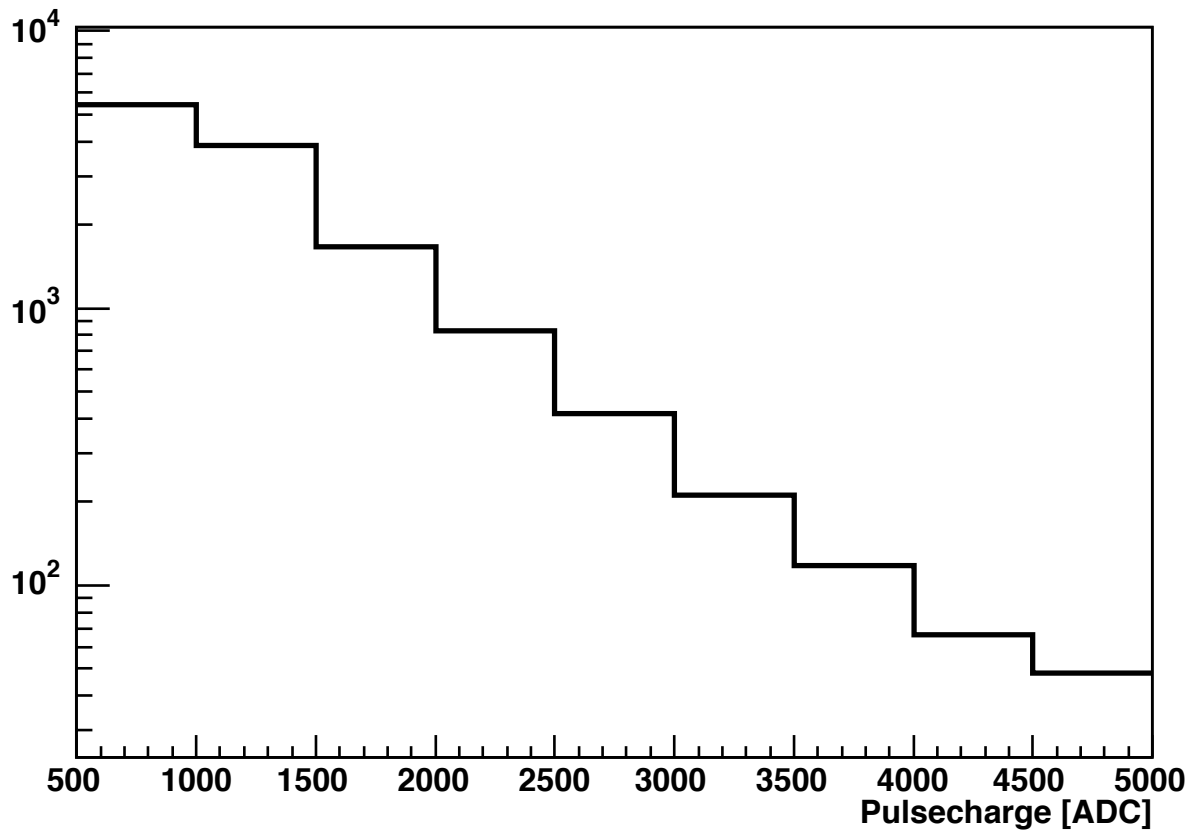

Total collected statistics of 12880 sufficient pulses in the testbeam data subdivided in the later used pulsecharge bins.

The drift time of the used test setup only fills half of the maximum acquisition time window of the ALTRO running at 10 MHz, so that only 512 timebins are recorded. Additionally, the acquisition window is reduced, since the first 12 timebins are zero due to the calculation delay in the ALTRO, as well as the gating grid has a big impact on at least the first 30 timebins. The ion tail lasts for roughly 40 µs that would lead to a hard cut in the pulse position which would dramatically reduce the statistics. The mean of the pulse was accepted from timebin 40 up to 200, so that not every pulse has the complete ion tail recorded. This leads to an acquisition end distribution as shown in the plot below. The results extracted after timebin 300 will therefore get more and more instable.

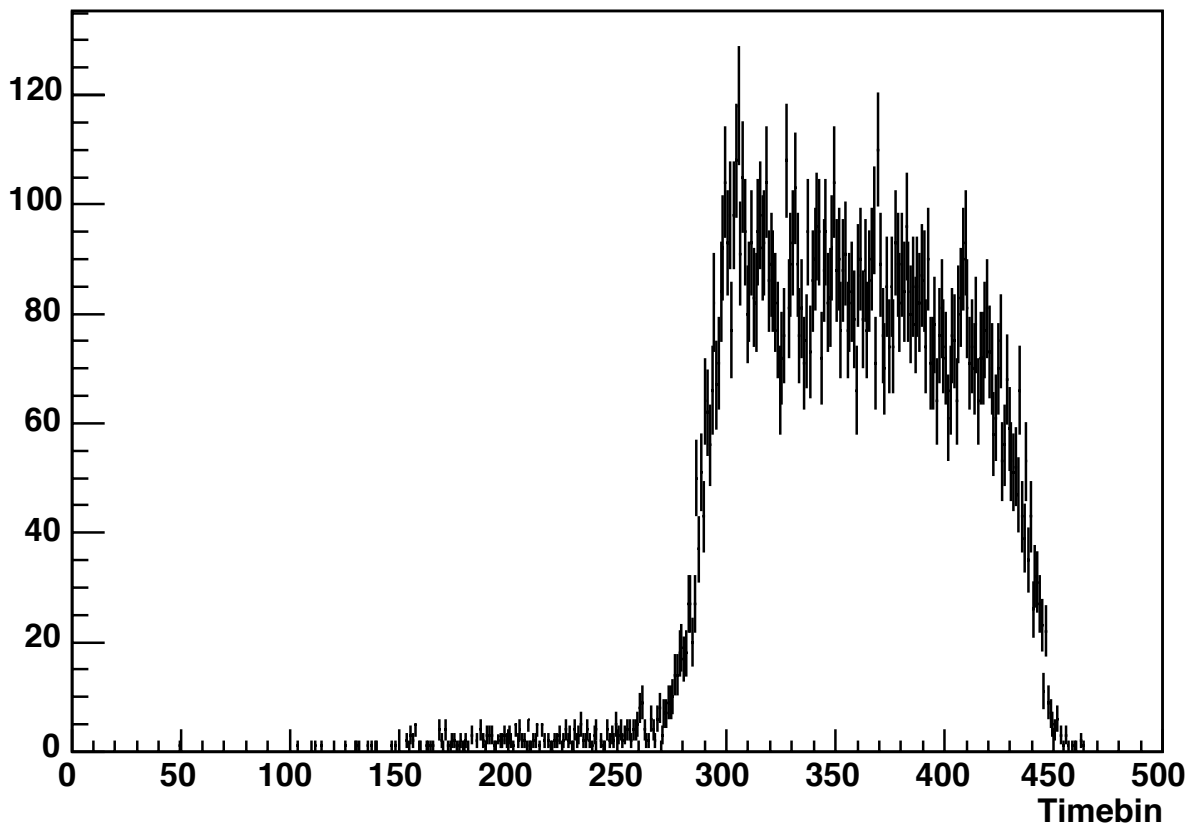

Distribution of the maximum time position recorded for all pulses.

Extracting the tail shape in its pulsecharge bins leads to the following plot, again the signal is inverted for better visibility.

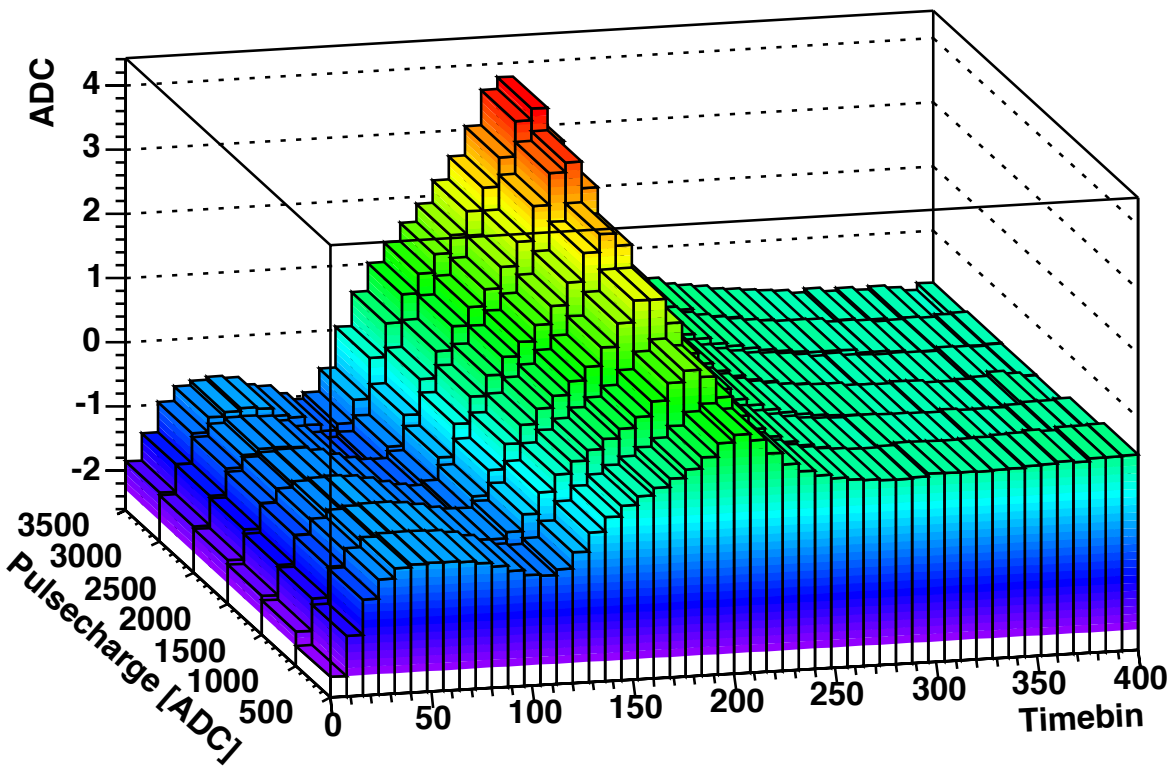

Shape of the inverted ion tail, separated in pulsecharge bins. Eight timebins are averaged to reduce the effect of the noise.

The shape of these ion tails is different compared to that of the cosmics data, the first minimum is above the baseline as well as the local maximum, that is followed by the second minimum. When comparing this cluster shape with the simulation shown on page 50 and the result of the cosmics data shown on page 49, it seems that in $Ne/N_2/CO_2$ more ions are drifting into the pad direction.

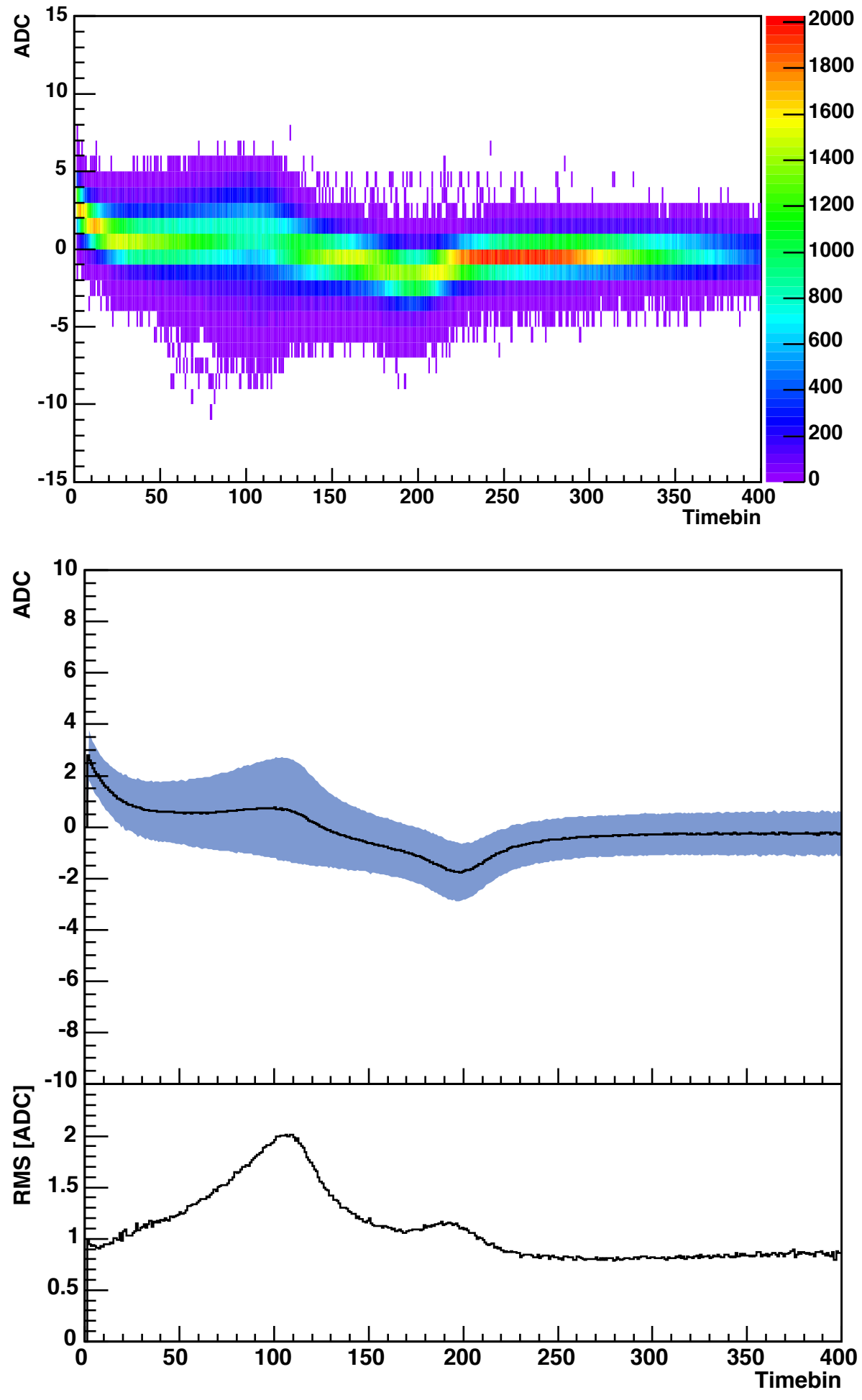

Hit graph (top) and extracted mean and RMS values (bottom) of all pulses with a clustercharge in between 1000 ADC and 1500 ADC. The spread of the signal is increased compared to the cosmics data.

In these data, the time profile of the ion tail also scales linearly with the cluster charge and no time position change is visible. When producing the hit graph out of the ADC values as shown in the previous plot, a change in the spread gets visible.
To determine more precisely the increase of the spread, in each timebin of the hit graph the mean and RMS value is calculated, as also done previously.
The noise of the acquisition chain was quantified at the pedestal calculation to RMS = 0.8 ADC, as described in »Baseline Dispersion« on page 38. When adding the additional uncertainty of the baseline calculation of $\sigma = 0.46$, this leads to a noise level of RMS = 0.92 ADC, which is consistent with looking at the end of the shape as shown in the previous plot. Additionally, it is clearly visible that the spread located at the two maxima is increased beyond the amount added by the inprecise baseline.

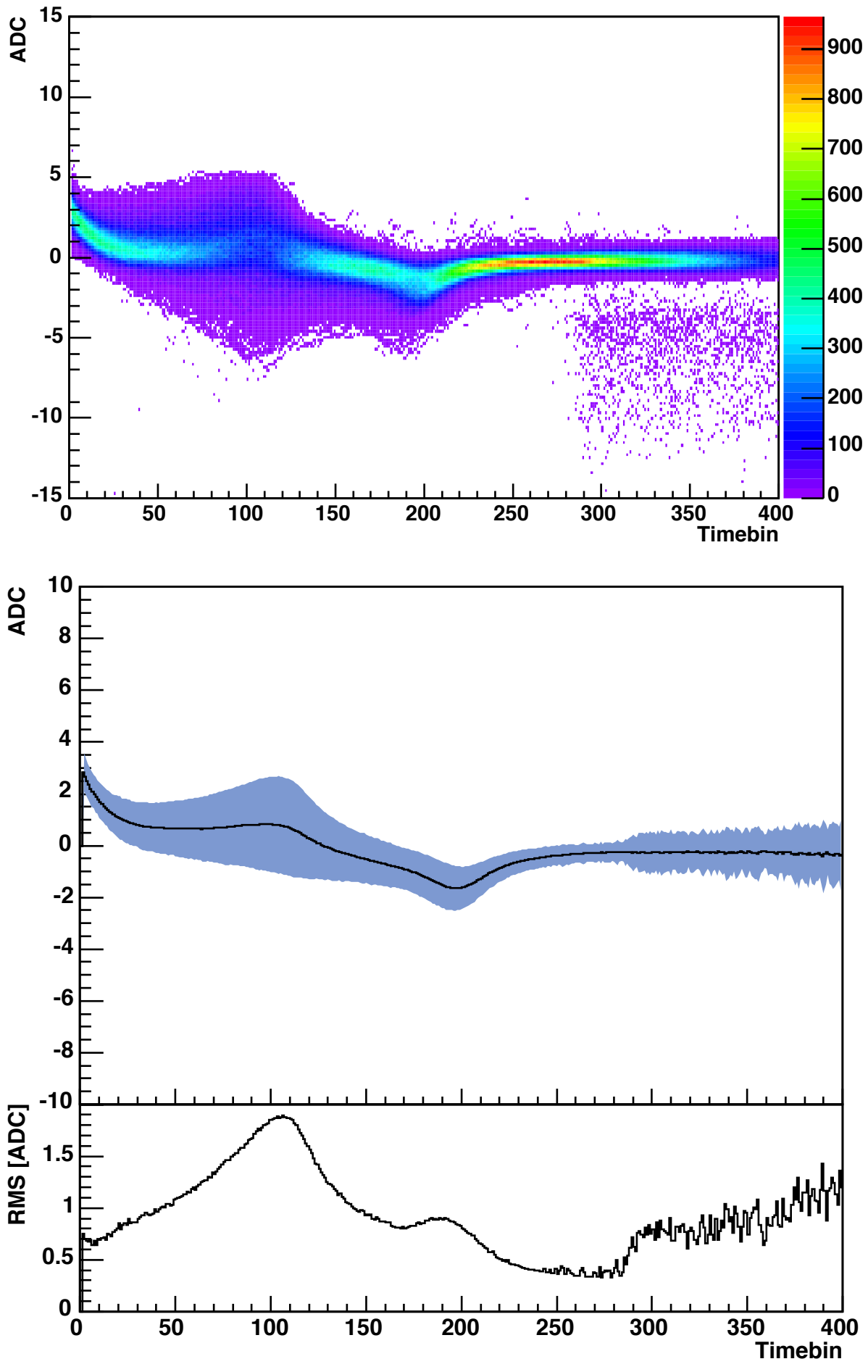

Hit graph (top) and extracted mean and RMS values (bottom) of all pulses after the moving average calculation with a clustercharge in between 1000 ADC and 1500 ADC. The increase in the spread beyond timebin 300 is due to the outlayers.

When producing the hit graph after the moving average filter the broadening of the signals concentrates more on the first maximum. The scattered entries beyond timebin 300 are due to pulses that end beforehand. The spread and the shape is quantified by using the mean and the RMS.
As expected, the spread is big in the area of the first maximum. The outlayers beyond the timebin 300 also increase the RMS, which is only an artefact by the imperfectness of this analysis and data. The increase of the RMS cannot be described by the additional error of the baseline calculation. It seems to be that the addition of Nitrogen changes the ion drift properties in the readout chamber. This analysis will be repeated as soon there is new data, what would cure the problem of the baseline and the short acquisition window.

### BCS2 Performance

These data is piped through the ALTRO++ (page 43) by the use of the BCS2 to check the correcting capability by the use of the same parameters as in the cosmics-based analysis. The result is shown in the plot below. For completeness, the spread of the signal out of the ADC based hit graph is calculated.

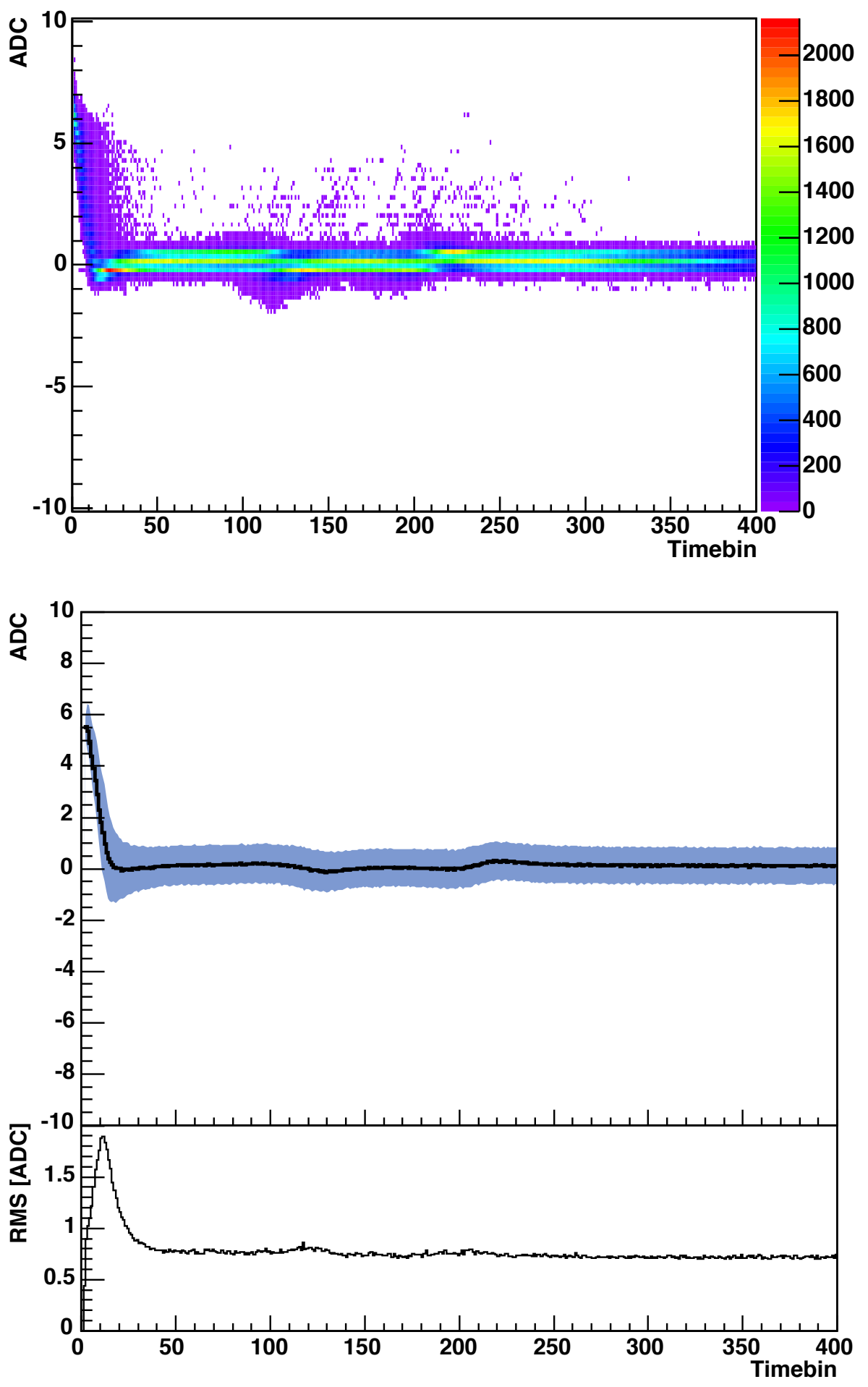

Hit graph (top) and extracted mean and RMS of all pulses after the ALTRO++ module and the moving average calculation. The ion tail structure is removed

The remaining structure is far below the noise level of the ALTRO, so that the ion tail influence can be cured by the BCS2 unit of the ALTRO also in the more inprecise data of the testbeam. Additionally, the influence of the inprecise baseline calculation is reduced, as shown in the plot below.

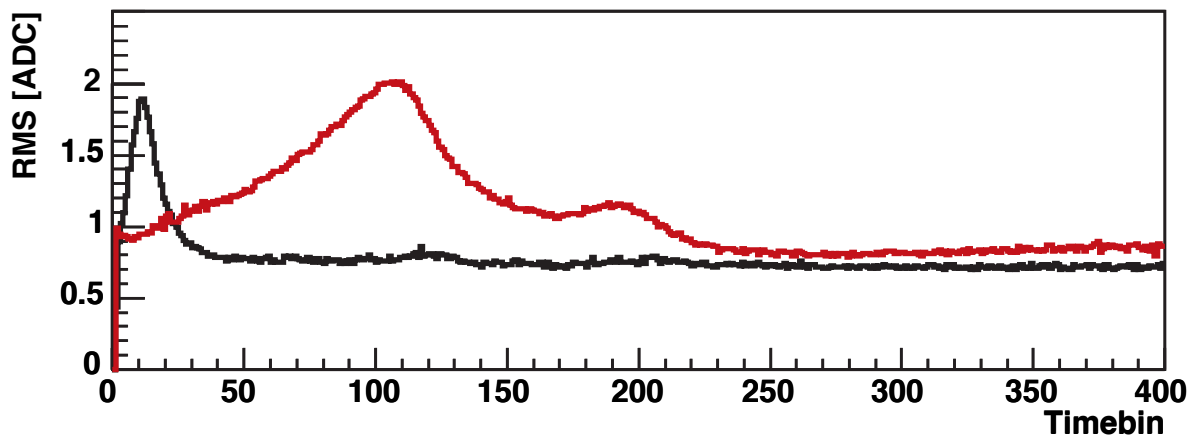

Comparison of the RMS before (red line) and after (black line) the ALTRO++ module. The ion tail structure is removed and the influence of the imperfect baseline calculation is reduced.

The big amount of noise in the start is due to some limitations in the ALTRO++. The problem starts with the tight starting point criteria described at the beginning of this analysis, because the BCS2 unit of the ALTRO++ needs at least eight samples in the acceptance window to properly get on track of the baseline. If this is not possible, the BCS2 creates additional perturbations. It could also happen that the first accepted bins are far away of the baseline so that the BCS2 starts with a wrong value.

# Résumé

# Résumé

The extraction of the configuration and the configuration procedure of the ALTRO (**AL**ICE **T**PC **R**ead**o**ut) chip was studied in this work beneath the analysis of the ion tail effect. For this purpose a software infrastructure was developed, tested and used.

In the examination of the jitter it was found that the clock scheme currently used on the front end electronics with a precision of 300 ps is accurate enough, since the introduced error of the jitter is negligible up to pulses of a maximum amplitude of 200 ADC (**A**nalog **D**igital **C**onverter). These pulses are already three orders of magnitude less probable then a MIP (**M**inimum **I**onising **P**article). This result removed the need of a more complex clock scheme, thus therefore saved a lot of work in implementing, debugging and testing, as well as the required money.

An online monitor system was developed for the TPC (**T**ime **P**rojection **C**hamber) prototype and adapted to the different data acquisition systems used. It can interface to the previous LabVIEW based data acquisition and to the DATE (**D**ate **A**cquisition **T**est **E**nvironment) system. An interface to the HLT (**H**igh **L**evel **T**rigger) system will be integrated. Several additional processing units were implemented to change the monitor according to the actual needs of the users. This includes a simple pulse- and clusterfinder, a moving average calculation, the ALTRO emulation and a dynamic baseline calculation. This monitor will also become the pad monitor for the TPC experts for the physics program of ALICE.

For the different processing units in the digital chain of the ALTRO a configuration scheme was developed. The pedestals are extracted from the real data. This includes also the calculation of the time dependent pedestals for the pedestal reference memory, which is used to remove systematic and constant perturbations. The capability of the correction as well as the quality was tested visually via the online monitor and quantitatively as presented in the according analysis. The measured instability of the baseline was smaller than the noise of the acquisition chain and this without the usage if the AUTOCAL circuit. For the extraction of the tail cancellation parameters only an algorithm implemented in MATLAB was existing. This algorithm was reimplemented in C++ which increased the speed by more then two orders of magnitude. Around this algorithm a complete infrastructure was built, which extracts sufficient pulses, optimises the parameters for the pulse, cross-correlates the coefficients and then searches for the set which shows the best performance. Two schemes were implemented and tested to find the best set. Additionally, the calculation time for the TCF (**T**ail **C**ancellation **F**ilter) parameters was measured. It turned out, that even in the worst case, when all 560000 channels need their own coefficient set the needed CPU time is manageable. Unfortunately, not enough data was available to address the question, if there is a difference in the sets depending on the ALTRO channel. For the remaining two processing units in the ALTRO, the BCS2 (**B**aseline **C**orrection and **S**ubtraction **2**) and ZSU (**Z**ero **S**uppression **U**nit), a scheme to extract the optimal set was developed and will be implemented. At the moment, only parts are existing like the bit-exact emulation of the ALTRO.

To encode the configuration data into the representation accepted by the ALTRO a set of classes were implemented which are already incooperated into the DCS (**D**etector **C**ontrol **S**ystem) system of the TPC.

The signal shape for ions reaching different electrodes was simulated, but neither the fraction of ions drifting towards their targets nor the variation from avalanche to avalanche is yet known. The presented analysis shows that the amplitude of the signal scales linear with the charge of a pulse, due to the small spread in the result this implies that the angle of incidence of the individual primary electrons does not play a significant role in the signal. Additionally the result of the two gas mixtures differs. The Ne/$CO_2$ 90/10 mixture shows no significant contribution of ions drifting to the pads, whereas in the Ne/$N_2$/$CO_2$ 90/5/5 mixture a plateau gets visible. This analysis will be refined if more data is available.

# Bibliography

## Introduction

## The Experiment

## Front End Electronic

### Jitter

### Prototype Environment

### ALTRO Parameter Optimisation

### Ion Tail Analysis

# Appendix

## .OM.config

```
<OM.Cv1.4>
            <General>
                        fixedBaseline            0
                        topviewMode              maxadc
                        displayFromPad           0
                        displayToPad             128
                        displayFromRow           0
                        displayToRow             62
                        screenresx               1600
                        screenresy               1200
                        encoding                 bigendian
                        autosavetopview          0
                        autosaveas               eps
                        debuglevel               4
            </General>
            <MovingAverage>
                        on                       1
                        normAndzoomed            1
                        zoomed                   0
                        left                     3
                        right                    4
                        direction                1
            </MovingAverage>
            <Clusterfinder>
                        padon                    1
                        rowon                    0
                        threshold                5
                        neededSuccesiveADCs      3
                        maxDiffinWeightedMeanofClusterSequences 4
            </Clusterfinder>
            <ClusterFit>
                        on                       0
                        Threshold                0.1
                        Order                    4
                        ShowFitParameters        0
            </ClusterFit>
            <AutoLastEvent>
                        frequency                5000
            </AutoLastEvent>
            <AltroEmulation>
                        on                       0
                        readdbaseline            1
            </AltroEmulation>
</OM.Cv1.4>
```

## run.table

```
<run.tablev1.2>
            <generalPlaces>
                        location    /Users/rbramm/Documents/Altro/AliceTPC/OM/trunk/
                        mapping     /Users/rbramm/Documents/Altro/AliceTPC/TPCMapping/trunk/MappingData/
                        rowmapping  /Users/rbramm/Documents/Altro/AliceTPC/TPCMapping/trunk/MappingData/
                        pedestals   /Users/rbramm/Documents/Altro/AliceTPC/Pedestals/trunk/Data/
                        altroconfig /Users/rbramm/Documents/Altro/AliceTPC/Altro/trunk/AltroConfigs/
            </generalPlaces>
            <DATEFile>
                        location    none /Users/rbramm/Documents/Altro/AliceTPC/DataFormat/trunk/
                        mapping     generalPlaces            MappingHWAdress.data
                        rowmapping  generalPlaces            MappingHWAdressRow.data
                        pedestals   generalPlaces            Pedestals.runDATE.data
                        altroconfig generalPlaces            AltroConfig.off.data
            </DATEFile>
            <DATEStream>
                        location    none @epaitbeam01:
                        mapping     generalPlaces            MappingHWAdress.data
                        rowmapping  generalPlaces            MappingHWAdressRow.data
                        pedestals   generalPlaces            Pedestals.norun.data
                        altroconfig generalPlaces            AltroConfig.off.data
            </DATEStream>
            ...
            ...
            <run0075>
                        location    generalPlaces            run0075
                        mapping     generalPlaces            mapping3231302910987.data
                        rowmapping  generalPlaces            mappingRowPad3231302910987.data
                        pedestals   generalPlaces            Pedestals.run0075.data
                        altroconfig generalPlaces            AltroConfig.NS.run0067.data
            </run0075>
            <run0077>
                        location    generalPlaces            run0077
                        mapping     generalPlaces            mapping3231302910987.data
                        rowmapping  generalPlaces            mappingRowPad3231302910987.data
                        pedestals   generalPlaces            Pedestals.run0077.data
                        altroconfig generalPlaces            AltroConfig.NS.run0067.data
            </run0077>
            <run0079>
                        location    generalPlaces            run0079
                        mapping     generalPlaces            mapping3231302910987.data
                        rowmapping  generalPlaces            mappingRowPad3231302910987.data
                        pedestals   generalPlaces            Pedestals.run0079.data
                        altroconfig generalPlaces            AltroConfig.NS.run0067.data
            </run0079>
            <run0081>
                        location    generalPlaces            run0081
                        mapping     generalPlaces            mapping3231302910987.data
                        rowmapping  generalPlaces            mappingRowPad3231302910987.data
                        pedestals   generalPlaces            Pedestals.run0081.data
                        altroconfig generalPlaces            AltroConfig.NS.run0067.data
            </run0081>
</run.tablev1.2>
```

## BCS1 Parameters Calculation log

```
 0 Event:   14 secondEquipment: 0 DefChannels: 1552 double Addresses: 1552 = 100 overflow Addresses:    0 =   0 Invalid Addresses:    0 =   0
 1 Event:   47 secondEquipment: 0 DefChannels:    4 double Addresses:    4 = 100 overflow Addresses:    0 =   0 Invalid Addresses:    0 =   0
 2 Event:   66 secondEquipment: 0 DefChannels:    4 double Addresses:    4 = 100 overflow Addresses:    0 =   0 Invalid Addresses:    0 =   0
 3 Event:   77 secondEquipment: 0 DefChannels: 2714 double Addresses: 2714 = 100 overflow Addresses:    0 =   0 Invalid Addresses:    0 =   0
 4 Event:   79 secondEquipment: 0 DefChannels: 2418 double Addresses: 2418 = 100 overflow Addresses:    0 =   0 Invalid Addresses:    0 =   0
 5 Event:  101 secondEquipment: 0 DefChannels: 1800 double Addresses: 1800 = 100 overflow Addresses:    0 =   0 Invalid Addresses:    0 =   0
 6 Event:  109 secondEquipment: 0 DefChannels: 1144 double Addresses: 1144 = 100 overflow Addresses:    0 =   0 Invalid Addresses:    0 =   0
 7 Event:  136 secondEquipment: 0 DefChannels:    4 double Addresses:    4 = 100 overflow Addresses:    0 =   0 Invalid Addresses:    0 =   0
 8 Event:  141 secondEquipment: 0 DefChannels:    2 double Addresses:    2 = 100 overflow Addresses:    0 =   0 Invalid Addresses:    0 =   0
 9 Event:  171 secondEquipment: 0 DefChannels:    4 double Addresses:    4 = 100 overflow Addresses:    0 =   0 Invalid Addresses:    0 =   0
10 Event:  178 secondEquipment: 0 DefChannels:  790 double Addresses:  790 = 100 overflow Addresses:    0 =   0 Invalid Addresses:    0 =   0
11 Event:  186 secondEquipment: 0 DefChannels: 3018 double Addresses: 3018 = 100 overflow Addresses:    0 =   0 Invalid Addresses:    0 =   0
12 Event:  188 secondEquipment: 0 DefChannels: 2268 double Addresses: 2268 = 100 overflow Addresses:    0 =   0 Invalid Addresses:    0 =   0
```

```
  13 Event:  189 secondEquipment: 0 DefChannels: 3154 double Addresses: 3072 =  97 overflow Addresses:    0 =   0 Invalid Addresses:   82 =   2
  14 Event:  190 secondEquipment: 0 DefChannels:  582 double Addresses:  582 = 100 overflow Addresses:    0 =   0 Invalid Addresses:    0 =   0
  15 Event:  197 secondEquipment: 0 DefChannels: 1624 double Addresses: 1624 = 100 overflow Addresses:    0 =   0 Invalid Addresses:    0 =   0
  16 Event:  206 secondEquipment: 0 DefChannels:    2 double Addresses:    2 = 100 overflow Addresses:    0 =   0 Invalid Addresses:    0 =   0
  17 Event:  229 secondEquipment: 0 DefChannels:    4 double Addresses:    4 = 100 overflow Addresses:    0 =   0 Invalid Addresses:    0 =   0
  18 Event:  235 secondEquipment: 0 DefChannels:    4 double Addresses:    4 = 100 overflow Addresses:    0 =   0 Invalid Addresses:    0 =   0
  19 Event:  236 secondEquipment: 0 DefChannels: 2158 double Addresses: 2158 = 100 overflow Addresses:    0 =   0 Invalid Addresses:    0 =   0
  20 Event:  239 secondEquipment: 0 DefChannels:    2 double Addresses:    2 = 100 overflow Addresses:    0 =   0 Invalid Addresses:    0 =   0
  21 Event:  251 secondEquipment: 0 DefChannels: 2588 double Addresses: 2588 = 100 overflow Addresses:    0 =   0 Invalid Addresses:    0 =   0
  22 Event:  269 secondEquipment: 0 DefChannels:    2 double Addresses:    2 = 100 overflow Addresses:    0 =   0 Invalid Addresses:    0 =   0
  23 Event:  310 secondEquipment: 0 DefChannels:    4 double Addresses:    4 = 100 overflow Addresses:    0 =   0 Invalid Addresses:    0 =   0
  24 Event:  314 secondEquipment: 0 DefChannels: 2056 double Addresses: 2056 = 100 overflow Addresses:    0 =   0 Invalid Addresses:    0 =   0
  25 Event:  335 secondEquipment: 0 DefChannels: 1730 double Addresses: 1730 = 100 overflow Addresses:    0 =   0 Invalid Addresses:    0 =   0
  26 Event:  339 secondEquipment: 0 DefChannels: 1432 double Addresses: 1432 = 100 overflow Addresses:    0 =   0 Invalid Addresses:    0 =   0
  27 Event:  359 secondEquipment: 0 DefChannels: 2038 double Addresses: 2038 = 100 overflow Addresses:    0 =   0 Invalid Addresses:    0 =   0
  28 Event:  361 secondEquipment: 0 DefChannels: 3008 double Addresses: 3008 = 100 overflow Addresses:    0 =   0 Invalid Addresses:    0 =   0
  29 Event:  384 secondEquipment: 0 DefChannels: 2744 double Addresses: 2744 = 100 overflow Addresses:    0 =   0 Invalid Addresses:    0 =   0
  30 Event:  396 secondEquipment: 0 DefChannels:    2 double Addresses:    2 = 100 overflow Addresses:    0 =   0 Invalid Addresses:    0 =   0
  31 Event:  400 secondEquipment: 0 DefChannels: 1814 double Addresses: 1814 = 100 overflow Addresses:    0 =   0 Invalid Addresses:    0 =   0
  32 Event:  404 secondEquipment: 0 DefChannels: 2908 double Addresses: 2908 = 100 overflow Addresses:    0 =   0 Invalid Addresses:    0 =   0
  33 Event:  407 secondEquipment: 0 DefChannels: 1792 double Addresses: 1792 = 100 overflow Addresses:    0 =   0 Invalid Addresses:    0 =   0
  34 Event:  424 secondEquipment: 0 DefChannels:  924 double Addresses:  924 = 100 overflow Addresses:    0 =   0 Invalid Addresses:    0 =   0
  35 Event:  432 secondEquipment: 0 DefChannels: 1748 double Addresses: 1748 = 100 overflow Addresses:    0 =   0 Invalid Addresses:    0 =   0
  36 Event:  433 secondEquipment: 0 DefChannels: 1052 double Addresses: 1052 = 100 overflow Addresses:    0 =   0 Invalid Addresses:    0 =   0
  37 Event:  478 secondEquipment: 0 DefChannels: 3132 double Addresses: 3072 =  98 overflow Addresses:    0 =   0 Invalid Addresses:   60 =   1
  38 Event:  492 secondEquipment: 0 DefChannels: 2574 double Addresses: 2574 = 100 overflow Addresses:    0 =   0 Invalid Addresses:    0 =   0
  39 Event:  522 secondEquipment: 0 DefChannels: 1794 double Addresses: 1794 = 100 overflow Addresses:    0 =   0 Invalid Addresses:    0 =   0
  40 Event:  531 secondEquipment: 0 DefChannels:    4 double Addresses:    4 = 100 overflow Addresses:    0 =   0 Invalid Addresses:    0 =   0
  41 Event:  533 secondEquipment: 0 DefChannels:    4 double Addresses:    4 = 100 overflow Addresses:    0 =   0 Invalid Addresses:    0 =   0
  42 Event:  578 secondEquipment: 0 DefChannels: 1350 double Addresses: 1350 = 100 overflow Addresses:    0 =   0 Invalid Addresses:    0 =   0
  43 Event:  580 secondEquipment: 0 DefChannels: 1596 double Addresses: 1596 = 100 overflow Addresses:    0 =   0 Invalid Addresses:    0 =   0
  44 Event:  609 secondEquipment: 0 DefChannels: 3117 double Addresses: 3072 =  98 overflow Addresses:    0 =   0 Invalid Addresses:   45 =   1
  45 Event:  619 secondEquipment: 0 DefChannels: 1844 double Addresses: 1844 = 100 overflow Addresses:    0 =   0 Invalid Addresses:    0 =   0
  46 Event:  620 secondEquipment: 0 DefChannels:  166 double Addresses:  166 = 100 overflow Addresses:    0 =   0 Invalid Addresses:    0 =   0
  47 Event:  637 secondEquipment: 0 DefChannels: 3113 double Addresses: 3072 =  98 overflow Addresses:    0 =   0 Invalid Addresses:   41 =   1
  48 Event:  650 secondEquipment: 0 DefChannels: 1832 double Addresses: 1832 = 100 overflow Addresses:    0 =   0 Invalid Addresses:    0 =   0
  49 Event:  666 secondEquipment: 0 DefChannels:    4 double Addresses:    4 = 100 overflow Addresses:    0 =   0 Invalid Addresses:    0 =   0
  50 Event:  677 secondEquipment: 0 DefChannels: 2132 double Addresses: 2132 = 100 overflow Addresses:    0 =   0 Invalid Addresses:    0 =   0
  51 Event:  692 secondEquipment: 0 DefChannels: 2472 double Addresses: 2472 = 100 overflow Addresses:    0 =   0 Invalid Addresses:    0 =   0
  52 Event:  695 secondEquipment: 0 DefChannels: 2458 double Addresses: 2458 = 100 overflow Addresses:    0 =   0 Invalid Addresses:    0 =   0
  53 Event:  696 secondEquipment: 0 DefChannels: 1760 double Addresses: 1760 = 100 overflow Addresses:    0 =   0 Invalid Addresses:    0 =   0
  54 Event:  701 secondEquipment: 0 DefChannels:    2 double Addresses:    2 = 100 overflow Addresses:    0 =   0 Invalid Addresses:    0 =   0
  55 Event:  703 secondEquipment: 0 DefChannels:    2 double Addresses:    2 = 100 overflow Addresses:    0 =   0 Invalid Addresses:    0 =   0
  56 Event:  706 secondEquipment: 0 DefChannels: 3296 double Addresses: 3296 = 100 overflow Addresses:    0 =   0 Invalid Addresses:    0 =   0
  57 Event:  714 secondEquipment: 0 DefChannels:    4 double Addresses:    4 = 100 overflow Addresses:    0 =   0 Invalid Addresses:    0 =   0
  58 Event:  718 secondEquipment: 0 DefChannels:  988 double Addresses:  988 = 100 overflow Addresses:    0 =   0 Invalid Addresses:    0 =   0
  59 Event:  722 secondEquipment: 0 DefChannels: 1292 double Addresses: 1292 = 100 overflow Addresses:    0 =   0 Invalid Addresses:    0 =   0
  60 Event:  778 secondEquipment: 0 DefChannels: 4338 double Addresses: 4338 = 100 overflow Addresses:    0 =   0 Invalid Addresses:    0 =   0
  61 Event:  779 secondEquipment: 0 DefChannels:    2 double Addresses:    2 = 100 overflow Addresses:    0 =   0 Invalid Addresses:    0 =   0
  62 Event:  790 secondEquipment: 0 DefChannels:    4 double Addresses:    4 = 100 overflow Addresses:    0 =   0 Invalid Addresses:    0 =   0
  63 Event:  796 secondEquipment: 0 DefChannels:    4 double Addresses:    4 = 100 overflow Addresses:    0 =   0 Invalid Addresses:    0 =   0
  64 Event:  813 secondEquipment: 0 DefChannels: 1834 double Addresses: 1834 = 100 overflow Addresses:    0 =   0 Invalid Addresses:    0 =   0
  65 Event:  826 secondEquipment: 0 DefChannels: 1496 double Addresses: 1496 = 100 overflow Addresses:    0 =   0 Invalid Addresses:    0 =   0
  66 Event:  885 secondEquipment: 0 DefChannels: 1504 double Addresses: 1504 = 100 overflow Addresses:    0 =   0 Invalid Addresses:    0 =   0
  67 Event:  889 secondEquipment: 0 DefChannels:  726 double Addresses:  726 = 100 overflow Addresses:    0 =   0 Invalid Addresses:    0 =   0
  68 Event:  890 secondEquipment: 0 DefChannels:  940 double Addresses:  940 = 100 overflow Addresses:    0 =   0 Invalid Addresses:    0 =   0
  69 Event:  910 secondEquipment: 0 DefChannels:    2 double Addresses:    2 = 100 overflow Addresses:    0 =   0 Invalid Addresses:    0 =   0
  70 Event:  914 secondEquipment: 0 DefChannels: 1936 double Addresses: 1936 = 100 overflow Addresses:    0 =   0 Invalid Addresses:    0 =   0
  71 Event:  989 secondEquipment: 0 DefChannels: 2642 double Addresses: 2642 = 100 overflow Addresses:    0 =   0 Invalid Addresses:    0 =   0
  72 Event:  991 secondEquipment: 0 DefChannels:    4 double Addresses:    4 = 100 overflow Addresses:    0 =   0 Invalid Addresses:    0 =   0
  73 Event:  994 secondEquipment: 0 DefChannels:    4 double Addresses:    4 = 100 overflow Addresses:    0 =   0 Invalid Addresses:    0 =   0
------------------
Count 'DefChannels: 2'         : 9
Count 'DefChannels > 2'        : 65
Count 'overflow Addresses > 0' : 0
Count 'Invalid Addresses > 0'  : 4
Total Errors                   : 74 = 7%
Total Events                   : 1000 Events Analysed.
```

## BCS1 Parameters Extraction

```
makeDatePedestalMem, extracts the pedestal memory content out of the testbeam date files
Parameters:
-t eventtype   : sets the eventtype: [mandatory]
               : DATEFile = Testbeam DATE events
-rn runnumber  : sets the runnumber [mandatory]
-rp runpath    : sets path to the run [mandatory]
-d             : sets the debuglevel  [default: 0]
-n #events     : number of events [default: 1000]
-as presamples : number of samples excluded at the start [default: 40]
               : This is needed to exclude the gating Grid influence
-am maxtimebin : maximum timebin for calculation [default: 500]
-aw window     : Acceptance window for the 2nd BSL pass calculatoin [default: 5]
-ca defchannels: maximum of allowed defunctioning channels [default: 0]
-f             : switch to turn on saving as float numbers
-b             : switch to turn on saving as binary, BEWARE:
               : this is not endianness save !!!
               : (My Pedestal Handler automaticly detects and swappes)
-c             : switch to turn on compression in gzip format

example: ./makeDatePedestalMem.app -t DATEFile -rn 820 -rp /Volumes/Daten/TestBeam -n 1000 -as 40 -am 500 -aw 5 -mw 10 -ca 0 -f
```

## Pulse Finder

```
pulseFinder, extracts selected pulses and denoises the tail.
Run Parameters :
    -t, --EventType :
        sets the eventtype: [mandatory]
        DATEFile = Testbeam DATE events
    -rn, --Runnumber :
        sets the runnumber [mandatory]
    -rp, --RunPath :
        sets path to the run [mandatory]
    -n, --EventCount :
        number of events [default: 1000]
    -o, --OutPlace :
        Path to the Output dir for the results [mandatory]
        BAWARE Folder MUST exist

Pulse Finder Parameters:
    -pa, --PreAquisitionSamples :
        Pre Aquisition Samples. [default: 40]
    -ps --PreSamples :
        Number of Presamples where no signal is allowed,
        to circumvent Pulses before event start. [default: 5]
    -pt --PreSampleThreshold :
        Threshold to define what is a pulse in the presample
        area [default: 20]
    -ah --MaxADCThreshold :
```

```
        Max ADCThreshold, upper boundarys to specify which
        clusters are to be found [800]
    -al --MinADCThreshold :
        Min ADCThreshold, lower boundarys to specify which
         clusters are to be found [600]
    -sp --MaxTimePosition :
        Maximum Time Position of Pulse [200]
        BEWARE ! The programm always assumes 1024 timebins max !
    -ct --PulseThreshold :
        Threshold, from where on somthing is called a
        pulse [default: 10]
    -cs --NeededSuccessiveADC :
        Number of consecutive Samples above PulseThreshold
        needed to define a Pulse [default: 3]
    -cf --FitThreshold :
        Factor to specify level of end of cluster. means
        maxadc*FitThreshold [default: 0.1]
    -ia --IntegralVSAmpThreshold :
        Integral vs Amplitude Threshold, to filter out double
        clusters [default: 4.5]

Moving Average Parameters/Smoothing parameters:
    -ml --MALeftSamples :
        MALeftSamples samples left to actual Point [default: 3]
    -mr --MARightSamples :
        MARightSamples samples right to actual Point [default: 4]
    -md --MADirection :
        Direction > 0 = from left to right; < 0 = vice versa [default: 1]
    -gt --AllowedGlitchesinSignal :
        Allowed Glitches in Signal threeshold [default: 5]

example: ./pulseFinder.app -t DATEFile -rn 820 -rp /Volumes/Daten/TestBeam/ -o ./run820/ -n 1000 -d 0 -ah 800 -al 600 -sp 200
```

## Parameter Set Finder

```
writePulsestoRoot, reads the results of the pulse finder and writes them into a root file
Run Parameters :
    -rn, --Runnumber :
        sets the runnumber [mandatory]
    -p, --PulsePlace :
        Path to the Output dir of the pulse finder [mandatory]

Stage 1 Prameters :
    -1e, --Stage1Epsilon :
        sets allowed variation [default: 0.0015]
    -1a, --Stage1AmplitudeTolerance :
        sets allowed amplitude tolerance [default: 0.1]
    -1s, --Stage1Start :
        sets the start timebin for the optimisation [default: 0]
    -1e, --pStage1End :
        sets the end timebin for the optimisation [default: 200]

Stage 2 Prameters :
    -2e, --Stage2Epsilon :
        sets allowed variation [default: 0.002]
    -2a, --Stage2AmplitudeTolerance :
        sets allowed amplitude tolerance [default: 0.1]
    -2s, --Stage2Start :
        sets the start timebin for the optimisation [default: 0]
    -2e, --pStage2End :
        sets the end timebin for the optimisation [default: 40]

Stage 3 / Equalisation Stage Prameters :
    -3s, --EqualStart :
        sets the start timebin for the optimisation [default: 0]
    -3e, --EqualEnd :
        sets the end timebin for the optimisation [default: endofpulse]

example: ./writePulsestoRoot.app -rn 820 -p ./teststart/
```

## Correlator

```
correlator, reads the results of the pulse finder and of the makeCoefficient
Programm and builds the correlation Matrix

Run Parameters :
    -rn, --Runnumber :
        sets the runnumber [mandatory]
    -p, --PulsePlace :
        Path to the Output dir of the pulse finder [mandatory]

Correlation Parameters :
    -w, --LevelofWidthofPulse :
        sets Level on wich the width of Pulse is calculated,
        (maxADC of Pulse)*LevelofWidthofPulse [default : 0.01]
    -u, --LevelofUndershootofPulse :
        sets Level on wich the undershoot after the Pulse is calculated,
        (maxADC of Pulse)*LevelofUndershootofPulse [default : 0.01]

example: ./correlator.app -rn 820 -p ./teststart/
```

## Best Set Finder

```
findBestSet, reads the correlation Matrix ans searches for the best set

Run Parameters :
    -rn, --Runnumber :
        sets the runnumber [mandatory]
    -p, --PulsePlace :
        Path to the Output dir of the pulse finder [mandatory]

    -m, --CorrelationMatrix :
        Path to the Output of the Correlation Matrix of the correlator [mandatory]

Quality wheight Parameters to find best set:
    -wq, --WheightQualityAlgorithmus :
        Turns on the WheightQualityAlgorithmus
    -wr, --AdditionalRMSWheighting :
        Turns on the RMS wheighting
    -wa, --WheightofAmplitude :
        sets the wheight of the amplitude difference on the overall
        Quality of the parameter set [default : 1]
    -ws, --WheightofShortening :
        sets the wheight of the shortening amount on the overall
        Quality of the parameter set [default : 1]
    -wu, --WheightofUndershootIntegral :
        sets the wheight of the Integral of the undershoot on the overall
```

```
        Quality of the parameter set [default : 1]
    -wd, --WheightQualityAlgorithmDetail :
        Adds a detailed Printout of the choosen Parameterset

Vote wheight parameters to find best set:
    -vw, --WheightedVoteAlgorithm :
        Turns on the Vote Algorithmus
    -vr, --VoteAdditionalRMSWheighting :
        Turns on the RMS wheighting
    -va, --VotesWheightofAmplitude :
        sets the wheight of the amplitude difference on the overall
        Quality of the parameter set [default : 1]
    -vs, --VotesWheightofShortening :
        sets the wheight of the shortening amount on the overall
        Quality of the parameter set [default : 1]
    -vl, --VotesAdditionalMalusonLengthening :
        Adds an additional Wheoght (if > 1) on ONLY the sets which
        lengthen the Pulse [default : 1]
    -vu, --VotesWheightofUndershootIntegral :
        sets the wheight of the Integral of the undershoot on the overall
        Quality of the parameter set [default : 1]
    -vp, --WheightedVoteArea :
        sets the area of best sets if sets are nearby. Parameter is set in
        Percent additional to the minimum vote [default : 1]
    -vc, --WheightedVoteAreaCount :
        Prints WheightedVoteAreaCount of the best results. If set
        overrides the WheightedVoteArea setting [default : 1]
    -vd, --WheightedVoteDetail :
        Adds a detailed Printout of the choosen Parameterset

Debug/check otions:
    -cm --CheckMonotony :
        Flag to add a monotony check

example: ./findBestSet.app -rn 820 -p ./teststart/ -m correlationMatrix.data -wq -vw -vr -vl 5 -va 5 -vs 5 -vu 1 -vc 10 -vd
```

## oldTail

```
CalcTail,
Run Parameters :
    -t, --EventType :
        sets the eventtype: [mandatory]
        DATEFile = Testbeam DATE events
    -rn, --Runnumber :
        sets the runnumber [mandatory]
    -rp, --RunPath :
        sets path to the run [mandatory]
    -rb, --PathtoPedestalMem :
    -n, --EventCount :
        number of events [default: 1000]
    -o, --Outfile :
        Output file of root [default: Runnumber]
Cluster Finder Parameters:
    -cs, --ClusterFinderStartPos [default: 30]
    -ct, --ClusterFinderThreshold [default: 5]
    -cn, --ClusterFinderSuccessiveADC [default: 3]
    -cf, --ClusterFinderFitThreshold [default: 0.1]
Cluster End Refinment Parameters:
    -er, --ClusterEndRefinmentOff [default: On]
    -el, --ClusterEndRefinmentLowThreshold [default: 6]
    -eh, --ClusterEndRefinmentHighThreshold [default: 2]
    -em, --ClusterEndRefinmentMaxCorrection [default: 10]
Moving Average Parameters:
    -ms, --MovingAverageStart [default: ClusterFinderStartPos]
    -ml, --MovingAverageLeft [default: 3]
    -mr, --MovingAverageRight [default: 4]
    -md, --MovingAverageDirection [default: 1]
Pulse Acceptance Parameters:
    -ps, --PulseAcceptanceWindowStart [default: 30]
    -pe, --PulseAcceptanceWindowEnd [default: 300]
    -pm, --PulseMinADC [default: 200]
    -px, --PulseMaxAquisitionTimebin [default: 1024]
Negative Signal Filter:
    -no, --NegativeSignalFilterOn [default: On]
    -nf, --NegativeSignalFilterOff [default: On]
    -ns, --NegativeSignalFilterStart [default: DynamicBaselineStartPos]
    -ne, --NegativeSignalFilterEnd [default: DynamicBaselineEndPos]
    -nt, --NegativeSignalFilterThreshold [default: 30]
    -na, --NegativeSignalFilterAllowedSamples [default: 2]
ALTRO++ Parameters:
    -ao, --Altro++On [default: Off]
    -as, --Altro++StartPosition [default: 40]
    -a1, --Altro++BCS1On [default: On]
    -at, --Altro++TCFOn [default: Off]
    -a2, --Altro++BCS2On [default: On]
    -ac, --Altro++ClippingOn [default: Off]
    -a1h, --Altro++BCS1HighThreshold [default: 5]
    -a1l, --Altro++BCS1LowThreshold [default: 5]
    -a1e, --Altro++BCS1PreSamples [default: 2]
    -a1o, --Altro++BCS1PostSamples [default: 2]
    -a1O, --Altro++BCS1Offset [default: 0]
Signal Debugging Checks and Parameters::
    -scp --SaveCompletePulse [default off]
example: ./oldTail.app -t run0052 -rn 52 -rp /Volumes/Daten/RUNS/ -cs 30 -ps 30 -pe 300 -px 1024 -n 4555 -o run0052
```

## dateTail

```
CalcTail,
Run Parameters :
    -t, --EventType :
        sets the eventtype: [mandatory]
        DATEFile = Testbeam DATE events
    -rn, --Runnumber :
        sets the runnumber [mandatory]
    -rp, --RunPath :
        sets path to the run [mandatory]
    -rb, --PathtoPedestalMem :
    -n, --EventCount :
        number of events [default: 1000]
    -o, --Outfile :
        Output file of root [default: Runnumber]
Dynamic Baseline Calculation Parameters:
    -ds, --DynamicBaselineStartPos [default: 40]
    -de, --DynamicBaselineEndPos [default: 500]
    -dl, --DynamicBaselineLowThreshold [default: 1]
    -dh, --DynamicBaselineHighThreshold [default: 3]
Cluster Finder Parameters:
    -cs, --ClusterFinderStartPos [default: 40]
    -ct, --ClusterFinderThreshold [default: 5]
```

```
    -cn, --ClusterFinderSuccessiveADC [default: 3]
    -cf, --ClusterFinderFitThreshold [default: 0.1]
Cluster End Refinment Parameters:
    -er, --ClusterEndRefinmentOff [default: On]
    -el, --ClusterEndRefinmentLowThreshold [default: 6]
    -eh, --ClusterEndRefinmentHighThreshold [default: 2]
    -em, --ClusterEndRefinmentMaxCorrection [default: 10]
Moving Average Parameters:
    -ms, --MovingAverageStart [default: ClusterFinderStartPos]
    -ml, --MovingAverageLeft [default: 3]
    -mr, --MovingAverageRight [default: 4]
    -md, --MovingAverageDirection [default: 1]
Pulse Acceptance Parameters:
    -ps, --PulseAcceptanceWindowStart [default: 50]
    -pe, --PulseAcceptanceWindowEnd [default: 200]
    -pm, --PulseMinADC [default: 200]
    -px, --PulseMaxAquisitionTimebin [default: 512]
Negative Signal Filter:
    -no, --NegativeSignalFilterOn [default: On]
    -nf, --NegativeSignalFilterOff [default: On]
    -ns, --NegativeSignalFilterStart [default: DynamicBaselineStartPos]
    -ne, --NegativeSignalFilterEnd [default: DynamicBaselineEndPos]
    -nt, --NegativeSignalFilterThreshold [default: 30]
    -na, --NegativeSignalFilterAllowedSamples [default: 2]
ALTRO++ Parameters:
    -ao, --Altro++On [default: Off]
    -as, --Altro++StartPosition [default: 40]
    -a1, --Altro++BCS1On [default: On]
    -at, --Altro++TCFOn [default: Off]
    -a2, --Altro++BCS2On [default: On]
    -ac, --Altro++ClippingOn [default: Off]
    -a1h, --Altro++BCS1HighThreshold [default: 5]
    -a1l, --Altro++BCS1LowThreshold [default: 5]
    -a1e, --Altro++BCS1PreSamples [default: 2]
    -a1o, --Altro++BCS1PostSamples [default: 2]
    -a1O, --Altro++BCS1Offset [default: 0]
Signal Debugging Checks and Parameters::
    -sb --StrangeBaselineCheck [default Off]
    -sbt --StrangeBaselineCheckThreshold [default : 1]
    -scp --SaveCompletePulse [default off]
example: ./dateTail.app -t DATEFile -rn 820 -rp /Volumes/Daten/TestBeam/ -n 200
```

# Acronyms

# Acronyms

0-9:

**0x:** Prefix for hexadecimal numbers in C/C++

A:

**AC:** Analog Current
**ADC:** Analog Digital Converter
**AGS:** Alternating Gradient Synchrotron
**ALEPH:** Apparatus for LEP Physics
**ALICE:** A Large Ion Collider Experiment
**ALTRO:** ALICE TPC readout

B:

**BC:** Board Controller
**BCS1:** Baseline Correction and Subtraction 1
**BCS2:** Baseline Correction and Subtraction 2
**BNL:** Brookhaven National Labs

C:

**CASTOR:** CERN Advances Storage
**CCL:** Common Control Logic
**CERN:** Conseil Européen pour la Recherche Nucléaire
**CHRDO:** Channel readout
**CINT:** C Interpreter
**CMOS:** Complementary Metal Oxide Semiconductor
**CPU:** Central Processing Unit
**CSA:** Charge Sensitive Amplifier

D:

**DAQ:** Data Acquisition
**DATE:** Date Acquisition Test Environment
**DC:** Digital Current
**DCS:** Detector Control System
**DDL:** Detector Data Link
**DFU:** Data Formatting Unit
**DIM:** Distributed Information Management
**DIU:** Destination Interface Unit

E:

**ENC:** Equivalent Noise Charge
**EPS:** Encapsulated Postscript

F:

**f(t):** LuT data
**FCB:** Front end Control Bus
**FEC:** Front End Card
**FeC2:** Front End Control and Configuration
**FEE:** Front End Electronics
**FMD:** Forword Multiplicity Detector
**fpd:** Fixed Pedestal Data
**FPGA:** Field Programmable Gate Array
**FWHM:** Full Width Half Maximum

G

**GDC:** Global Data Concentrator
**GIF:** Graphics Interchange Format
**GTL:** Gunning Transceiver Logic
**GUI:** Graphical User Interface

H:

**HBT:** Hanbury-Brown Twiss
**HLT:** High Level Trigger
**HMPID:** High Momentum Particle Identification Detector

I:

**$I^2C$:** Inter-IC
**IC:** Integrated Circuit
**IIR:** Infinite Impulse Response
**IROC:** Inner Readout Chamber
**ISBN:** International Standard Book Number
**ITS:** Inner Tracking System

L:

**L0:** Level 0 Trigger
**L1:** Level 1 Trigger
**L2:** Level 2 Trigger
**LBL:** Lawrence Berkeley Labs
**LDC:** Local Data Concentrator
**LEP:** Large Electron Positron Collider
**LHC:** Large Hadron Collider
**LHCC:** LHC Committee
**LuT:** Look up Table
**LVCMOS:** Low Voltage CMOS

M:

**MBZ:** Must Be Zero
**MEB:** Multi Event Buffer
**MIP.** Minimum Ionising Particle
**MSPS:** Million Samples Per Second
**MWPC:** Multi Wire Proportional Chamber

O:

**OpenGL:** Open Graphics Library
**OROC:** Outer Readout Chamber
**OS:** Operating System

P:

**PASA:** Preamplifier/Shaper
**PCB:** Printed Circuit Board
**PCI:** Peripheral Component Interconnect
**PHOS:** Photon Spectrometer
**PMD:** Photon Multiplicity Detector
**PPR:** Physics Performance Report
**PS:** Postscritp
**PS:** Proton Synchrotron

Q:

**QGP** Quark Gluon Plasma

R:

**RCU:** **R**eadout **C**ontrol **U**nit
**RHIC:** **R**elativistic **H**eavy **I**on **C**ollider
**RMS:** **R**oot **M**ean **S**quare
**ROC:** **R**ead**o**ut **C**hamber
**ROI:** **R**egion **o**f **I**nterest
**RORC:** **R**ead**o**ut **R**eceiver **C**ard

S:

**SIU:** **S**ource **I**nterface **U**nit
**SO:** **S**hared **O**bject
**SPS:** **S**uper **P**roton **S**ynchrotron
**STAR:** **S**olenoidal **T**racker **at** **R**HIC
**SVG:** **S**calable **V**ector **G**raphics

T:

**TCF:** **T**ail **C**ancellation **F**ilter
**TDR:** **T**echnical **D**esign **R**eport
**TOF:** **T**ime **o**f **F**light
**TPC:** **T**ime **P**rojection **C**hamber
**TRD:** **T**ransition **R**adiation **D**etector
**TTCRX:** **TT**: Trigger and **C**: Control and **Rx**: Receiver

Z:

**ZDC:** **Z**ero **D**egree **C**alorimeters
**ZSU:** **Z**ero **S**uppression **U**nit